\documentclass{eptcs}

\usepackage[utf8]{inputenc}
\usepackage[english]{babel}

\usepackage{mathtools}
\usepackage{amsmath}
\usepackage{amsfonts}
\usepackage{amssymb}
\usepackage{bussproofs}

\usepackage{thmtools}

\usepackage{appendix}

\usepackage{enumitem}
\usepackage{array}

\usepackage[font=footnotesize]{subcaption}

\usepackage{tikz}
\usepackage{color}
\hypersetup{hidelinks}
\usepackage{doi}
\usepackage{latexsym}

\usepackage{style}
\usepackage{shortcuts}

\usepackage[normalem]{ulem}

\title{Axiomatising Infinitary Probabilistic Weak Bisimilarity\linebreak[3] of Finite-State Behaviours}
\def\titlerunning{Axiomatising Infinitary Probabilistic Weak Bisimilarity of Finite-State Behaviours}
\author{Nick Fischer
\institute{Universität des Saarlandes, Saarbrücken, Germany}
\and
Rob van Glabbeek
\institute{Data61, CSIRO, Sydney, Australia}
\institute{University of New South Wales, Sydney, Australia}
}
\def\authorrunning{N. Fischer \& R.J. van Glabbeek}
\providecommand{\publicationstatus}{To appear in JLAMP, doi:\href{https://doi.org/10.1016/j.jlamp.2018.09.006}{10.1016/j.jlamp.2018.09.006}}

\begin{document}
\maketitle

\begin{abstract}
In concurrency theory, weak bisimilarity is often used to relate processes exhibiting the same observable behaviour. The probabilistic environment gives rise to several generalisations; we study the infinitary semantics, which abstracts from a potentially unbounded number of internal actions being performed when something observable happens. Arguing that this notion yields the most desirable properties, we provide a sound and complete axiomatisation capturing its essence. Previous research has failed to achieve completeness in the presence of unguarded recursion, as only the finitary variant has been axiomatised, yet.
\\[2ex]
\end{abstract}

\section{Introduction} \label{sec:introduction}

\noindent
A major branch of concurrency theory is concerned with categorising semantic resemblances of processes modelled as labelled transition systems, by studying the interconnections and differences of a diverse set of  equivalence relations on processes. In his papers on \emph{the linear time -- branching time spectrum} \cite{vanGlabbeek1990,vanGlabbeek1993a}, the second author arranged many relevant equivalences in a lattice structure: A `horizontal' dimension of the lattice of \cite{vanGlabbeek1993a} distinguishes the relevant relations according to divergence sensitivity while the `vertical' dimension orders them with respect to their discriminating power concerning the processes' branching structure, ranging from branching bisimilarity as the finest relation to trace equivalences. Several of these relations, especially bisimilarities, have been captured axiomatically~\cite{vanGlabbeek1993b,Lohrey2002}, rooted in the seminal work of Robin Milner~\cite{Milner1989b} in the setting of a process language involving recursion.

In this work, we consider a more general setting involving probabilities, so as to be able to reason about randomised concurrent systems. The linear time -- branching time spectrum carries over; however, in this more general context it makes sense to introduce another  dimension into the spectrum. As depicted in \autoref{fig:probabilistic-spectrum}, each equivalence comes in four different flavours: On the one hand, we distinguish between \emph{convex} ($\convex$) and non-convex relations and, on the other hand, we discriminate \emph{finitary} ($\finitary$) and \emph{infinitary} ($\infinitary$) probabilistic behaviours. The convexity property is thoroughly investigated and is mentioned in various axiomatisations~\cite{Bandini2001,Deng2007}, so far exclusively interwoven with the finitary probabilistic character, while the infinitary version has not been axiomatised yet, to the best of our knowledge.

\begin{figure}[b!]
\begin{minipage}[b]{.48\textwidth}
\begin{center}
\begin{tikzpicture}[thick, rounded corners=.01cm, line cap=round]
\def\alpha{16}
\def\a{3cm}
\def\b{.6cm}
\def\c{.5cm}

\draw[name path=border, fill=black!3] (0, 0) coordinate (0) -- ++(-\alpha:\a) coordinate (1) -- ++(-90 + \alpha:\a) coordinate (2) -- ++(180 - \alpha:\a) coordinate (3) -- cycle;

\node (r0) at ($(0) + (-\alpha:\b) + (-90 + \alpha:\c)$) {${{\wbis}^\finitary}$};
\node (r1) at ($(1) + (180 - \alpha:\b) + (-90 + \alpha:\c)$) {${{\wbis}^\finitary_\convex}$};
\node (r2) at ($(2) + (180 - \alpha:\b) + (90 + \alpha:\c)$) {${{\wbis}^\infinitary_\convex}$};
\node (r3) at ($(3) + (-\alpha:\b) + (90 + \alpha:\c)$) {${{\wbis}^\infinitary}$};
\draw (r0) -- (r1) -- (r2) -- (r3) -- (r0);

\def\d{3.2cm}
\def\e{2.4cm}
\def\beta{10}
\def\gamma{84}

\coordinate (m) at ($(0)!.5!(2)$);
\path[draw, thick, ->, >=latex] (m) -- ($(m) + (\beta:\d)$) node[above left=0cm, font=\footnotesize, align=center] {divergence\\sensitivity};
\path[name path=ds] (m) -- ($(m) + (\beta:-\e)$);
\draw[name intersections={of=border and ds}, dash pattern={on 1.2pt off 3.2pt}] (m) -- (intersection-1);
\draw[name intersections={of=border and ds}, ->, >=latex] (intersection-1) -- ($(m) + (\beta:-\e)$);

\coordinate (m) at ($(0)!.5!(2)$);
\path[draw, thick, ->, >=latex] (m) -- ($(m) + (\gamma:\d)$) node[below left=.1cm, font=\footnotesize, align=center] {branching\\structure};
\path[name path=bs] (m) -- ($(m) + (\gamma:-\e)$);
\draw[name intersections={of=border and bs}, dash pattern={on 1.2pt off 3.2pt}] (m) -- (intersection-1);
\draw[name intersections={of=border and bs}, ->, >=latex] (intersection-1) -- ($(m) + (\gamma:-\e)$);
\end{tikzpicture}
\end{center}
\caption{The probabilistic dimension of the branching time spectrum. Each relation is included in all connected, lower positioned ones.} \label{fig:probabilistic-spectrum}
\end{minipage}
\hfill
\begin{minipage}[b]{.48\textwidth}
\begin{center}
\begin{tikzpicture}[transition diagram, state/.append style={node distance=1.4cm, on grid}]
\node[initial, state] (0) {};
\node[state, below=1.8cm of 0] (1) {};
\node[state, below=of 1] (2) {};

\node[splitter] (*) at ($(0)!.5!(1)$) {};
\path (0) edge["$\tau$", auto=right] (*);
\path (*) edge[transition, bend right=40, "$\frac 1 2$", auto=right] (0);
\path (*) edge[transition, "$\frac 1 2$", auto=right] (1);

\path (1) edge[transition, "$a$", auto=right] (2);

\node at ($(0) + (1.2cm, -.5cm)$) {${}^{\phantom\infinitary}{\wbis^\infinitary}$};
\node at ($(1) + (1.2cm, .5cm)$) {${}^{\phantom\finitary}{{\not\wbis}^\finitary}$};

\node[initial, initial where=right, state] (0) at ($(0) + (2.4cm, 0)$) {};
\node[state, below=1.8cm of 0] (1) {};
\node[state, below=of 1] (2) {};

\path (0) edge[transition, "$\tau$"] (1);
\path (1) edge[transition, "$a$"] (2);
\end{tikzpicture}
\end{center}
\caption{Exemplary processes distinguishing\newline $\wbis^\infinitary$ and $\wbis^\finitary$.} \label{fig:finitary-vs-infinitary}
\end{minipage}
\end{figure}

The finitary and infinitary semantics arise from different definitions of the underlying weak transitions. As an example consider the two processes depicted in \autoref{fig:finitary-vs-infinitary}; $\tau$ denotes the silent action that is assumed to be invisible for external observers. The finitary semantics yields weak transitions that can only hide a finite number of $\tau$ actions. Here, after a bounded number of steps, the process on the right-hand side is guaranteed to perform an observable $a$ action, while the left-hand side still resides in its initial state with positive probability. Hence, the processes are distinguished by $\wbis^\finitary$. However, if we allow infinitary transitions there is no step bound and the probability of residing in the initial state of the left process, i.e.\ the probability of looping forever, becomes $0$. Consequently, the $a$ action happens almost surely as well and $\wbis^\infinitary$ relates both systems. 

The infinitary semantics is not to be confused with divergence sensitivity; both processes are said to be convergent. Even though we allow weak transitions of unbounded size, there essentially exists no diverging trace, i.e.\ an infinite sequence of $\tau$ transitions occurring with positive probability.

Both transition variants frequently occur in literature: Finitary weak transitions form the basis of \cite{Bandini2001,Deng2007,DengPP05}. Infinitary weak transitions have been proposed in different, (nearly) equivalent shapes, mainly defined by means of schedulers \cite{Segala1995,Cattani2002,Eisentraut2010,Hermanns2012}, derivations \cite{Deng2009} and transition trees \cite{Eisentraut2013}. Note that a distinction between finitary and infinitary behaviours only arises when dealing with recursion (or infinite processes). This insight may justify a simplified formalisation in some settings, as for instance in \cite{Bandini2001}.

In this paper, we focus on the most prominent equivalence of the linear time -- branching time spectrum, namely weak bisimilarity, $\wbis$. We argue that the convex infinitary relation $\wbis_\convex^\infinitary$, as the coarsest variant, yields the most natural and desirable semantics. We choose the convex form of weak bisimilarity, because the non-convex one either leads to a non-transitive bisimilarity, or to a notion that fails to be a congruence, or fails to satisfy some natural axioms. Additionally, the convex form is more in line with the way randomised algorithms resolve nondeterminism. We choose the infinitary form for two reasons: Firstly, systems involving lossy channels or Las Vegas algorithms that keep resending information until successfully delivered (conceptually similar to the left process of \autoref{fig:finitary-vs-infinitary}) are most commonly not intended to be differentiated from their respective deterministic versions. Secondly, the authors of \cite{DengPP05} have established an axiomatisation of $\wbis_\convex^\finitary$ that turns out to be incomplete for expressions containing unguarded variables. They even conjecture that the relation is undecidable, so that a finite axiom set cannot exist.

Even though decision \cite{Cattani2002,Hermanns2012} and minimisation \cite{Eisentraut2013} algorithms have been developed, there is no axiomatisation of infinitary probabilistic weak bisimilarity, yet. An axiomatisation is interesting from a theoretical point of view, since it allows a concise characterisation of the respective relation, here $\wbis_\convex^\infinitary$. We propose a conservative extension of Milner's foundational work \cite{Milner1989b}, generalising some laws appropriately and introducing a fresh axiom expressing the nature of infinitary semantics.

As to be expected, the main challenge turns out to be the proof of completeness of the axiom set. We follow Milner's proof steps closely, but encounter some additional obstacles: Several proofs by induction on the length or the number of weak transitions become infeasible. We manage to overcome this by exploiting some facts of Markov Decision Theory. Often we need to leap from induction to coinduction (at least conceptually), while the proof ideas remain similar.

This paper is organised as follows: \hyperref[sec:definitions]{Section~\ref*{sec:definitions}} introduces all necessary preliminary definitions, \hyperref[sec:axiomatisation]{section~\ref*{sec:axiomatisation}} presents the axiom set and the sections~\ref{sec:soundness} and \ref{sec:completeness} prove the axiomatisation sound and complete, respectively. \hyperref[sec:conclusion]{Section~\ref*{sec:conclusion}} concludes and gives an outlook on further work.

\section{Definitions} \label{sec:definitions}

\subsection{Probability theory} \label{sec:definitions:sec:probability-theory}

\noindent
A (discrete) \emph{subprobability distribution} $\mu$ over a set $S$ is a mapping $\mu: S \to \Real_{\geq 0}$, such that $\sum_{s \in S} \mu(s) \leq 1$. The \emph{support} $\Supp\mu$ of a subprobability distribution $\mu$ is a set containing an element $s \in S$ iff $\mu(s) > 0$ and the \emph{size} $|\mu| = \sum_{s \in S} \mu(s)$ of $\mu$ states the probability mass distributed over all supported elements. If $|\mu| = 1$, then we call $\mu$ a \emph{probability distribution}. We shall denote the set of all subprobability distributions (resp.\ probability distributions) over $S$ by $\SubDistr S$ (resp.\ $\Distr S$), which is usually ranged over by Greek letters $\mu, \nu, \gamma, \eta, \pi, \ldots$. Let $\emptydistr$ denote the \emph{empty distribution} supporting no elements and for $s \in S$, we define $\dirac s$ as a \emph{Dirac distribution}, assigning $1$ to $s$ and $0$ to all other elements. Occasionally, we write $\distr{s_i \mapsto p_i}$ to express a distribution assigning $p_i$ to $s_i$, for each $i$.

Moreover, we define some straightforward operations on (sub-)probability distributions: Let $\leq$ be introduced as a pointwise comparison of distributions, i.e.\ $\mu \leq \nu$, if $\mu(s) \leq \nu(s)$ for all $s \in S$. We find natural notions for addition and subtraction as well: For $\mu, \nu$ with $|\mu| + |\nu| \leq 1$, let $\mu + \nu$ denote the distribution assigning $\mu(s) + \nu(s)$ to $s \in S$ and given $\nu \leq \mu$, let $\mu - \nu$ be defined in the obvious way. Finally, for any scalar \raisebox{0pt}[0pt][0pt]{$0 \mathbin< \lambda \mathbin< \tfrac{1}{|\mu|}$}, $\lambda \mu$ denotes the pointwise rescaling of $\mu$.

When not considering probabilities, we argue about processes represented by transition systems. In our context, (simple) \emph{probabilistic automata} \cite{Segala1995} generalise this concept naturally and therefore form the semantics of the process algebra to be established. Our understanding of probabilistic automata agrees with the probabilistic transition systems of \cite{Deng2009}.

\begin{definition}[Probabilistic automaton]
Let $\Act$ be a fixed set of action labels. A \emph{probabilistic automaton} (PA) is a tuple $(S, \trans{}, s_0)$, where
\begin{itemize}
\item $S$ is a non-empty set of states,
\item ${\trans{}} \subseteq S \times \Act \times \Distr S$ is the transition relation,
\item $s_0 \in S$ is the initial state.
\end{itemize}
\end{definition}

\noindent
We will often use transition diagrams to describe probabilistic automata. The empty circles represent states, while the filled circles `split' a transition leading to multiple target states.

\subsection{Process algebra} \label{sec:definitions:process-algebra}

\noindent
Our process algebra extends the process algebra of \cite{Milner1989b} by a binary operator for probabilistic choice. Equivalently it extends the process algebra of \cite{Stark2000} with an operator $\nchoice$ for nondeterministic choice, and it extends the process algebra of \cite{Bandini2001} by a recursion operator, such that every finite-state system involving finitely many nondeterministic and probabilistic choices becomes expressible.
We are not considering a prefix operator with the same generality as in \cite{Deng2007}, since this operator leads to generalised probabilistic automata as described in \cite{Segala1995}.

\subsubsection{Syntax} \label{sec:definitions:sec:process-algebra:sec:syntax}

\noindent
We distinguish between two syntactic categories, namely \emph{nondeterministic expressions} (collected in $\Exp$ and ranged over by $E, F, \ldots$) and \emph{probabilistic expressions} (collected in $\PExp$ and ranged over by $P, Q, \ldots$). Let $\Var$ denote a set of process variables and let $\Act$ be a set of action labels, where $\tau \in \Act$ represents the internal action. Then the syntax is given by:
\begin{align*}
	E \Coloneqq{}
	&\nil					&&\text{(inaction)}\\
	&X						&&\text{(variable, $X \in \Var$)}\\
	&\alpha\prefix P		&&\text{(prefixing, $\alpha \in \Act$)}\\
	&\rec X E				&&\text{(recursion, $X \in \Var$)}\\
	&E \nchoice E,			&&\text{(nondeterministic choice)}\\
	P \Coloneqq{}
	&\psingle E				&&\text{(Dirac choice)}\\
	&P \pchoice p P.		&&\text{(probabilistic choice, $0 < p < 1$)}
\end{align*}

\noindent
The nondeterministic expressions stem from Milner's CCS~\cite{Milner1989a}. As usual, $\nil$ denotes the inactive process yielding no outgoing transitions. We use process variables in addition to the recursion operator to express arbitrary loops and the $\nchoice$ operator expresses choice. In contrast to \cite{Milner1989a}, a prefixing expression consists of an action $\alpha$ and a probabilistic expression, which may be read as a probability distribution over nondeterministic expressions. $\psingle E$ represents a Dirac distribution supporting $E$, while $P \pchoice p Q$ constructs a convex combination of the distributions corresponding to $P$ and $Q$. Here, to avoid corner cases, we require $0 < p < 1$. We sometimes abbreviate $\alpha\prefix \psingle E$ with $\alpha\prefix E$.

Hereinafter, we will often write $\sum_{i \in I} E_i$ as a notation for the nondeterministic sum $E_1 \nchoice \ldots \nchoice E_n$ for some index set $I = \{1, \ldots, n\}$. The set nature will be justified by the commutativity and associativity of $\nchoice$. Moreover, let $\psum_{i \in I} {p_i} P_i$, always assuming $\sum_{i \in I} p_i = 1$, denote $P_1 \pchoice{q_1} (P_2 \pchoice{q_2} (\ldots \pchoice{q_{n-1}} P_n) \ldots )$ for some appropriate values $q_i$, such that $p_i$ indicates the probability of choosing $P_i$, respectively, and let $\psum_{i \in I} {p_i} E_i$ denote $\psum_{i \in I} {p_i} \psingle{E_i}$.

An occurrence of a variable $X$ in an expression $E$ is called \emph{free} if it does not lay
within some subexpression $\rec X F$; we gather all variables occurring free in $E$ in $\Var(E)$. $E$ is called \emph{closed} if $\Var(E) = \emptyset$ and \emph{open} otherwise; let $\CExp \subset \Exp$ denote the set of all closed expressions. For expressions $\vec F = (F_1, \ldots, F_n)$ and variables $\vec X = (X_1, \ldots, X_n)$, $E \subst{\vec F}{\vec X}$ denotes the expression $E$ after simultaneously replacing each free occurrence of $X_i$ in $E$ with $F_i$, while renaming bound variables in $E$ as needed to avoid free variables in $\vec F$ to become bound in $E \subst{\vec F}{\vec X}$. Finally, let $\synid$ denote syntactical equality of expressions.

\subsubsection{Semantics} \label{sec:definitions:sec:process-algebra:sec:semantics}

\noindent
As usual, the semantics of an expression $E$ is given by the PA $(\Exp, {\trans{}}, E)$, where ${\trans{}} \subseteq \Exp \times \Act \times \Distr\Exp$ is the smallest relation satisfying the inference rules given in \autoref{tab:semantics}, where $\alpha$ ranges over $\Act$.

\begin{table}
\caption{Operational semantics.} \label{tab:semantics}
\hsep
\begin{center}
\AxiomC{$P \ptrans \mu$}
\RuleName{prefix}
\UnaryInfC{$\alpha\prefix P \trans\alpha \mu$}
\DisplayProof\\[1.8ex]
\AxiomC{$E \subst{\rec X E} X \trans\alpha \mu$}
\RuleName{rec}
\UnaryInfC{$\rec X E \trans\alpha \mu$}
\DisplayProof\\[1.8ex]
\hfill
\AxiomC{$E \trans\alpha \mu$\vphantom{$F \trans\alpha \nu$}}
\RuleName{choice-l}
\UnaryInfC{$E \nchoice F \trans\alpha \mu$\vphantom{$E \nchoice F \trans\alpha \nu$}}
\DisplayProof
\hfill
\AxiomC{$F \trans\alpha \nu$\vphantom{$E \trans\alpha \mu$}}
\RuleName{choice-r}
\UnaryInfC{$E \nchoice F \trans\alpha \nu$\vphantom{$E \nchoice F \trans\alpha \mu$}}
\DisplayProof
\hspace*{\fill}
\end{center}
\hsep
\begin{center}
\hfill
\AxiomC{\vphantom{$P \ptrans \mu$}\vphantom{$Q \ptrans \nu$}}
\RuleName{idle}
\UnaryInfC{$\psingle E \ptrans \dirac E$\vphantom{$P \pchoice p Q \ptrans p \cdot \mu + (1\mathord-p) \cdot \nu$}}
\DisplayProof
\hfill
\AxiomC{$P \ptrans \mu$}
		\AxiomC{$Q \ptrans \nu$}
\RuleName{pchoice}
\BinaryInfC{$P \pchoice p Q \ptrans p \cdot \mu + (1\mathord-p) \cdot \nu$\vphantom{$\psingle E \ptrans \dirac E$}}
\DisplayProof
\hspace*{\fill}
\end{center}
\hsep
\end{table}

The operational rules characterise the auxiliary relation ${\ptrans} \subseteq \PExp \times \Distr\Exp$ as follows: Given some probabilistic expression $P$, it points to the probability distribution over all nondeterministic expressions occurring in $P$.

\subsection{Weak transitions} \label{sec:definitions:sec:weak-transitions}

\noindent
The nonprobabilistic setting yields a single, obvious definition of weak transitions: $\wtrans{}$ is chosen to be the transitive, reflexive closure of $\trans\tau$, and $\wtrans\alpha$ denotes $\wtrans{}\trans\alpha\wtrans{}$.

However, the probabilistic case is somehow ambiguous, since one could think of multiple different generalisations. We are going to analyse two important properties before fixing our final understanding of weak transitions.

\subsubsection{Convexity} \label{sec:definitions:sec:weak-transitions:sec:convexity}

\noindent
First of all, we design $\wtrans{}_\convex$ to be \emph{convex}, meaning that whenever $E \wtrans{}_\convex \mu$ and $E \wtrans{}_\convex \nu$, then there is also a transition leading to every convex combination of $\mu$ and $\nu$: $E \wtrans{}_\convex p \cdot \mu + (1 - p) \cdot \nu$ for any $p \in [0, 1]$. The motivation behind this convexity property arises from the concept of schedulers: Essentially, any weak transition performs a run through a PA following the transitions chosen by some scheduler in every state. If we consider randomised schedulers, i.e.\ schedulers that may randomly choose a transition instead of deterministically fixing a decision, then the resulting weak transitions happen to be convex. Analogously, the non-convex variant corresponds to deterministic schedulers. It seems desirable to assume randomised schedulers, such that potentially randomised algorithms may resolve nondeterminism \cite{Segala1995}.

In \autoref{sec:definitions:sec:nonconvex-weak-bisimilarity} we will revisit this decision, and
try to define a notion of weak bisimilarity based on non-convex weak transitions. The technical
problems that ensue are another argument for using convex weak transitions.

Preliminarily, we introduce the \emph{convex lifting} of a relation ${\trans{}} \subseteq S \times \Act \times \Distr S$ to subdistributions, denoted by ${\liftconvex{\trans{}}} \subseteq \SubDistr S \times \Act \times \SubDistr S$, which is the smallest relation satisfying:
\begin{enumerate}[label=(\roman*)]
\item $s \trans\alpha \mu$ implies $\dirac s \liftconvex{\trans\alpha} \mu$, and
\item $\nu_i \liftconvex{\trans\alpha} \mu_i$ for all $i \in I$ implies $(\sum_{i \in I} p_i \nu_i) \liftconvex{\trans\alpha} (\sum_{i \in I} p_i \mu_i)$, for all $p_i \in \Real_{\geq 0}$ with $\sum_{i \in I} p_i \leq 1$.
\end{enumerate}
Sometimes, we refer to $\trans\alpha_\convex$ as \emph{combined} strong transitions.
\subsubsection{Finitary vs.\ Infinitary} \label{sec:definitions:sec:weak-transitions:sec:finitary-vs-infinitary}

\noindent
Next, we continue discriminating further between \emph{finitary} and \emph{infinitary} weak transitions. The difference is best explained by an example: Consider the probabilistic automaton depicted in \autoref{fig:infinitary}; it does not introduce nondeterministic choices. We raise the question, whether there exists an $a$-labelled weak transition leading from $s$ to $u$. Certainly, the PA would be required to move silently from $s$ to $t$ with probability $1$. We simulate a run through the automaton by the transition tree depicted on the right: Each internal node represents an intermediate state of the run and each leaf represents a state in which the transition stops. Now distinguish two cases:

\begin{table}
\caption{Inference rules for finitary weak transitions.} \label{tab:finitary}
\hsep
\begin{center}
\hfill
	\AxiomC{}
	\UnaryInfC{$s \wtrans{}_\convex^\finitary \dirac s$}
	\bottomAlignProof
	\DisplayProof
\hfill
	\AxiomC{$\dirac s \trans\tau_\convex \mu$}
	\UnaryInfC{$s \wtrans{}_\convex^\finitary \mu$}
	\bottomAlignProof
	\DisplayProof
\hfill
	\AxiomC{$s \wtrans{}_\convex^\finitary \distr{s_i \mapsto p_i}$}
			\AxiomC{$\forall i.\; s_i \wtrans{}_\convex^\finitary \mu_i$}
	\BinaryInfC{$s \wtrans{}_\convex^\finitary \sum_i p_i \mu_i$}
	\bottomAlignProof
	\DisplayProof
\hspace*{\fill}
\end{center}
\vspace{-1.5ex}
\hsep
\end{table}

If the transition tree is finite, i.e.\ of arbitrary, yet limited depth $k$, the desired transition cannot exist! Clearly, there is a remaining probability of $(\tfrac 1 2)^k > 0$ of still residing in the initial state $s$. This semantics corresponds exactly to finitary transitions $\wtrans{}_\convex^\finitary$, being defined inductively in \autoref{tab:finitary}. As long as the transition tree is of finite depth, it can be constructed by the inference rules of $\wtrans{}_\convex^\finitary$, but a transition $s \wtrans{}_\convex^\finitary \dirac t$ cannot exist.

\begin{figure}
\begin{center}
\begin{tikzpicture}[transition diagram, state/.append style={node distance=1.2cm, on grid}]
\node[initial, state] (0) {$s$};
\node[state, below=1.8cm of 0] (1) {$t$};
\node[state, below=of 1] (2) {$u$};

\node[splitter] (*) at ($(0)!.5!(1)$) {};
\path (0) edge["$\tau$", auto=right] (*);
\path (*) edge[transition, bend right=32, "$\frac 1 2$", auto=right] (0);
\path (*) edge[transition, "$\frac 1 2$", auto=right] (1);

\path (1) edge[transition, "$a$", auto=right] (2);
\end{tikzpicture}
\hspace{2.0cm}
\begin{tikzpicture}[transition diagram, state/.append style={node distance=1.2cm and 1.6cm, on grid}]
\node[initial, state] (0) {$s$};
\node[state, below right=of 0] (00) {$s$};
\node[state, below left=of 0] (01) {$t$};
\node[state, below right=of 00] (000) {$s$};
\node[state, below left=of 00] (001) {$t$};
\node[state, below=of 01] (012) {$u$};
\node[state, below right=of 000, draw=none] (0000) {};
\node[state, below left=of 000, draw=none] (0001) {};
\node[state, below=of 001] (0012) {$u$};

\node[splitter] (*0) at ($(0) + (0, -.7cm)$) {};
\path (0) edge["$\tau$"] (*0);
\path (*0) edge[transition, "$\frac 1 2$"] (00);
\path (*0) edge[transition, "$\frac 1 2$", auto=right] (01);

\node[splitter] (*00) at ($(00) + (0, -.7cm)$) {};
\path (00) edge["$\tau$"] (*00);
\path (*00) edge[transition, "$\frac 1 2$"] (000);
\path (*00) edge[transition, "$\frac 1 2$", auto=right] (001);
\path (01) edge[transition, "$a$", auto=right] (012);

\node[splitter] (*000) at ($(000) + (0, -.7cm)$) {};
\path (000) edge["$\tau$"] (*000);
\path (*000) edge[transition, "$\frac 1 2$"] (0000);
\path (*000) edge[transition, "$\frac 1 2$", auto=right] (0001);
\path (001) edge[transition, "$a$", auto=right] (0012);

\path let \p1 = ($(0001) - (*000)$) in node[rotate={atan(\y1/\x1)}] at (0001) {$\cdots$};
\path let \p1 = ($(0000) - (*000)$) in node[rotate={atan(\y1/\x1)}] at (0000) {$\cdots$};
\end{tikzpicture}
\end{center}
\caption{Visualises the exemplary PA on the left and the infinitary transition $s \wtrans{}_\convex^\infinitary \dirac t \trans{a}_\convex \dirac u$ as an infinite transition tree on the right.} \label{fig:infinitary}
\end{figure}

However, if we allow infinite trees, the probability of staying in $s$ becomes $\lim_{k \to \infty} (\frac 1 2)^k = 0$, i.e.\ with probability $1$ the transition leads to $t$: $s \wtrans{}_\convex^\infinitary \dirac t$. We continue defining $\wtrans{}_\convex^\infinitary$ formally, which requires more effort than for finitary transitions. The following construction is borrowed from \cite{Deng2009}.

\begin{definition}[Derivation]
A sequence $(\mu_i^\proc, \mu_i^\halt)_{i \in \Nat}$ of subdistributions $\mu_i^\proc,\allowbreak\mu_i^\halt$ is called a derivation, if $\mu_i^\proc \liftconvex{\trans\tau} \mu_{i+1}$ for all $i \geq 0$, where $\mu_i = \mu_i^\proc + \mu_i^\halt$.
\end{definition}

\noindent
Think of a derivation $(\mu_i^\proc, \mu_i^\halt)_{i \in \Nat}$ as an infinite sequence of $\tau$ transitions, which is allowed to stop partially after each step:
\[
\def\arraystretch{1.3}
\begin{array}{c@{\;\;}c@{\;\;}c@{\;\;}c@{\;\;}c@{\:\:}c@{\:\:}c}
	& & \mu_0 & = & \mu_0^\proc & + & \mu_0^\halt \\
	\mu_0^\proc & \trans\tau_\convex & \mu_1 & = & \mu_1^\proc & + & \mu_1^\halt \\
	\mu_1^\proc & \trans\tau_\convex & \mu_2 & = & \mu_2^\proc & + & \mu_2^\halt \\
	& & \vdots & & & & \\
\end{array}
\]
Each $\mu_i^\proc$ component continues moving silently, while each $\mu_i^\halt$ stops. So overall, the derivation stops in $\sum_{i \in \Nat} \mu_i^\halt$. We say there is an infinitary weak transition $\mu \wtrans{}_\convex^\infinitary \nu$ if there is a derivation $(\mu_i^\proc, \mu_i^\halt)_{i \in \Nat}$ with $\mu = \mu_0 = \mu_0^\proc + \mu_0^\halt$ and $\nu = \sum_{i \in \Nat} \mu_i^\halt$.

\subsubsection{Basic properties} \label{sec:definitions:sec:weak-transitions:sec:properties}

\noindent
As argued before, we focus on convex infinitary transitions from now on. Thus, we assign ${\wtrans{}} = {\wtrans{}_\convex^\infinitary}$ and discontinue explicitly annotating $\convex$, $\finitary$ and $\infinitary$. The distinction between $\trans\alpha$ and $\liftconvex{\trans\alpha}$ will be clear from the context.

For some $\alpha \in \Act$, let $\mu \wtrans\alpha \nu$ denote $\mu \wtrans{} \trans\alpha \wtrans{} \nu$ and sometimes, we write $s \wtrans\alpha \nu$ in order to express $\dirac s \wtrans\alpha \nu$. For some sequence $\sigma \in \Act^*$, $\hat\sigma$ shall denote the sequence $\sigma$ after removing every occurrence of $\tau$; we are often writing \raisebox{0pt}[0pt][0pt]{$\wtrans{\hat\alpha}$} to express a weak transition that allows idling if $\alpha = \tau$.

It is vital to assert some basic properties about $\wtrans{}$ as defined by derivations. Most of the results are established by \cite{Deng2009}; note that even seemingly easy statements have led to exhausting proofs.

\begin{proposition}[\cite{Deng2009}]
$\wtrans{}$ is convex.
\end{proposition}

\begin{proposition}[\cite{Deng2009}]
If $\mu \wtrans{} \nu$, then $|\mu| \geq |\nu|$.
\end{proposition}

\noindent
There are actually transitions $\mu \wtrans{} \nu$ with $|\mu| > |\nu|$, e.g.\ $\rec X \tau\prefix X \wtrans{} \emptydistr$ \cite{Deng2009}. One could understand such a transition to be \emph{divergent} with probability mass $|\mu| - |\nu|$ (or just \emph{divergent} if $\nu = \emptydistr$); if $|\mu| = |\nu|$ we call $\mu \wtrans{} \nu$ \emph{convergent}.

\begin{proposition}[Linearity and Decomposition of $\wtrans{}$, \cite{Deng2009}] \label{prp:linearity-decomposition}
For an index set $I$, let $p_i \in \Real_{\geq 0}$ for all $i \in I$ with $\sum_{i \in I} p_i \leq 1$. Then:
\begin{enumerate}[label={\normalfont(\roman*)}]
\item $\mu_i \wtrans{} \nu_i$ for all $i \in I$ implies $(\sum_{i \in I} p_i \mu_i) \wtrans{} (\sum_{i \in I} p_i \nu_i)$, \label{prp:linearity-decomposition:itm:linearity}
\item $(\sum_{i \in I} p_i \mu_i) \wtrans{} \nu$ implies $\nu = \sum_{i \in I} p_i \nu_i$ for some $\nu_i$, such that $\mu_i \wtrans{} \nu_i$ for each $i \in I$. \label{prp:linearity-decomposition:itm:decomposition}
\end{enumerate}
\end{proposition}

\begin{proposition}[Reflexivity and Transitivity of $\wtrans{}$, \cite{Deng2009}]\label{prp:preorder}
\hspace{0cm}
\begin{enumerate}[label={\normalfont(\roman*)}]
\item $\mu \wtrans{} \mu$,
\item $\mu \wtrans{} \gamma$ and $\gamma \wtrans{} \nu$ implies $\mu \wtrans{} \gamma$.
\end{enumerate}
\end{proposition}

\begin{definition}
A transition $\mu \wtrans{} \gamma$ is called an \emph{initial segment} of $\mu \wtrans{} \nu$, if there are derivations $(\nu_i^\proc, \nu_i^\halt)_{i \in \Nat}$ and $(\gamma_i^\proc, \gamma_i^\halt)_{i \in \Nat}$ justifying $\mu \wtrans{} \nu$ and $\mu \wtrans{} \gamma$, respectively, and satisfying for all $i$
\begin{align*}
	\gamma_i^\proc &\leq \nu_i^\proc, &
	\gamma_i &\leq \nu_i, &
	\nu_i^\proc - \gamma_i^\proc &\trans\tau \nu_{i+1} - \gamma_{i+1}.
\end{align*}
\end{definition}

\begin{proposition}[\cite{Deng2009}] \label{prp:initial-segment}
Let $\mu \wtrans{} \gamma$ be an initial segment of $\mu \wtrans{} \nu$. Then $\gamma \wtrans{} \nu$.
\end{proposition}

\noindent
We state some more propositions that turn out to be particularly helpful in the proofs to come.

\begin{proposition} \label{prp:wtrans-tau}
The following statements are equivalent:
\begin{enumerate}[label=\normalfont(\alph*)]
\item $\mu \wtrans\tau \nu$, \label{prp:wtrans-tau:itm:wtrans}
\item $\mu \trans\tau \wtrans{} \nu$, \label{prp:wtrans-tau:itm:trans}
\item there is a derivation $(\nu_i^\proc, \nu_i^\halt)_{i \in \Nat}$ with $\mu = \nu_0$ and $\nu = \sum_{i \in \Nat} \nu_i^\halt$, where $\nu_0^\halt = \emptydistr$. \label{prp:wtrans-tau:itm:derivation}
\end{enumerate}
\end{proposition}
\begin{proof}
We begin by proving \ref{prp:wtrans-tau:itm:wtrans} implies \ref{prp:wtrans-tau:itm:trans}, so assume that $\mu \wtrans\tau \nu$, i.e.\ $\mu \wtrans{} \gamma \trans\tau \eta \wtrans{} \nu$ for some $\gamma, \eta$. Hence, we obtain a derivation $(\gamma_i^\proc, \gamma_i^\halt)_{i \in \Nat}$ with $\mu = \gamma_0$ and $\gamma = \sum_{i \in \Nat} \gamma_i^\halt$. Next, $\sum_{i \in \Nat} \gamma_i^\halt \trans\tau \eta$ can be decomposed to find values $\eta_{i+1}^\halt$, such that $\gamma_i^\halt \trans\tau \eta_{i+1}^\halt$ for each $i$. One might easily check that $(\eta_i^\proc, \eta_i^\halt)_{i \in \Nat}$, with
\begin{align*}
	\eta_i^\proc &= \gamma_i^\proc + \gamma_i^\halt, & \eta_0^\halt &= \emptydistr,
\end{align*}
is a derivation justifying $\mu \wtrans{} \eta$. Moreover, $\mu = \eta_0^\proc \trans\tau \gamma_1$ is an initial segment, so we are given $\gamma_1 \wtrans{} \eta$ by \autoref{prp:initial-segment}. All in all, we have $\mu \trans\tau \gamma_1 \wtrans{} \eta \wtrans{} \nu$, thus, by transitivity we are finished.

The direction \ref{prp:wtrans-tau:itm:trans} implies \ref{prp:wtrans-tau:itm:wtrans} is immediate by reflexivity of $\wtrans{}$. Moreover, \ref{prp:wtrans-tau:itm:trans} and \ref{prp:wtrans-tau:itm:derivation} are easily proven equivalent by some rewriting.
\end{proof}

\begin{proposition} \label{prp:wtrans-decompose-tau}
Let $\mu \wtrans{} \nu$. Then there is some $\gamma$ with $\gamma \leq \mu$ and $\gamma \leq \nu$, such that $\mu - \gamma \wtrans\tau \nu - \gamma$.
\end{proposition}
\begin{proof}
We are given a derivation $(\mu_i^\proc, \mu_i^\halt)_{i \in \Nat}$ with $\mu = \mu_0^\proc + \mu_0^\halt$ and $\nu = \sum_{i \in \Nat} \mu_i^\halt$. Simply take $\gamma = \mu_0^\halt$, then obviously $\gamma \leq \mu$ and $\gamma \leq \nu$. Now think of the derivation from before, after assigning $\mu_0^\halt = \emptydistr$; \autoref{prp:wtrans-tau} yields $\mu - \gamma \wtrans\tau \nu - \gamma$.
\end{proof}

\begin{proposition} \label{prp:wtrans-weak-construction}
Let $(\mu_i^\proc, \mu_i^\halt)_{i \in \Nat}$ be a sequence with $\mu_i = \mu_i^\proc + \mu_i^\halt$ and $\mu_i^\proc \wtrans{} \mu_{i+1}$ for all $i$, where both $\mu_i^\proc$ and $\mu_i^\halt$ converge to $\emptydistr$. Then $\mu_0 \wtrans{} \sum_{i \in \Nat} \mu_i^\halt$.
\end{proposition}
\begin{proof}
Although the statement should hold in general, our context allows us to assume a finite state space
$S$, with finitely many transitions. Then \cite{Deng2009} shows that the set $\{ \nu \mid \mu_0 \wtrans{} \nu \}$ is closed. Hence, for an arbitrary convergent sequence $(\gamma_i)_{i \in \Nat}$: If $\mu_0 \wtrans{} \gamma_i$ for all $i$, then also $\mu_0 \wtrans{} \lim_{i \to \infty} \gamma_i$. So we choose $(\mu_i^\proc + \sum_{j \leq i} \mu_j^\halt)_{i \in \Nat}$: Obviously, we find $\mu_0 \wtrans{} \mu_i^\proc + \sum_{j \leq i} \mu_j^\halt$ for all $i$, exploiting reflexivity, transitivity and linearity of $\wtrans{}$. Moreover, $(\mu_i^\proc + \sum_{j \leq i} \mu_j^\halt)_{i \in \Nat}$ is convergent by assumption with limit $\lim_{i \in \Nat} (\mu_i^\proc + \sum_{j \leq i} \mu_j^\halt) = \sum_{i \in \Nat} \mu_i^\halt$, so indeed, $\mu_0 \wtrans{} \sum_{i \in \Nat} \mu_i^\halt$.
\end{proof}

\subsection{Weak bisimilarity and weak congruence} \label{sec:definitions:sec:weak-bisimilarity}

\noindent
Given a relation $\mathcal R \subseteq S \times S$, we \emph{lift} it to a relation ${\liftequ{\mathcal R}} \subseteq \SubDistr S \times \SubDistr S$ on subdistributions as follows \cite{Deng2009}: Let $\liftequ{\mathcal R}$ be the smallest relation satisfying
\begin{enumerate}[label=(\roman*)]
\item if $s \mathrel{\mathcal R} t$, then $\dirac s \liftequ{\mathcal R} \dirac t$, and
\item if $\mu_i \liftequ{\mathcal R} \nu_i$ for all $i \in I$, then $\sum_{i \in I} p_i \mu_i \liftequ{\mathcal R} \sum_{i \in I} p_i \nu_i$, for any $p_i \in \Real_{\geq 0}$ with $\sum_{i \in I} p_i \leq 1$.
\end{enumerate}

\begin{definition}[Weak bisimilarity]  \label{def:weak-bisimilarity}\rm
A relation $\mathcal R \subseteq \CExp \times \CExp$ is called a \emph{weak bisimulation} if, for every pair $(E, F) \in \mathcal R$, the following conditions hold:
\begin{enumerate}[label=(\roman*)]
\item whenever $E \trans\alpha \mu$, then there is some $\nu$, such that $F \wtrans{\hat\alpha} \nu$ and $\mu \liftequ{\mathcal R} \nu$,
\item whenever $F \trans\alpha \nu$, then there is some $\mu$, such that $E \wtrans{\hat\alpha} \mu$ and $\mu \liftequ{\mathcal R} \nu$.
\end{enumerate}
Two closed expressions $E, F$ are called \emph{weakly bisimilar}, denoted $E \wbis F$, if there is a weak bisimulation relating $E$ and $F$.
\end{definition}

\noindent
This definition matches Segala's weak probabilistic bisimilarity~\cite{Segala1995}.

It is well-known that $\wbis$ is not a congruence relation with respect to $\nchoice$. So instead, we prefer dealing with the coarsest congruence relation contained in $\wbis$, which is equivalently defined as follows:

\begin{definition}[Weak congruence] \label{def:weak-congruence}\rm
Two closed expressions $E, F$ are called \emph{weakly congruent}, denoted $E \wcong F$, if the following conditions hold:
\begin{enumerate}[label=(\roman*)]
\item whenever $E \trans\alpha \mu$, then there is some $\nu$, such that $F \wtrans\alpha \nu$ and $\mu \liftequ\wbis \nu$,
\item whenever $F \trans\alpha \nu$, then there is some $\mu$, such that $E \wtrans\alpha \mu$ and $\mu \liftequ\wbis \nu$.
\end{enumerate}
\end{definition}

Formally, weak bisimilarity and weak congruence only differ in their respective choice of weak transitions (\raisebox{0px}[0px][0px]{$\wtrans{\hat\alpha}$} vs.\ \raisebox{0px}[0px][0px]{$\wtrans{\alpha}$}). For $\alpha \neq \tau$ both characterisations agree, the difference is spotted in the case that $\alpha = \tau$: If $E \wcong F$ and \raisebox{0pt}[0pt][0pt]{$E \trans\tau \mu$}, $F$ is guaranteed to evolve into some distribution $\nu \wbis \mu$ by actually making progress ($F \wtrans\tau \nu$), while in the case of $E \wbis F$, the only guarantee is $F \wtrans{} \nu$, that is, $F$ could also idle.

From now on, we will no longer explicitly annotate $\liftequ{\mathcal R}$, when addressing the probabilistic lifting of some relation $\mathcal R$.

We proceed with some useful facts concerning weak bisimilarity and weak congruence.

\begin{lemma} \label{lem:wbis-wtrans}
Let $\nu \wbis \mu \trans{\alpha}_\convex \mu'$. Then there exists a $\nu'$, such that $\nu \wtrans{\hat\alpha} \nu' \wbis \mu'$.
\end{lemma}
\begin{proof}
By the definition of convex lifting $\mu$ and $\mu'$ must have the forms $\sum_{i \in I} p_i \dirac{E_i}$ and $\sum_{i \in I} p_i \mu'_i$ for certain $p_i \in \Real_{\geq 0}$ with $\sum_{i \in I} p_i \leq 1$, such that $E_i \trans{\alpha} \mu'_i$ for all $i$. The definition of the lifting of $\wbis$ from nondeterministic expressions to subdistributions implies that $\nu$ has the form $\sum_{i \in I} p_i \nu_i$ with $\nu_i \wbis \dirac{E_i}$ for all $i$. This, in turn, implies that each $\nu_i$ has the form $\sum_{j \in J_i} q_{i,j} \dirac{F_{i,j}}$ with $\sum_{j \in J_i} q_{i,j} = p_i$ and $F_{i,j} \wbis E_i$. By \autoref{def:weak-bisimilarity} there are $\nu'_{i,j}$ for all $i$ and $j$ such that $F_{i,j} \wtrans{\hat\alpha} \nu'_{i,j}$ and $\mu'_i \wbis \nu'_{i,j}$. In case $\alpha=\tau$ we have $F_{ij} \wtrans{} \nu'_{i,j}$  for all $i$ and $j$, so $\nu \synid \sum_{i \in I,\,j\in J_i} q_{i,j} F_{i,j} \wtrans{} \nu' :\synid \sum_{i \in I,\,j\in J_i} q_{i,j} \nu'_{i,j}$ by \autoref{prp:linearity-decomposition}\ref{prp:linearity-decomposition:itm:linearity}. Using that $\mu' \synid \sum_{i \in I,\,j\in J_i} q_{i,j} \mu_{i}$, the lifting of $\wbis$ yields $\mu'\wbis \nu'$. In case $\alpha\neq\tau$ we have $F_{i,j} \wtrans{} \trans{\alpha}\wtrans{} \nu'_{i,j}$ for all $i$ and $j$, and we reach the same conclusion in three steps.
\end{proof}
In the exact same way one obtains:
\begin{lemma}\label{lem:wbis-cong-wtrans}
  Let $\nu \wcong \mu \trans{\alpha}_\convex \mu'$.
  Then there exists a $\nu'$, such that $\nu \wtrans{\alpha} \nu' \wbis \mu'$.
\end{lemma}

\noindent
Visualise the next proposition as the following completable diagram. A similar statement holds for weak congruence, as well.
\begin{center}
\begin{tikzpicture}[relation diagram, baseline={([yshift=-.8ex]current bounding box.center)}]
\matrix[relation matrix, row sep={1.2cm,between origins}, column sep={1.6cm,between origins}, font=\footnotesize] (m)
{
	\mu & \nu \\
	\mu' & \\
};

\path (m-1-1.center) edge[weak transition, "$\hat\alpha$", auto=right] (m-2-1.center);

\node at ($(m-1-1.east)!.5!(m-1-2.west)$) {${\wbis}$};
\end{tikzpicture}
\hspace{1.2cm}implies\hspace{1.2cm}
\begin{tikzpicture}[relation diagram, baseline={([yshift=-.8ex]current bounding box.center)}]
\matrix[relation matrix, row sep={1.2cm,between origins}, column sep={1.6cm,between origins}, font=\footnotesize] (m)
{
	\mu & \nu \\
	\mu' & \nu' \\
};

\path (m-1-1.center) edge[weak transition, "$\hat\alpha$", auto=right] (m-2-1.center);
\path (m-1-2.center) edge[weak transition, "$\hat\alpha$"] (m-2-2.center);

\node at ($(m-1-1.east)!.5!(m-1-2.west)$) {${\wbis}$};
\node at ($(m-2-1.east)!.5!(m-2-2.west)$) {${\wbis}$};
\end{tikzpicture}
\end{center}

\begin{proposition} \label{prp:wbis-wtrans}
Let $\nu \wbis \mu \wtrans{\hat\alpha} \mu'$ be convergent. Then there exists some $\nu'$, such that $\nu \wtrans{\hat\alpha} \nu' \wbis \mu'$.
\end{proposition}
\begin{proof}
We find it convenient to first prove the statement for the case $\alpha = \tau$. Now $\mu \wtrans{} \mu'$ yields a derivation $(\mu_i^\proc, \mu_i^\halt)_{i \in \Nat}$ with $\mu \synid \mu_0$ and $\mu' \synid \sum_{i \in \Nat} \mu_i^\halt$. The sequence $(\nu_i^\proc, \nu_i^\halt)_{i \in \Nat}$ is constructed inductively: Initially, note that $\nu \wbis \mu \synid \mu_0^\proc + \mu_0^\halt$, so we split $\nu \synid \nu_0 \synid \nu_0^\proc + \nu_0^\halt$ appropriately, such that $\mu_i^\proc \wbis \nu_i^\proc$ and $\mu_i^\halt \wbis \nu_i^\halt$ for $i = 0$; from now on we ensure that this invariant keeps maintained for all $i$ that have already been considered. So let $i \geq 0$, assuming $\nu_i^\proc \wbis \mu_i^\proc \trans\tau \mu_{i+1}$, thus, $\nu_i^\proc \wtrans{} \nu_{i+1} \wbis \mu_{i+1}$. It remains to split $\nu_{i+1}$ as before. Finally, $\mu \wtrans{} \mu'$ being convergent allows the application of \autoref{prp:wtrans-weak-construction}, which implies $\nu \wtrans{} \sum_{i \in \Nat} \nu_i^\halt \wbis \sum_{i \in \Nat} \mu_i^\halt \synid \mu'$.

The case $\alpha \neq \tau$ is now straightforward: We are given $\nu \wbis \mu \wtrans{} \trans\alpha \wtrans{} \mu'$, which may be simulated by $\nu \wtrans{} \wtrans\alpha \wtrans{} \nu' \wbis \mu'$, exploiting \autoref{lem:wbis-wtrans} and transitivity of $\wtrans{}$ and the previous result only involving $\tau$ actions. 
\end{proof}

\noindent
The following example shows that in \autoref{prp:wbis-wtrans} the condition that $\mu \wtrans{\hat\alpha} \mu'$ be convergent cannot be skipped.

\begin{example}\rm
Consider two expressions $E \synid \rec X \tau\prefix X$ and $F \synid \nil$. Then clearly $\dirac{E} \approx \dirac{F}$. We find $\dirac{E} \wtrans{} \frac{1}{2} \cdot \dirac{E}$, but, using that $\mu \wbis \nu$ implies $|\mu| = |\nu|$, there is no matching transition from $\dirac{F}$. In fact, the only weak transition leaving $\dirac{F}$ is $\dirac{F} \wtrans{} \dirac{F}$.
\end{example}

\begin{proposition} \label{prp:wcong-wtrans}
Let $\nu \wcong \mu \wtrans\alpha \mu'$ be convergent. Then there exists some $\nu'$, such that $\nu \wtrans\alpha \nu' \wbis \mu'$.
\end{proposition}
\begin{proof}
The case $\alpha \neq \tau$ is immediately solved by
\autoref{prp:wbis-wtrans}, so let $\alpha = \tau$. $\mu \wtrans\tau
\mu'$ may be rewritten to $\nu \wcong \mu \trans\tau \mu'' \wtrans{}
\mu'$, by \autoref{prp:wtrans-tau}. Now $\nu \wtrans\tau \nu'' \wbis
\mu''$ by \autoref{lem:wbis-cong-wtrans} and it suffices to apply \autoref{prp:wbis-wtrans}, again.
\end{proof}

\noindent
Note that we have defined $\wbis$ as well as $\wcong$ only for closed
expressions. Both relations $\mathcal R \mathbin\in \{\wbis, \wcong\}$
are lifted to general (non-closed) expressions $E, F$ with free
variables $\vec X$ by requiring $E \subst{\vec G}{\vec X}
\mathrel{\mathcal R} F \subst{\vec G}{\vec X}$ for all closed
expressions $\vec G$.\!\!\!

\subsection{Non-convex weak bisimilarity} \label{sec:definitions:sec:nonconvex-weak-bisimilarity}

\noindent
Let us further motivate why we only focus on an axiomatisation of the convex weak bisimilarity.

A non-convex version $\wtrans{}^\finitary$ of the finitary weak transition relation can be defined as in \autoref{tab:finitary}, but changing the second rule into
\begin{center}
	\AxiomC{$s \trans\tau \mu$}
	\UnaryInfC{$s \wtrans{}^\finitary \mu$}
	\bottomAlignProof
	\DisplayProof;
\end{center}
this yields the non-convex finitary transition relation proposed in \cite{DengPP05}. The non-convex version of the infinitary transitions is then defined in such a way that for finite (recursion-free) processes it coincides with the finitary version.

Suppose we would define non-convex weak bisimilarity exactly as in \autoref{def:weak-bisimilarity}, but using the (finitary or infinitary) non-convex weak transition relation instead of the infinitary convex one. Then, as documented in \cite{DengPP05}, the resulting notion fails to be transitive.\footnote{The counterexample appears in the technical report version of \cite{DengPP05} but is suppressed in the published version.} The following is a mild simplification of the counterexample of \cite{DengPP05}.
\begin{example}\rm
Let
\begin{align*}
	R &\synid \tau\prefix a\prefix \nil \nchoice \tau\prefix b\prefix \nil, \\
	E &\synid \tau\prefix (\psingle R \pchoice{1/2} \psingle{R \nchoice \nil}) + \tau\prefix (\psingle{a\prefix\nil} \pchoice{1/2} \psingle{b\prefix\nil}), \\
	F &\synid \tau\prefix (\psingle R \pchoice{1/2} \psingle{R \nchoice \nil}), \\
	G &\synid \tau\prefix R.
\end{align*}
A crucial difference between convex and non-convex finitary transitions is that $R \wtrans{}_\convex^\finitary \mu$ whereas $R \not\wtrans{}^\finitary\mu$, for $\mu = \distr{a\prefix \nil \mapsto \frac 1 2, b\prefix \nil \mapsto \frac 1 2}$. However, since $R \wtrans{}^\finitary a\prefix\nil$ and $R \nchoice \nil \wtrans{}^\finitary b\prefix \nil$ it follows that $\tau\prefix (\psingle R \pchoice{1/2} \psingle{R \nchoice \nil}) \wtrans{}^\finitary \mu$. Using this, it is easy to see that $E$ is non-convex weakly bisimilar to $F$. Furthermore, since $R$ is non-convex (weakly) bisimilar to $R \nchoice \nil$, it follows immediately from the definition of lifting that $\distr{R \mapsto \frac 1 2, R \nchoice \nil \mapsto \frac 1 2}$ is non-convex (weakly) bisimilar to $\dirac R$. Hence also $F$ and $G$ are non-convex weakly (even strongly) bisimilar. Yet, $E$ and $G$ are \emph{not} non-convex weakly bisimilar, for $E \wtrans{}^\finitary \mu$, and this transition cannot be matched by $G$ (or $R$).
\end{example}
One might wish to give a different definition of non-convex weak bisimilarity that avoids this problem. The only reasonable way appears to make $F$ and $G$ non-bisimilar. This can be done by reformulating the condition of \autoref{def:weak-bisimilarity}:
\begin{enumerate}[label={}]
\item whenever $E \wtrans\alpha \mu$, then there is some $\nu$, such that $F \wtrans{\hat\alpha} \nu$ and $\mu \liftequ{\mathcal R} \nu$.
\end{enumerate}
The price for this would be, however, that non-convex weak bisimilarity does not include non-convex strong bisimilarity as defined in \cite{Segala1995}, and moreover, either axioms \axiom{N4} or \axiom{P3} of \autoref{sec:axiomatisation}, or the congruence property, needs to be given up.

\subsection{Guardedness} \label{sec:definitions:sec:guardedness}

\noindent
Milner's complete axiomatisation of CCS~\cite{Milner1989b} crucially depends on the \emph{guardedness} property of CCS expressions. He defines $X$ to be unguarded in $E$ if there is some occurrence of $X$ not being in the scope of some $\alpha$-prefixing with $\alpha \neq \tau$, and guarded otherwise. We refer to this definition as \emph{nonprobabilistic guardedness}. However, this notion turns out to be too restrictive in our setting, so instead we define a probabilistic lifting:

\begin{table}
\caption{Inference rules for probabilistic unguardedness; here $V, W \subset \Var$.} \label{tab:guardedness}
\hsep
\begin{center}
\hfill
	\AxiomC{\vphantom{$E \unguarded V$}}
	\UnaryInfC{$X \unguarded \{X\}$}
	\DisplayProof
\hfill
	\AxiomC{$P \unguarded V$}
	\UnaryInfC{$\tau\prefix P \unguarded V$}
	\DisplayProof
\hfill
	\AxiomC{$E \unguarded V$}
	\RightLabel{$V \neq \{X\}$}
	\UnaryInfC{$\rec X E \unguarded V \setminus \{X\}$}
	\DisplayProof
\hspace*{\fill}\null\\[1.8ex]
\hfill
	\AxiomC{$E \unguarded V$}
	\UnaryInfC{$E \nchoice F \unguarded V$}
	\DisplayProof
\hfill
	\AxiomC{$F \unguarded W$}
	\UnaryInfC{$E \nchoice F \unguarded W$}
	\DisplayProof
\hspace*{\fill}
\end{center}
\hsep
\begin{center}
\hfill
	\AxiomC{$E \unguarded V$}
	\UnaryInfC{$\psingle E \unguarded V$\vphantom{$P \pchoice p Q \unguarded V \union W$}}
	\DisplayProof
\hfill
	\AxiomC{$P \unguarded V$}
			\AxiomC{$Q \unguarded W$}
	\BinaryInfC{$P \pchoice p Q \unguarded V \union W$\vphantom{$\psingle E \unguarded V$}}
	\DisplayProof
\hspace*{\fill}
\end{center}
\hsep
\end{table}

\begin{definition}[Guardedness]\label{def:guardedness}\rm
A free variable $X$ is called \emph{(probabilistically) unguarded} in $E$, denoted $E \unguarded X$, if $E \unguarded \{X\}$ according to \autoref{tab:guardedness}, and \emph{(probabilistically) guarded} otherwise. An expression $E$ is called (probabilistically) guarded, if for every recursion $\rec X F$, occurring as a subterm of $E$, $X$ is guarded in $F$.
\end{definition}

\noindent
When considering the fragment of our process algebra without the operator $\pchoice{p}$, one has $E \unguarded V$ only when $V$ is a singleton. In that case $E \unguarded \{X\}$ holds exactly when $X$ occurs free and unguarded in $E$ as defined by Milner \cite{Milner1989b,Milner1989a}, and our definition yields the same notion of a guarded expression as used in CCS.

To understand \autoref{def:guardedness} better, call an occurrence of a variable $X$ in a nondeterministic expression $E$ \emph{strongly unguarded} if it is free and does not occur within a subexpression $\alpha\prefix P$, including the case that $\alpha = \tau$.  Now $E \unguarded V$, for $V$ a set of variables, guarantees that, by executing a weak transition, $E$ can evolve into a probability distribution $\mu$, such that, for each expression $F \in \Supp{\mu}$, one of the variables in $V$ occurs strongly unguarded in $F$. In case $V$ has the form $\{X\}$ this means that $X$ occurs strongly unguarded in each expression $F \in \Supp{\mu}$. Here we mention this for the sake of intuition only; in \autoref{apx:soundness-r2} we will make this more precise and use it to obtain the soundness of our only axiom that depends on guardedness.

\begin{example}\rm
Consider the expression
\[
	E \synid \tau\prefix (\psingle X \pchoice{1/3} \psingle{Y \nchoice a\prefix \nil}).
\]
It holds that $X$ is probabilistically guarded in $E$, $E \not\unguarded \{X\}$, although $X$ occurs nonprobabilistically unguarded in $E$! However, we have $E \unguarded \{X, Y\}$. This confirms the explanation above, as there is a silent move from $E$ to the distribution $\distr{X \mapsto \frac 1 3, Y \nchoice a\prefix \nil \mapsto \frac 2 3}$. From $E \unguarded \{X, Y\}$, infer that also $\rec Y E \unguarded \{X\}$. The correspondence still holds: $\rec Y E \wtrans{} \dirac X$.
\end{example}

\section{Axiomatisation} \label{sec:axiomatisation}

\begin{table}[t]
\caption{Sound and complete axioms for $\wcong$.} \label{tab:axioms}

\newcommand\tstrut{\noalign{\vspace{.9ex}}}
\newcommand\bstrut{\noalign{\vspace{1ex}}}

\setlength\extrarowheight{.8ex}
\setlength\arrayrulewidth{1pt}
\begin{tabular*}{\textwidth}{>{\centering\arraybackslash\axiom\bgroup}p{.1\textwidth}<{\egroup}@{\extracolsep{\fill}}>{\centering\arraybackslash}p{.85\textwidth}}
\hline\tstrut
	N1 & $E \nchoice F = F \nchoice E$ \\
	N2 & $E \nchoice (F \nchoice G) = (E \nchoice F) \nchoice G$ \\
	N3 & $E \nchoice E = E$ \\
	N4 & $E \nchoice \nil = E$ \\
\bstrut\hline\tstrut
	P1 & $P \pchoice p Q = Q \pchoice{1-p} P$ \\
	P2 & $P \pchoice p (Q \pchoice{\frac q {1-p}} R) = (P \pchoice{\frac p {p+q}} Q) \pchoice{p+q} R$ \\
	P3 & $P \pchoice p P = P$ \\
\bstrut\hline\tstrut
	T1 & $\alpha\prefix(\psingle{\tau\prefix\psingle E} \pchoice p P) = \alpha\prefix(\psingle E \pchoice p P)$ \\
	T2 & $\tau\prefix \psum_{i \in I} {p_i} (E_i \nchoice F) \nchoice F = \tau\prefix \psum_{i \in I} {p_i} (E_i \nchoice F)$ \\
	T3 & $\tau\prefix\psum_{i \in I} {p_i} (E_i \nchoice \alpha\prefix P_i) \nchoice \alpha\prefix\psum_{i \in I} {p_i} P_i = \tau\prefix\psum_{i \in I} {p_i} (E_i \nchoice \alpha\prefix P_i)$ \\
	T4 & $\alpha\prefix\psum_{i \in I} {p_i} (E_i \nchoice \tau\prefix P_i) \nchoice \alpha\prefix\psum_{i \in I} {p_i} P_i = \alpha\prefix\psum_{i \in I} {p_i} (E_i \nchoice \tau\prefix P_i)$ \\
\bstrut\hline\tstrut
	C & $\alpha\prefix P \nchoice \alpha\prefix Q = \alpha\prefix P \nchoice \alpha\prefix (P \pchoice p Q) \nchoice \alpha\prefix Q$ \\
\bstrut\hline\tstrut
	R1 & $\rec X E = E \subst {\rec X E} X$ \\
	R2 & if $F = E \subst F X$ and $E \not\unguarded X$, then $F = \rec X E$ \\
	R3 & $\rec X (\tau\prefix (\psingle{X \nchoice E} \pchoice p P) \nchoice F) = \rec X (\tau\prefix (\psingle{X \nchoice E} \pchoice p P) \nchoice \tau\prefix P \nchoice F)$ \\
	R4 & $\rec X (X \nchoice E) = \rec X E$ \\
	R5 & $\rec X (\tau\prefix\psingle X \nchoice E) = \rec X \tau\prefix\psingle E$ \\
	R6 & $\rec X (\tau\prefix\psum_{i \in I} {p_i} (X \nchoice E_i) \nchoice F) = \rec X (\tau\prefix\psingle X \nchoice \sum_{i \in I} E_i \nchoice F)$ \\
\bstrut\hline
\end{tabular*}
\end{table}

\noindent
We give equational laws that precisely capture the essence of $\wcong$. More formally, we introduce a relation $=$, which relates two expressions $E$ and $F$ if and only if $E$ may be transformed into $F$ by applying the axioms of \autoref{tab:axioms} only. Importantly, $=$ is designed to be a congruence relation, meaning that an axiom may be applied to some subexpression of $E$, as well. Besides, $=$ always allows appropriate renaming of variables. In the end, our goal is proving that $\wcong$ and $=$ coincide. Given this use of the symbol $=$, we use $\synid$ for syntactic identity. In this section, we have a closer look at the provided axioms.

First of all, \axiom{N1}--\axiom{N4} express basic algebraic properties of nondeterministic choice: commutativity, associativity, idempotence and $\nil$ being the neutral element of $\nchoice$. These axioms on their own are sound and complete for strong bisimilarity in the absence of recursion~\cite{Milner1989a,Milner1989b}.

The axioms \axiom{P1}--\axiom{P3} stem from \cite{Stark2000} and state simple properties of probabilistic choice. Normally, we do not tag applications of these laws. Instead, we often implicitly assume the convenient shape $\psum_{i \in I} {p_i} E_i$.

Next, focus on \axiom{T1}--\axiom{T4}. The axioms \axiom{T1} and \axiom{T4} are conservative generalisations of Milner's \cite{Milner1989b} axioms $\alpha\prefix \tau\prefix E = \alpha\prefix E$ and $\alpha\prefix (E \nchoice \tau\prefix F) \nchoice \alpha\prefix F = \alpha\prefix (E \nchoice \tau\prefix F)$, respectively. However, we are required to introduce two extensions of $E \nchoice \tau\prefix E = \tau\prefix E$. On the one hand, if $E$ is a prefix expression, then \axiom{T3} allows some transformations potentially involving quantities. On the other hand, we still rely on \axiom{T2} if $E$ is a variable, for instance. Remark that each closed instance of \axiom{T2} becomes derivable in the presence of \axiom{T3} when not considering recursion \cite{Bandini2001}. The axioms \axiom{T1}, \axiom{T3} and \axiom{T4} first appeared in \cite{Bandini2001} and \axiom{T1}--\axiom{T4} occurred in \cite{DengPP05}, too, in almost equal forms.

The convexity axiom \axiom{C} captures the convex property of combined transitions: Whenever $E \trans\alpha \mu$ and $E \trans\alpha \nu$, \axiom{C} essentially allows to add or remove transitions leading to a convex combination of $\mu$ and $\nu$.

Finally, we analyse the recursion axioms. \axiom{R1} and \axiom{R2} establish existence and uniqueness of equation's solutions; \axiom{R2} premises the equation $E$ to probabilistically guard $X$. Fortunately, \axiom{R2} can be shown to remain sound even after having loosened the guardedness notion. The axiom \axiom{R3} is new. It expresses exactly the infinitary character of weak transitions: Suppose that $P$ is of the shape $\psum_{i=1}^n {p_i} G_i$ and consider the following depiction:
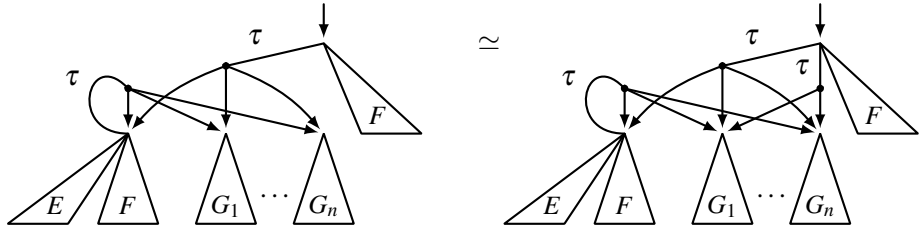
\vspace{-1ex}
\begin{center}
\begin{tikzpicture}[process diagram]

\def\pwidth{.8cm}
\def\pheight{1.2cm}

\node[initial, initial where=above, state] (0) at (2.2cm, 0) {};
\node[state] (1) at ($(0) + (-2.6cm, -\pheight)$) {};
\node[state] (2) at ($(0) + (-1.3cm, -\pheight)$) {};
\node[state] (3) at ($(0) + (0cm, -\pheight)$) {};

\pic["$F$"] at (0) {process={.9cm, \pwidth, \pheight}};

\pic["$E$"] at (1) {process={-1.2cm, \pwidth, \pheight}};
\pic["$F$"] at (1) {process={0cm, \pwidth, \pheight}};

\pic["$G_1$"] at (2) {process={0, \pwidth, \pheight}};

\pic["$G_n$"] at (3) {process={0, \pwidth, \pheight}};

\node at ($(2)!.5!(3) + (0, .7 * -\pheight)$) {${\cdots}$};

\node[splitter] (*0) at ($(1)!.5!(3) + (0, .9cm)$) {};
\path (0) edge[shorten <=0, line cap=round, "$\tau$", auto=right] (*0);
\path (*0) edge[transition, bend right=15] (1);
\path (*0) edge[transition] (2);
\path (*0) edge[transition, bend left=15] (3);

\node[splitter] (*1) at ($(1) + (0, .6cm)$) {};
\path (1) edge[shorten <=0, line cap=round, out=180, in=135, looseness=3.6, "$\tau$"] (*1);
\path (*1) edge[transition] (1);
\path (*1) edge[transition] (2);
\path (*1) edge[transition] (3);

\node at ($(0) + (2.2cm, 0)$) {${\wcong}$};

\node[initial, initial where=above, state] (0) at ($(0) + (6.6cm, 0)$) {};
\node[state] (1) at ($(0) + (-2.6cm, -\pheight)$) {};
\node[state] (2) at ($(0) + (-1.3cm, -\pheight)$) {};
\node[state] (3) at ($(0) + (0cm, -\pheight)$) {};

\pic["$F$"] at (0) {process={.9cm, \pwidth, \pheight}};

\pic["$E$"] at (1) {process={-1.2cm, \pwidth, \pheight}};
\pic["$F$"] at (1) {process={0cm, \pwidth, \pheight}};

\pic["$G_1$"] at (2) {process={0, \pwidth, \pheight}};

\pic["$G_n$"] at (3) {process={0, \pwidth, \pheight}};

\node at ($(2)!.5!(3) + (0, .7 * -\pheight)$) {${\cdots}$};

\node[splitter] (*0) at ($(1)!.5!(3) + (0, .9cm)$) {};
\path (0) edge[shorten <=0, line cap=round, "$\tau$", auto=right] (*0);
\path (*0) edge[transition, bend right=15] (1);
\path (*0) edge[transition] (2);
\path (*0) edge[transition, bend left=15] (3);

\node[splitter] (*1) at ($(1) + (0, .6cm)$) {};
\path (1) edge[shorten <=0, line cap=round, out=180, in=135, looseness=3.6, "$\tau$"] (*1);
\path (*1) edge[transition] (1);
\path (*1) edge[transition] (2);
\path (*1) edge[transition] (3);

\node[splitter] (*2) at ($(3) + (0, .6cm)$) {};
\path (0) edge[shorten <=0, line cap=round, "$\tau$", auto=right] (*2);
\path (*2) edge[transition] (2);
\path (*2) edge[transition] (3);

\end{tikzpicture}\vspace{1ex}
\end{center}
The processes $E, F$ may be ignored for now; they are only required to fit in a most general context. Now, there is a transition targeting $G_1, \ldots, G_n$ with some probability $< 1$. The remaining probability mass reaches some state providing the exact same transition, again. So intuitively, \axiom{R3} allows unrolling this transition infinitely many times, until some $G_i$ is certainly reached. This weak transition corresponds to a scheduler that always schedules the aforenamed transition while no $G_i$ is visited, and halts otherwise.

We continue presenting the remaining axioms: In Milner's context, \axiom{R4}--\axiom{R6} serve the purpose of transforming an arbitrary expression into a guarded one. Here, the law \axiom{R5} will also be required again in a later part of the completeness proof. Both \axiom{R4} and \axiom{R5} agree with their original versions exactly~\cite{Milner1989b}, whereas \axiom{R6} describes an interesting generalisation:
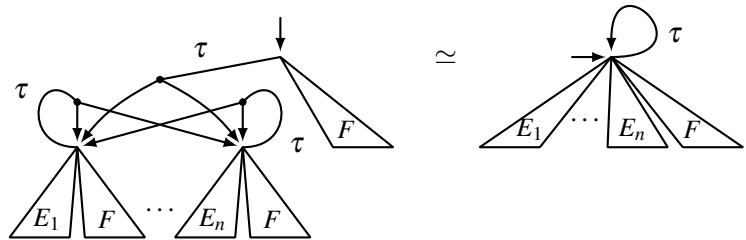
\vspace{-4ex}
\begin{center}
\begin{tikzpicture}[process diagram]

\def\pwidth{.8cm}
\def\pheight{1.2cm}

\node[initial, initial where=above, state] (0) at (1.6cm, 0) {};
\node[state] (1) at (-1.1cm, -\pheight) {};
\node[state] (2) at (1.1cm, -\pheight) {};

\pic["$F$"] at (0) {process={1.1cm, \pwidth, \pheight}};

\pic["$E_1$"] at (1) {process={-.5cm, \pwidth, \pheight}};
\pic["$F$"] at (1) {process={.5cm, \pwidth, \pheight}};

\pic["$E_n$"] at (2) {process={-.5cm, \pwidth, \pheight}};
\pic["$F$"] at (2) {process={.5cm, \pwidth, \pheight}};

\node at ($(1)!.5!(2) + (0, .7 * -\pheight)$) {${\cdots}$};

\node[splitter] (*0) at ($(1)!.5!(2) + (0, .9cm)$) {};
\path (0) edge[shorten <=0, line cap=round, "$\tau$", auto=right] (*0);
\path (*0) edge[transition, bend right=13] (1);
\path (*0) edge[transition, bend left=13] (2);

\node[splitter] (*1) at ($(1) + (0, .6cm)$) {};
\path (1) edge[shorten <=0, line cap=round, out=180, in=135, looseness=3.6, "$\tau$"] (*1);
\path (*1) edge[transition] (1);
\path (*1) edge[transition] (2);

\node[splitter] (*2) at ($(2) + (0, .6cm)$) {};
\path (2) edge[shorten <=0, line cap=round, out=0, in=45, looseness=3.6, "$\tau$" pos=.33, auto=right] (*2);
\path (*2) edge[transition] (1);
\path (*2) edge[transition] (2);

\node at ($(0) + (2.2cm, 0)$) {${\wcong}$};

\node[initial, initial where=left, state] (0) at ($(0) + (4.4cm, 0)$) {};

\pic["$E_1$"] at (0) {process={-1.35cm, \pwidth, \pheight}};
\pic["$E_n$"] at (0) {process={.35cm, \pwidth, \pheight}};
\pic["$F$"] at (0) {process={1.35cm, \pwidth, \pheight}};

\path let \p1 = ($(-1.35cm + .5 * \pwidth, -\pheight)!.5!(.35cm - .5 * \pwidth, -\pheight)$) in node at ($(0) + (.7 * \x1, .7 * \y1)$) {${\cdots}$};

\path (0) edge[transition, shorten <=0, line cap=round, out=0, in=90, looseness=1400, "$\tau$" pos=.33, auto=right] (0);

\end{tikzpicture}\vspace{1ex}
\end{center}
Think of a random experiment with outcomes $E_1, \ldots, E_n$; we ignore $F$ for now. Now \axiom{R6} states: If we are allowed to repeat the experiment unboundedly often and halt whenever convenient, then in fact, the actual probabilities $p_i$ do not matter, as we can nondeterministically choose which outcome is selected.

More formally, suppose you want to visit some state offering $E_i$ on the left-hand side. Then we follow the weak transition corresponding to a scheduler that always loops, until the respective target state is reached. This transition does not diverge, as the probability of not visiting the desired state converges to $0$. So \axiom{R6} exploits the infinitary property, again.

Deng and Palamidessi \cite{Deng2007,DengPP05} established axiomatisations taking the finitary transition variant as a basis. Consequently, the law \axiom{R6} is not sound in their context, and there is no other obvious generalisation of Milner's original. Thus, that axiomatisation turns out to be incomplete for nonprobabilistically unguarded expressions. In fact, the authors conjecture that the problem even becomes undecidable when considering arbitrary expressions.

The next sections are concerned with proving the soundness and completeness of $=$.

\section{Soundness} \label{sec:soundness}
\newcommand{\rwbu}{rooted weak bisimulation up to}

\noindent
Soundness asserts that whenever some expressions $E, F$ are equal according to the axioms, then indeed, $E \wcong F$. Once we establish that $\wcong$ is a congruence (\autoref{thm:wcong-cong}), each axiom may be considered on its own and a soundness proof typically finds a weak bisimulation relating the desired expressions, before additionally checking the root condition of weak congruence. Thanks to the definition of $\wcong$, it suffices to show the soundness of these laws for closed expressions only.

Most axioms are proven correct easily: We are not giving proper proofs for the axioms \axiom{N1}--\axiom{N4}, the interested reader may refer to \cite{Milner1989a}. The axioms \axiom{P1}--\axiom{P3} essentially express some rewriting identities that lead back to basic mathematical properties. \axiom{T1}--\axiom{T4} rely on some characteristics of probabilistic weak transitions, however, they do not depend on $\wtrans{}$ being infinitary! Therefore, the proofs of \cite{Deng2007} carry over naturally. Moreover, the axiom \axiom{C} is immediate by the convexity property of \raisebox{0pt}[0pt][0pt]{$\wtrans\alpha$}. We distinguish further between the recursion axioms; the axioms \axiom{R1}, \axiom{R4} and \axiom{R5} are identical to Milner's versions, thus, we omit the proofs and, again, refer to \cite{Milner1989a}. However, the remaining axioms \axiom{R3} and \axiom{R6} turn out to be particularly interesting, as they strongly rely on the infinitary property of $\wtrans{}$. Lastly, \axiom{R2} takes a special role just like in the nonprobabilistic setting, which results in an exhausting proof, but interesting insights. We continue focusing on these challenging laws in this section.

Before, we shall lift some already known terminology to (sub-)probability distributions: We say a distribution $\mu$ is (probabilistically) guarded, if every expression supported by $\mu$ is probabilistically guarded. Moreover, let $\mu \subst E X$ denote the subdistribution $\mu$ after substituting $E$ for $X$ in every supported expression.

Some of the following proofs rely on Milner's `bisimulation up to' technique~\cite{Milner1989a}, merely adapted to our context:
\begin{definition}[Rooted weak bisimulation up to $\wbis$] \label{def:wcong-up-to}\rm
A relation $\mathcal R \subseteq \CExp \times \CExp$ is called a \emph{one-sided {\rwbu} $\wbis$}, if the following conditions hold for all $(E, F) \in \mathcal R$:
\begin{enumerate}[label=(\roman*)]
\vspace{-1ex}
\item whenever $E \trans\alpha \mu$, then there is some $\nu$, such that $F \wtrans\alpha \nu$ and $\mu \mathrel{\mathcal R}\wbis \nu$, \label{def:wcong-up-to:itm:left-sided-left}
\vspace{-1ex}
\item whenever $F \trans\alpha \nu$, then there is some $\mu$, such that $E \wtrans\alpha \mu$ and $\mu \wbis\mathrel{\mathcal R} \nu$. \label{def:wcong-up-to:itm:left-sided-right}
\vspace{-1ex}
\end{enumerate}
Call $\mathcal R$ a \emph{two-sided {\rwbu} $\wbis$}, if the following conditions hold for all $(E, F) \in \mathcal R$:
\begin{enumerate}[label=(\roman*)]
\vspace{-1ex}
\item whenever $E \wtrans\alpha \mu \in \Distr\Exp$, then there is some $\nu$, such that $F \wtrans\alpha \nu$ and $\mu \wbis\mathrel{\mathcal R}\wbis \nu$, and \label{def:wcong-up-to:itm:two-sided-left}
\vspace{-1ex}
\item whenever $F \wtrans\alpha \nu \in \Distr\Exp$, then there is some $\mu$, such that $E \wtrans\alpha \mu$ and $\mu \wbis\mathrel{\mathcal R}\wbis \nu$. \label{def:wcong-up-to:itm:two-sided-right}
\vspace{-1ex}
\end{enumerate}
\end{definition}

\noindent
Here, $\mathrel{\mathcal R} \wbis$ (${\wbis \mathrel{\mathcal R} \wbis}$) denotes the composition of ($\liftequ\wbis$ and) $\liftequ{\mathcal R}$ and $\liftequ\wbis$. For the two-sided variant, we add the convenient assumption $|\mu| = 1$, such that we are only required to deal with convergent transitions \raisebox{0pt}[0pt][0pt]{$E \wtrans\alpha \mu$}.

It turns out that any two expressions related by a {\rwbu} $\wbis$ are in fact weakly congruent, as shown by the following lemmas. We benefit from this approach by not being obliged to find an exact weak bisimulation relating two given processes, but rather proving the relaxed statement tolerating additional $\wbis$-relations in between.

\begin{lemma} \label{lem:one-sided-wcong-up-to}
Any one-sided {\rwbu} $\wbis$ is also a two-sided {\rwbu} $\wbis$. 
\end{lemma}
\begin{proof}
Let $\mathcal R$ be a one-sided {\rwbu} $\wbis$ and let $(E, F) \in \mathcal R$. We assume $E \wtrans\alpha \mu$ (convergent); by symmetry, it suffices to show that $F \wtrans\alpha \nu$ for some $\nu$ with $\mu \wbis\mathrel{\mathcal R}\wbis \nu$. For now, assume the following claim:
\begin{enumerate}[label=($*$)]
\item if $\gamma \mathrel{\mathcal R}\wbis \eta$ and $\gamma \wtrans{} \gamma'$ (convergent), then $\eta \wtrans{} \eta'$ for some $\eta'$ with $\gamma' \mathrel{\mathcal R}\wbis \eta'$. \label{lem:wcong-up-to:itm:weak-property}
\end{enumerate}
In case that $\alpha = \tau$, we have $E \trans\tau \mu' \wtrans{} \mu$ for some $\mu'$ by \autoref{prp:wtrans-tau}, thus, $F \wtrans\tau \nu'$ for some $\nu'$ with $\mu' \mathrel{\mathcal R}\wbis \nu'$. Conclude by \ref{lem:wcong-up-to:itm:weak-property}.

On the other hand, if $\alpha \neq \tau$, then $E \wtrans{} \mu' \trans\alpha \mu'' \wtrans{} \mu$ for some $\mu', \mu''$. \ref{lem:wcong-up-to:itm:weak-property} yields $F \wtrans{} \nu'$ for some $\pi', \nu'$ with $\mu' \mathrel{\mathcal R} \pi' \wbis \nu'$. Exploit that $\mathcal R$ is a {\rwbu} $\wbis$ to find that $\pi' \wtrans\alpha \pi''$ for some $\pi''$, s.t.\ $\mu'' \mathrel{\mathcal R} \wbis \pi''$. By \autoref{prp:wbis-wtrans}, $\nu' \wtrans\alpha \nu''$, where $\mu'' \mathrel{\mathcal R} \wbis \pi'' \wbis \nu''$. It suffices to apply \ref{lem:wcong-up-to:itm:weak-property} again.

We are left to prove \ref{lem:wcong-up-to:itm:weak-property}, so let $\gamma \wtrans{} \gamma'$ be derived by $(\gamma_i^\proc, \gamma_i^\halt)_{i \in \Nat}$; our goal is to construct a sequence $(\eta_i^\proc, \eta_i^\halt)_{i \in \Nat}$ as evidence for $\eta \wtrans{} \eta'$. Invariantly, assert that\vspace{-2ex}
\begin{align*}
	&\gamma_i^\proc \mathrel{\mathcal R} \xi_i^\proc \wbis \eta_i^\proc, &\gamma_i^\halt \mathrel{\mathcal R} \xi_i^\halt \wbis \eta_i^\halt,
\end{align*}
with intermediate values $\xi_i^\proc, \xi_i^\halt$. Initially, $\gamma \synid \gamma_0^\proc + \gamma_0^\halt \mathrel{\mathcal R} \xi \wbis \eta$, for some $\xi$, so split $\xi$ and $\eta$ into $\xi_0^\proc, \xi_0^\halt$ and $\eta_0^\proc, \eta_0^\halt$ with $\gamma_0^\proc \mathrel{\mathcal R} \xi_0^\proc \wbis \eta_0^\proc$ and $\gamma_0^\halt \mathrel{\mathcal R} \xi_0^\halt \wbis \eta_0^\halt$. Now let $i \geq 0$. By $(\gamma_i^\proc, \gamma_i^\halt)_{i \in \Nat}$ being a derivation, there is a transition $\gamma_i^\proc \trans\tau \gamma_{i+1}^\proc + \gamma_{i+1}^\halt$, which, in turn, implies $\xi_i^\proc \wtrans\tau \xi_{i+1}$ for some $\xi_{i+1}$ with $\gamma_{i+1}^\proc + \gamma_{i+1}^\halt \mathrel{\mathcal R}\wbis \xi_{i+1}$, by $\mathcal R$ being a one-sided {\rwbu} $\wbis$. Again, split $\xi_{i+1}$ into $\xi_{i+1}^\proc$ and $\xi_{i+1}^\halt$, s.t.\ $\gamma_{i+1}^\proc \mathrel{\mathcal R}\wbis \xi_{i+1}^\proc$ and $\gamma_{i+1}^\halt \mathrel{\mathcal R}\wbis \xi_{i+1}^\halt$. Recall that $\xi_i^\proc \wbis \eta_i^\proc$ and apply \autoref{prp:wbis-wtrans} to find that $\eta_i^\proc \wtrans{} \eta_{i+1}$ for some $\eta_{i+1}$ with $\xi_{i+1} \wbis \eta_{i+1}$. Finally, split $\eta_{i+1}$ into $\eta_{i+1}^\proc$ and $\eta_{i+1}^\halt$ to maintain the invariant.

Observe that we constructed a sequence $(\eta_i^\proc, \eta_i^\halt)_{i \in \Nat}$ with $\eta_i^\proc \wtrans{} \eta_{i+1}^\proc + \eta_{i+1}^\halt$ for all $i$, thus, by \autoref{prp:wtrans-weak-construction}, there is a weak transition $\eta \wtrans{} \eta'$ for $\eta' :\synid \sum_{i \in \Nat} \eta_i^\halt \wbis\mathrel{\mathcal R} \sum_{i \in \Nat} \gamma_i^\halt \synid \gamma'$. The convergence criterion is fulfilled by $\gamma \wtrans{} \gamma'$ being convergent.
\end{proof}

\begin{lemma} \label{lem:wcong-up-to}
Let $\mathcal R$ be a {\rwbu} $\wbis$ with $E \mathrel{\mathcal R} F$. Then $E \wcong F$.
\end{lemma}
\begin{proof}
By \autoref{lem:one-sided-wcong-up-to}, we can assume that $\mathcal R$ is a two-sided {\rwbu} $\wbis$. We first show that ${\wbis \mathrel{\mathcal R} \wbis}$ is a weak bisimulation.
So let $E \wbis E' \mathrel{\mathcal R} F' \wbis F$ and $E \trans\alpha \mu$. First, assume $\alpha = \tau$. We find $E' \wtrans{} \mu' \wbis \mu$ for some $\mu'$. \autoref{prp:wtrans-decompose-tau} allows us to split $\mu' \synid p \cdot \dirac{E'} + (\mu' - p \cdot \dirac{E'})$, such that $(1 - p) \cdot \dirac{E'} \wtrans\tau \mu' - p \cdot \dirac{E'}$. By $\mathcal R$ being a two-sided {\rwbu} $\wbis$, it follows that $(1 - p) \cdot \dirac{F'} \wtrans\tau \nu' \wbis \mathrel{\mathcal R}^{-1} \wbis \mu' - p \cdot \dirac{E'}$, which itself implies $(1 - p) \cdot \dirac F \wtrans{} \nu \wbis \mathrel{\mathcal R}^{-1} \wbis \mu' - p \cdot \dirac{E'}$. Now, by reflexivity of $\wtrans{}$, we obtain $p \cdot \dirac F \wtrans{} p \cdot \dirac F \wbis p \cdot \dirac{F'} \mathrel{\mathcal R}^{-1} p \cdot \dirac{E'}$ and the desired transition is constructed as a convex combination. Next, let $\alpha \neq \tau$. Then trivially $E' \wtrans\alpha \mu'$, which causes $F' \wtrans\alpha \nu' \wbis \mathrel{\mathcal R}^{-1} \wbis \mu'$. Finally, $F \wtrans\alpha \nu \wbis \nu' \wbis \mathrel{\mathcal R}^{-1} \wbis \mu' \wbis \mu$, so $\wbis \mathrel{\mathcal R} \wbis$ is a weak bisimulation.

We still need to consider the root condition of congruence (\autoref{def:weak-congruence}): Assume that $E \mathrel{\mathcal R} F$ and $E \trans\alpha \mu$. Then also $E \wtrans\alpha \mu$, so we find $F \wtrans\alpha \nu \wbis \mathrel{\mathcal R} \wbis \mu$, which provably means $\nu \wbis \mu$.
\end{proof}

\begin{theorem} \label{thm:wcong-cong}
\hspace{0cm}
\begin{enumerate}[label={\normalfont(\arabic*)}]
\item $\wcong$ is an equivalence relation, \label{thm:wcong-cong:itm:equ}
\vspace{-1ex}
\item $\wcong$ is a congruence relation, i.e.\ for all expressions $E, F, G \mathbin\in \Exp$ with $E \wcong F$:
	\begin{enumerate}[label={\normalfont(\alph*)}]
	\item $E \nchoice G \wcong F \nchoice G$ and $G \nchoice E \wcong G \nchoice F$, \label{thm:wcong-cong:itm:cong:itm:nchoice}
	\item $\rec X E \wcong \rec X F$, and \label{thm:wcong-cong:itm:cong:itm:rec}
	\item $\psingle E \wcong \psingle F$; \label{thm:wcong-cong:itm:cong:itm:psingle}
	\end{enumerate}
	and for all probabilistic expressions $P, Q, R\in \PExp$ with $P\wcong Q$ we have:
	\begin{enumerate}[label={\normalfont(\alph*)}, resume]
	\item $\alpha\prefix P \wcong \alpha\prefix Q$, \label{thm:wcong-cong:itm:cong:itm:prefix}
	\item $P \pchoice{p} R \wcong Q \pchoice{p} R$ and $R \pchoice{p} P \wcong R \pchoice{p} Q$. \label{thm:wcong-cong:itm:cong:itm:pchoice}
	\end{enumerate}
	Here we write $P\wcong Q$ when the unique distributions pointed to by $P$ and $Q$ are weakly congruent. \label{thm:wcong-cong:itm:cong}
\end{enumerate}
\end{theorem}
\begin{proof}
The transitivity of $\wcong$ follows immediately by \autoref{prp:wcong-wtrans}. Reflexivity and symmetry are trivial, thus, $\wcong$ is an equivalence relation.

The proofs of
\ref{thm:wcong-cong:itm:cong}\ref{thm:wcong-cong:itm:cong:itm:nchoice},
\ref{thm:wcong-cong:itm:cong:itm:psingle},
\ref{thm:wcong-cong:itm:cong:itm:prefix} and
\ref{thm:wcong-cong:itm:cong:itm:pchoice} are straightforward, and the proof of
\ref{thm:wcong-cong:itm:cong}\ref{thm:wcong-cong:itm:cong:itm:rec} is similar to \cite{Milner1989a}, using the one-sided `up to' technique of \autoref{def:wcong-up-to} and \autoref{lem:wcong-up-to}.
\end{proof}

\begin{lemma}[Soundness of \axiom{R3}] \label{lem:soundness-r3}
Let $E, F \in \Exp$ and $P \in \PExp$.\newline
Then $\rec X (\tau\prefix (\psingle{X \nchoice E} \pchoice p P) \nchoice F) \wcong \rec X (\tau\prefix (\psingle{X \nchoice E} \pchoice p P) \nchoice \tau\prefix P \nchoice F)$.
\end{lemma}
\begin{proof}
Without loss of generality we may assume that $\Var(E), \Var(F),\allowbreak \Var(P) \subseteq \{X\}$.\\ Let $G \synid \rec X (\tau\prefix (\psingle{X \nchoice E} \pchoice p P) \nchoice F)$ and $H \synid \rec X (\tau\prefix (\psingle{X \nchoice E} \pchoice p P) \nchoice \tau\prefix P \nchoice F)$. We show that the relation $\mathcal R$, containing a pair\vspace{-1ex}
\[
	(\; K \subst{G}{X} \;,\; K \subst{H}{X} \;)
\]
for all expressions $K$ with free variables in  $\{X\}$, is a (one-sided) rooted weak bisimulation (up to $\wbis$). By choosing $K \synid X$, we are finished.

Assuming $K \subst H X \trans\alpha \mu$, we prove that $K \subst G X \wtrans\alpha \nu \mathrel{\mathcal R} \mu$ for some $\nu$; the other direction then becomes obvious. We perform an induction on the inference tree deriving $K \subst H X \trans\alpha \mu$ while making a case distinction on the shape of $K$:

$K \synid \beta\prefix P$, where $P \mapsto \nu$. Then $\beta = \alpha$ and $P \subst H X \ptrans \nu \subst H X \synid \mu$. Trivially $K \subst G X \trans\alpha \nu \subst G X \mathrel{\mathcal R} \mu$.

$K \synid L \nchoice M$, where $L \subst H X \trans\alpha \mu$ (resp. $M \subst H X \trans\alpha \mu$). By induction.

$K \synid \rec X M$. Then there is no unbound occurrence of $X$ in $K$, so $K \synid K \subst H X \synid K \subst G X$ and we are immediately finished.

$K \synid \rec Y M$ for some $Y \neq X$.
Then $K\subst H X = \rec Y (M \subst H X) \trans\alpha \mu$,
so $M \subst{\rec Y M}{Y} \subst H X \equiv M \subst H X \subst{(\rec Y M \subst H X)}{Y} \trans\alpha \mu$. Conclude by induction.

The only remaining case is $K \synid X$, thus, $K \subst H X \synid H$.
Consequently $\tau\prefix (\psingle{X \nchoice E} \pchoice p P)\subst H X \nchoice \tau\prefix P\subst H X \nchoice F\subst H X \trans\alpha \mu$. If $\tau\prefix(\psingle{X \nchoice E} \pchoice p P) \subst H X$ or $F \subst H X$ derive the transition, then we are easily finished, by induction. So let $\tau\prefix P \subst H X \trans\alpha \mu$. It follows that $\alpha = \tau$ and with $P$ being of the shape $\psum_{i \in I} {p_i} N_i$, infer that $\mu \synid \distr{N_i \subst H X \mapsto p_i \mid i \in I}$. So we proceed by checking that
\[
	G \wtrans\tau \nu \synid \distr{N_i \subst G X \mapsto p_i \mid i \in I},
\]
which suffices, since $\mu \mathrel{\mathcal R} \nu$. The derivation $(\nu_i^\proc, \nu_i^\halt)_{i \in \Nat}$ is constructed as follows:
\begin{align*}
	\nu_0^\proc &\synid \dirac G, &
	\nu_0^\halt &\synid \emptydistr, \\
	\nu_{i+1}^\proc &\synid p^{i+1} \cdot \dirac{G \nchoice E}, &
	\nu_{i+1}^\halt &\synid p^i \cdot (1 - p) \cdot \nu;
\end{align*}
for any $i$. The desired transition $\nu_i^\proc \trans\tau \nu_{i+1} \synid \nu_{i+1}^\proc + \nu_{i+1}^\halt$ is immediate and $\nu \synid \sum_{i \in \Nat} p^i \cdot (1 - p) \cdot \nu$. \autoref{prp:wtrans-tau} yields $\wtrans\tau$ instead of $\wtrans{}$.
\end{proof}

\begin{lemma}[Soundness of \axiom{R6}] \label{lem:soundness-r6}
Let $E, F \in \Exp$.\newline Then $\rec X (\tau\prefix \psum_{i \in I} {p_i} (X \nchoice E_i) \nchoice F) \wcong \rec X (\tau\prefix \psingle X \nchoice \sum_{i \in I} E_i \nchoice F)$.
\end{lemma}
\begin{proof}
Without loss of generality we may assume that $\Var(E), \Var(F) \subseteq \{X\}$. Let
\begin{center}
  $G \synid \rec X (\tau\prefix \psum_{i \in I} {p_i} (X \nchoice E_i) \nchoice F)$  \qquad and \qquad $H \synid \rec X (\tau\prefix \psingle X \nchoice \sum_{i \in I} E_i \nchoice F)$.
\end{center}
Define $\mathcal R$ as the relation containing a pair
\[
	(\; K \subst G X \;,\; K \subst H X \;)
\]
for all expressions $K$ with free variables in $\{X\}$. It is enough show that $\mathcal R$ is a one-sided {\rwbu} $\wbis$.

First, pick some pair $(K \subst G X, K \subst H X) \in \mathcal R$ and assume $K \subst G X \trans\alpha \mu$. We prove that $K \subst H X\linebreak[2]\trans\alpha \nu \wbis\mathrel{\mathcal R}^{-1} \mu$ for some $\nu$ by carrying out an induction on the inference tree deriving the transition; proceed as in \autoref{lem:soundness-r3}. Again, the case $K \synid X$ turns out to be most interesting: Then either $\tau\prefix \psum_{i \in I} {p_i} (G \nchoice E_i \subst G X) \trans\alpha \mu$ or $F \subst G X \trans\alpha \mu$. The latter case is easy; conclude by induction. The former case entails $\alpha = \tau$ and $\mu \synid \distr{G \nchoice E_i \subst G X \mapsto p_i \mid i \in I}$, so the self-loop of $H$ proves itself vital: $H \trans\tau \dirac H$. Finally, note that for all $i$, $H \wcong H \nchoice E_i \subst H X$ by the soundness of \axiom{R1} and \axiom{N1}--\axiom{N3}, hence, $\dirac H \wbis \distr{H \nchoice E_i \subst H X \mapsto p_i \mid i \in I} \mathrel{\mathcal R}^{-1} \mu$.

Now let $K \subst H X \trans\alpha \mu$. We perform a similar induction and focus on the case $K \synid X$, where $E_j \subst H X \trans\alpha \mu$ for some $j \in I$. Obviously, it suffices to show that $G$ can move silently to $G \nchoice E_j \subst G X$. Construct a derivation $(\gamma_i^\proc, \gamma_i^\halt)$ as follows:
\begin{talign*}
	\gamma_0^\proc &\synid \dirac G, & \gamma_{i+1}^\proc &\synid (1 - p_j)^i \cdot \sum_{k \in I \setminus \{j\}} p_k \cdot \dirac{G \nchoice E_k \subst G X}, \\
	\gamma_0^\halt &\synid \emptydistr, & \gamma_{i+1}^\halt &\synid (1 - p_j)^i \cdot p_j \cdot \dirac{G \nchoice E_j \subst G X}.
\end{talign*}
The transitions $\gamma_i^\proc \trans\tau \gamma_{i+1}$ obtain naturally, so indeed, there is a transition
\\[2ex]\mbox{}\hfill
$	\dirac G \wtrans{} \sum_{i \in \Nat} (1 - p_j)^i \cdot p_j \cdot \dirac{G \nchoice E_j \subst G X} \synid \dirac{G \nchoice E_j \subst G X} \trans\alpha \nu \mathrel{\mathcal R} \mu$.
\end{proof}

\begin{restatable}[Soundness of \axiom{R2}]{lemma}{soundnessrtwo} \label{lem:soundness-r2}
Let $E, F \in \Exp$. If $F \wcong E \subst F X$ and $E \not\unguarded X$, then $F \wcong \rec X E$.
\end{restatable}
\begin{proof}
See \autoref{apx:soundness-r2}.
\end{proof}

\begin{theorem}[Soundness]
If $E = F$, then $E \wcong F$.
\end{theorem}
\begin{proof}
By induction on the proof of $E = F$. Such a proof can be formalised as a finite tree of which the nodes are labelled with equations $G = H$ with $G, H \in \Exp$ or $G, H \in \PExp$, the root is labelled with $E = F$, and each node applies either reflexivity, symmetry or transitivity of $=$, the congruence property of \autoref{thm:wcong-cong}, or one of the axioms of \autoref{tab:axioms}. Axioms are applied only in leaf nodes, except for \axiom{R2}. The induction is a straightforward application of \autoref{thm:wcong-cong} and the soundness of the individual axioms.
\end{proof}

\section{Completeness} \label{sec:completeness}

\begin{figure}[t]
\begin{minipage}[b]{.49\textwidth}
\begin{center}
\begin{tikzpicture}[
	node distance=.9cm,
	on grid,
	phase/.style={minimum size=.5cm},
	left label/.style={align=right, font=\footnotesize, left=of #1},
	right label/.style={align=left, font=\footnotesize, right=of #1},
	connection/.style={thick, >=latex, ->}
]

\node[phase] (E) {$E$};
\node[phase, below=of E] (E') {$E'$}; \node[left label=E'] {guarded};
\node[phase, below=of E'] (S) {$\mathcal S$}; \node[left label=S] {guarded};
\node[phase, below=of S] (S') {$\mathcal S'$}; \node[left label=S'] {guarded,\\[-.3ex]saturated};

\node[phase, right=2cm of E] (F) {$F$};
\node[phase, right=2cm of E'] (F') {$F'$}; \node[right label=F'] {guarded};
\node[phase, right=2cm of S] (T) {$\mathcal T$}; \node[right label=T] {guarded};
\node[phase, right=2cm of S'] (T') {$\mathcal T'$}; \node[right label=T'] {guarded,\\[-.3ex]saturated};

\node[phase, below=of $(S')!0.5!(T')$] (U) {$\mathcal U$}; \node[right label=U] {guarded};

\node[phase, left=2cm of U] (obs) {$E \wcong F$};
\node[phase, below=of U] (equ) {$E = F$};

\draw[connection] (E) edge[] (E');
\draw[connection] (E') edge[] (S);
\draw[connection] (S) edge[] (S');
\draw[connection] (F) edge[] (F');
\draw[connection] (F') edge[] (T);
\draw[connection] (T) edge[] (T');
\draw[connection] (S') edge[] (U);
\draw[connection] (T') edge[] (U);
\draw[connection] (obs) edge[] (U);
\draw[connection] (U) edge[] (equ);
\end{tikzpicture}
\end{center}
\vspace{-.25cm}
\caption{Milner's completeness proof~\cite{Milner1989b} visualised. $E$ and $F$ are expressions satisfying the equation systems $\mathcal S$ and $\mathcal T$, respectively.} \label{fig:completeness-roadmap}
\end{minipage}
\hfill
\begin{minipage}[b]{.49\textwidth}
\begin{center}
\tikzset{
	state/.append style={node distance=1.5cm and .8cm, on grid},
	baseline=(1.base)
}
\begin{tikzpicture}[transition diagram]
\node[initial, state] (0) {$E$};
\node[state, below=of 0] (1) {};
\node[state, below left=of 1] (2) {};
\node[state, below right=of 1] (3) {};

\node[splitter] (*) at ($(0)!0.5!(1) + (.5cm, 0)$) {};
\path (0) edge[bend right, "$\tau$" pos=.6, auto=right] (*);
\path (*) edge[transition, bend left=35, "$\frac 1 2$" pos=.35] (1);
\path (*) edge[transition, bend right=35, "$\frac 1 2$" pos=.35, auto=right] (0);
\path (1) edge[transition, bend left, "$\tau$"] (0);

\path (1) edge[transition, "$\tau$", auto=right] (2);
\path (2) edge[transition, "$a$"] (3);
\path (1) edge[transition, "$b$"] (3);

\node[initial, state] (0) at ($(0) + (2.6cm, 0)$) {$F$};
\node[state, below=of 0] (1) {};
\node[state, below left=of 1] (2) {};
\node[state, below right=of 1] (3) {};

\path (0) edge[transition, "$\tau$"] (1);
\path (0) edge[transition, bend left, "$b$"] (3);

\node[splitter] (*) at ($(1)!0.5!(2) + ({atan(1.5/.8) + 90}:.5cm)$) {};
\path (1) edge[bend right, "$\tau$", auto=right] (*);
\path (*) edge[transition, bend right, "$\frac 1 3$", auto=right] (2);
\path (*) edge[transition, bend right, "$\frac 2 3$", auto=right] (1);
\path (2) edge[transition, "$a$"] (3);
\path (1) edge[transition, "$b$"] (3);
\end{tikzpicture}
\end{center}
\caption{The PA induced by $E, F$.} \label{fig:example-completeness}
\end{minipage}
\end{figure}

\noindent
The completeness proof follows Milner's proof~\cite{Milner1989b} as far as possible. It might be helpful to be familiar with his fundamental work (\autoref{fig:completeness-roadmap} outlines the major steps), however, we feel confident that the prospective proofs are well-understandable without prior knowledge, too.

\begin{example} \label{exm:completeness}\rm
The upcoming steps are illustrated by the example depicted in \autoref{fig:example-completeness}; here
\begin{align*}
	E &\synid \rec X \tau\prefix (\psingle X \pchoice{1/2} \psingle{\tau\prefix X \nchoice \tau\prefix a\prefix\nil \nchoice b\prefix\nil}), \\
	F &\synid b\prefix\nil \nchoice \tau\prefix \rec Y (\tau\prefix (\psingle Y \pchoice{2/3} \psingle{a\prefix\nil}) \nchoice b\prefix\nil).
\end{align*}
Note that $E$ and $F$ are in fact weakly congruent. It is easy to find a weak bisimulation (relate $E$ to $F$ and to both $\tau$ successors of $E$ and $F$); the root condition is trivial.
\end{example}

\subsection{Reduction to guarded expressions} \label{sec:completeness:sec:guardedness}

\noindent
The initial step of the completeness proof transforms an arbitrary expression into an equal, yet guarded one. It turns out that guardedness is a particularly useful property when dealing with equation systems later on. Before we begin, let us briefly discuss which guardedness notion suits our context.

Recall that we introduced probabilistic and nonprobabilistic
guardedness; it might not be obvious why we cannot reduce to nonprobabilistically guarded expressions in general! To see this, consider the PA depicted in \autoref{fig:guardedness-counterexample}, which is induced by the expression $G$ with
\begin{align*}
	G &\synid \rec X \tau\prefix (\psingle*{\rec Y b\prefix Y} \pchoice{1/2} \psingle*{\tau\prefix (\psingle*{\rec Y a\prefix Y} \pchoice{1/2} \psingle X)}), \\
	H &\synid \tau\prefix (\psingle*{\rec Y a\prefix Y} \pchoice{1/2} \psingle G).
\end{align*}
$G$ happens to be probabilistically guarded, but nonprobabilistically unguarded. Now, a nonprobabilistically guarded, equal PA is required to collapse the states $G$ and $H$ that currently establish the problematic $\tau$-loop. Clearly this is not possible; the state $G$ is able to reach the left state with maximal probability $< \frac 1 2$, whereas $H$ may reach the same state with probability $> \frac 1 2$. Hence, $G$ and $H$ are not bisimilar and there cannot exist a nonprobabilistically guarded expression equal to $G$.

Fortunately, it is possible to transform any expression into a probabilistically guarded expression. We begin by asserting the following lemma:

\begin{lemma}
Let $E$ be some expression and let $E \unguarded X$. Then $E = E \nchoice X$.
\label{lem:unguarded-variable}
\end{lemma}
\begin{proof}
The statement is shown by proving a stronger claim. Let $E \unguarded V$ for some $V \subset \Var$. Then one of the following statements holds:
\begin{enumerate}[label=(\arabic*), leftmargin=3em, topsep=0.9ex, itemsep=0.1ex]
\item $V = \{X\}$ for some $X$ and $E = E \nchoice X$, \label{lem:unguarded-variable:itm:single}
\item $V = \{X_1, \ldots, X_n\}$ with $n > 1$ and $E = E \nchoice \tau\prefix \psum_{i=1}^n {p_i} \psum_{j \in J_i} {q_{i, j}} (F_{i, j} \nchoice X_i)$ for some expressions $F_{i, j}$ and probabilities $p_i$ and $q_{i, j}$. \label{lem:unguarded-variable:itm:multiple}
\end{enumerate}
First, analyse the two cases: \ref{lem:unguarded-variable:itm:single} primitively extends the non-probabilistic case, whereas \ref{lem:unguarded-variable:itm:multiple} is new in the context of probabilistic prefixing. Intuitively, this case states that even if there are two or more variables unguarded in $E$, then there has to be some $\tau$ prefixing in $E$, that may be dragged out. We show the claim by an induction on the structure of $E$.

$E \synid \nil$: This case yields a contradiction, since $E \unguarded V$.

$E \synid Y$ for some variable $Y$: First, let $V = \{X\}$, thus $X = Y$. We obtain \ref{lem:unguarded-variable:itm:single} by axiom \axiom{N3}. The other case implies $|V| > 1$ and hence leads to a contradiction.

$E \synid \alpha\prefix \psum_{k \in I} {p_k} F_k$: We have an unguarded occurrence in $E$, hence $\alpha = \tau$. First, assume that $E \unguarded \{X\}$, so we additionally obtain $F_k \unguarded \{X\}$ for all $k \in I$. Then:
\begin{tflalign*}
	& E & \\
	\synid{} & \tau\prefix \psum_{k \in I} {p_k} F_k & \\
	={} & \tau\prefix \psum_{k \in I} {p_k} (F_k \nchoice X) & \text{induction} \\
	={} & \tau\prefix \psum_{k \in I} {p_k} (F_k \nchoice X) \nchoice X & \text{\axiom{T2}} \\
	={} & \tau\prefix \psum_{k \in I} {p_k} F_k \nchoice X & \text{induction} \\
	\synid{} & E \nchoice X.
\end{tflalign*}
So assume that $V = \{X_1, \ldots, X_n\}$ and let each $F_k \unguarded V_k \subseteq V$, such that $V = \bigcup_{k \in I} V_k$. Now consider each $k$ individually: If $|V_k| = 1$, then by induction $F_k = F_k \nchoice X$. Thus, for an arbitrary probabilistic expression $P$:
\begin{tflalign*}
	& \tau\prefix (\psingle{F_k} \pchoice {p_k} P) & \\
	={} & \tau\prefix (\psingle{F_k \nchoice X} \pchoice {p_k} P) & \text{induction} \\
	={} & \tau\prefix (\psingle{\tau\prefix \psingle{F_k \nchoice X}} \pchoice {p_k} P) & \text{\axiom{T1}} \\
	={} & \tau\prefix (\psingle{F_k \nchoice \tau\prefix \psingle{F_k \nchoice X}} \pchoice {p_k} P). & \text{\axiom{T2}}
\end{tflalign*}
If $|V_k| > 1$ instead, then by induction there are expressions $G_{k, i, j}$, such that\\ $F_k = F_k \nchoice \tau\prefix \psum_{i=1}^n {q_{k, i}} \psum_{j \in J_{k, i}} {r_{k, i, j}} (G_{k, i, j} \nchoice X_i)$. In any case:
\begin{tflalign*}
	& E & \\
	\synid{} & \tau\prefix \psum_{k \in I} {p_k} F_k & \\
	={} & \tau\prefix \psum_{k \in I} {p_k} (F_k \nchoice \tau\prefix \psum_{i=1}^n {q_{k, i}} \psum_{j \in J_{k, i}} {r_{k, i, j}} (G_{k, i, j} \nchoice X_i)) & \\
	={} & E \nchoice \tau\prefix \psum_{k \in I} {p_k} \psum_{i=1}^n {q_{k, i}} \psum_{j \in J_{k, i}} {r_{k, i, j}} (G_{k, i, j} \nchoice X_i) & \text{\axiom{T3}} \\
	={} & E \nchoice \tau\prefix \psum_{i=1}^n {\sum_{k \in I} q_{k, i}} \psum_{k \in I, j \in J_{k, i}} {p_k \cdot s_{k, i, j}} (G_{k, i, j} \nchoice X_i), & \text{re-indexing}
\end{tflalign*}
where $s_{k, i, j} = r_{k, i, j} \cdot q_{k, i} / (\sum_{k \in I} q_{k, i})$.\pagebreak[3]

$E \synid \rec Y F$: First, assume that $V = \{X\}$, so $X \neq Y$. We either obtain $F \unguarded \{X\}$ or $F \unguarded \{X, Y\}$, where the former case is easier, so we focus only on the latter one. Then $F = F \nchoice \tau\prefix (\psum_{j \in I} {q_j} (G_j \nchoice X) \pchoice p \psum_{j \in J} {r_j} (H_j \nchoice Y))$ for some expressions $G_j$ and $H_j$. Now:
\begin{tflalign*}
	& E & \\
	\synid{} & \rec Y F & \\
	={} & \rec Y (F \nchoice \tau\prefix (\psum_{j \in I} {q_j} (G_j \nchoice X) \pchoice p \psum_{j \in J} {r_j} (H_j \nchoice Y))) & \text{induction} \\
	={} & \rec Y (F \nchoice \tau\prefix \psum_{j \in I} {q_j} (G_j \nchoice X)) & \text{\axiom{R3}} \\
	={} & \rec Y (F \nchoice X \nchoice \tau\prefix \psum_{j \in I} {q_j} (G_j \nchoice X)) & \text{\axiom{T2}} \\
	={} & \rec Y (F \nchoice X) & \text{undo expansion of $F$} \\
	={} & (X \nchoice F) \subst{\rec Y (F \nchoice X)}{Y} & \text{\axiom{R1}} \\
	={} & F \subst{\rec Y F}{Y} \nchoice X & \\
	={} & \rec Y F \nchoice X & \text{\axiom{R1}} \\
	\synid{} & E \nchoice X. &
\end{tflalign*}
The case \ref{lem:unguarded-variable:itm:multiple} is solved analogously.

$E \synid F \nchoice G$: W.l.o.g.\ let $F \unguarded V$. Case \ref{lem:unguarded-variable:itm:single} yields $F = F \nchoice X$ for $V = \{X\}$. We immediately conclude by $E \synid F \nchoice G = F \nchoice X \nchoice G = F \nchoice G \nchoice X \synid E \nchoice X$. The second case is shown trivially, too.
\end{proof}

\noindent
Again, the transformations performed by the next theorem are basically identical to Milner's ideas: First, we collapse all strongly connected components (sets of states in which any two states may reach each other according to $\wtrans{}$) into single states using \axiom{R6}. Finally, all $\tau$-transitions looping in a state with probability $1$ are removed by \axiom{R5}.

Before stating the general proof, recall the example from before:

\begin{figure}
\begin{minipage}[b]{.34\textwidth}
\begin{center}
\begin{tikzpicture}[transition diagram, state/.append style={node distance=1.3cm and 1.3cm, on grid}]
\node[initial, initial where=above, state] (0) {$G$};
\node[state, below=of 0] (1) {$H$};
\node[state, left=of $(0)!0.5!(1)$] (2) {};
\node[state, right=of $(0)!0.5!(1)$] (3) {};

\node[splitter] (*0) at ($(1)!0.5!(3) + ({atan(.8/1.6) + 90}:.22cm)$) {};
\path (0) edge[bend left, "$\tau$" pos=.35] (*0);
\path (*0) edge[transition, bend left, "$\frac 1 2$"] (1);
\path (*0) edge[transition, bend right, "$\frac 1 2$" pos=.2] (3);

\node[splitter] (*1) at ($(0)!0.5!(2) + ({atan(.8/1.6) + 90}:-.22cm)$) {};
\path (1) edge[bend left, "$\tau$" pos=.35] (*1);
\path (*1) edge[transition, bend left, "$\frac 1 2$"] (0);
\path (*1) edge[transition, bend right, "$\frac 1 2$" pos=.2] (2);

\path (2) edge[transition, loop above, "$a$"] (2);
\path (3) edge[transition, loop above, "$b$"] (3);
\end{tikzpicture}
\end{center}
\vspace{-.3cm}
\caption{The PA induced by some nonprobabilistically unguarded $G$, for which there is no nonprobabilistically guarded $G'$ with $G \wcong G'$.} \label{fig:guardedness-counterexample}
\end{minipage}
\hfill
\begin{minipage}[b]{.64\textwidth}
\begin{center}
\tikzset{
	transition diagram,
	state/.append style={node distance=1.5cm and .8cm, on grid},
	baseline=(1.base)
}
\begin{tikzpicture}
\node[initial, state] (0) {$E$};
\node[state, below=of 0] (1) {};
\node[state, below left=of 1] (2) {};
\node[state, below right=of 1] (3) {};

\node[splitter] (*) at ($(0)!0.5!(1) + (.5cm, 0)$) {};
\path (0) edge[bend right, "$\tau$" pos=.6, auto=right] (*);
\path (*) edge[transition, bend left=35, "$\frac 1 2$" pos=.35] (1);
\path (*) edge[transition, bend right=35, "$\frac 1 2$" pos=.35, auto=right] (0);
\path (1) edge[transition, bend left, "$\tau$"] (0);

\path (1) edge[transition, "$\tau$", auto=right] (2);
\path (2) edge[transition, "$a$"] (3);
\path (1) edge[transition, "$b$"] (3);

\path (1) ++(1.3cm, 0) node {$=$} node[above=.25ex] {\scriptsize\axiom{R6}};

\node[state, initial] (1) at ($(1) + (2.6cm, 0)$) {};
\node[state, below left=of 1] (2) {};
\node[state, below right=of 1] (3) {};

\path (1) edge[transition, loop above, "$\tau$"] (1);

\path (1) edge[transition, "$\tau$", auto=right] (2);
\path (2) edge[transition, "$a$"] (3);
\path (1) edge[transition, "$b$"] (3);

\path (1) ++(1.3cm, 0) node {$=$} node[above=.25ex] {\scriptsize\axiom{R5}};

\node[state] (1) at ($(1) + (2.6cm, 0)$) {};
\node[initial, state, above=of 1] (0) {$E'$};
\node[state, below left=of 1] (2) {};
\node[state, below right=of 1] (3) {};

\path (0) edge[transition, "$\tau$"] (1);

\path (1) edge[transition, "$\tau$", auto=right] (2);
\path (2) edge[transition, "$a$"] (3);
\path (1) edge[transition, "$b$"] (3);
\end{tikzpicture}
\end{center}
\caption{$E$ is transformed into a guarded expression $E'$.} \label{fig:example-guardedness}
\end{minipage}
\end{figure}

\begin{example}\rm
The expression $F$ is already probabilistically guarded, so consider $E$; the transformation is shown in \autoref{fig:example-guardedness}. \autoref{lem:unguarded-variable} yields $E = \rec X \tau\prefix (\psingle X \linebreak[2]\pchoice{1/2} \psingle{X \nchoice \tau\prefix X \nchoice \tau\prefix a\prefix\nil \nchoice b\prefix\nil})$, which is transformed into $\rec X (\tau\prefix X \nchoice \tau\prefix a\prefix\nil \nchoice b\prefix\nil)$ by \axiom{R6}. Here $I = \{1, 2\}$, $E_1 \synid \nil$, $E_2 \synid \tau\prefix X \nchoice{} \tau\prefix a\prefix \nil \nchoice b\prefix \nil$ and $F \synid \nil$. We achieve guardedness after applying \axiom{R5} to obtain $E' \synid \rec X \tau\prefix (\tau\prefix a\prefix\nil \nchoice b\prefix\nil)$.
\end{example}

\begin{theorem} \label{thm:guardedness}
For every expression $E$, there exists a guarded expression $E'$ with $E = E'$.
\end{theorem}
\begin{proof}
To transform $E$ into a guarded expression, we need to transform every recursion $\rec X F$ occurring as a subterm in $E$. We perform an induction over the nesting depth of the recursion $\rec X F$ to assert that there is some expression $F'$ satisfying:
\begin{enumerate}[label=(\arabic*)]
\item $\rec X F'$ is guarded, \label{thm:guardedness:itm:rec-guarded}
\item no free, unguarded occurrence of a variable $Y$ in $F'$ lies within a recursion in $F'$, \label{thm:guardedness:itm:no-unguarded}
\item $\rec X F = \rec X F'$. \label{thm:guardedness:itm:rec-equal}
\end{enumerate}
If there is no recursion occurring as a subterm of $F$, we may skip one paragraph by choosing $F' \synid F$.

Otherwise, we assume inductively that every top-level nested recursion $\rec Y G$ yields some expression $G'$ satisfying \ref{thm:guardedness:itm:rec-guarded}, \ref{thm:guardedness:itm:no-unguarded} and \ref{thm:guardedness:itm:rec-equal}. Here, a top-level recursion shall denote a recursion contained in $F$, but not contained in any recursion in between. By \ref{thm:guardedness:itm:rec-equal} and \axiom{R1} we have $\rec Y G = \rec Y G' = G' \subst{\rec Y G'} Y$. Let $F'$ be the expression $F$ after simultaneously replacing every top-level recursion $\rec Y G$ by $G' \subst{\rec Y G'} Y$. Then obviously $F = F'$. Moreover, we observe that \ref{thm:guardedness:itm:no-unguarded} is also satisfied. To see that, consider a free, unguarded occurrence of a variable $Z$ in $F'$. Suppose that $Z$ occurs within a recursion, i.e.\ within some term $\rec Y G'$, where $\rec Y G'$ only occurs within some term $G' \subst{\rec Y G'} Y$. However, due to \ref{thm:guardedness:itm:rec-guarded} we know that each occurrence of $Y$ within $G'$ is guarded, leading to the contradiction that every occurrence of $Z$ within $\rec Y G'$ within $F'$ is guarded.

In both cases we find an expression $F' = F$ satisfying \ref{thm:guardedness:itm:no-unguarded}. It remains to transform $F'$ into an expression $F''$ that preserves \ref{thm:guardedness:itm:no-unguarded} and additionally satisfies \ref{thm:guardedness:itm:rec-guarded} and \ref{thm:guardedness:itm:rec-equal}. If $X$ only occurs guarded in $F'$, then we trivially choose $F'' \synid F'$ and conclude.

Thus, assume that $X$ occurs unguarded in $F'$. We repeat the following steps, until each unguarded occurrence of $X$ in $F'$ is eliminated. Recall that none of these occurrences is contained within a recursion, due to \ref{thm:guardedness:itm:no-unguarded}.

If each unguarded occurrence of $X$ does not lie within a prefix expression, then $F' = X \nchoice F''$, where $X$ does not occur unguarded in $F''$. By \axiom{R4}, we are allowed to remove the $X$ term, which establishes \ref{thm:guardedness:itm:rec-equal}.

Otherwise, $F' = \tau\prefix \psum_{i \in I} {p_i} G_i \nchoice H$ for some expressions $G_i$ not guarding $X$. Thus:
\begin{tflalign*}
	& \rec X F' & \\
	={} & \rec X (\tau\prefix \psum_{i \in I} {p_i} G_i \nchoice H) & \\
	={} & \rec X (\tau\prefix \psum_{i \in I} {p_i} (G_i \nchoice X) \nchoice H) & \text{\autoref{lem:unguarded-variable}} \\
	={} & \rec X (\tau\prefix \psingle X \nchoice \sum_{i \in I} G_i \nchoice H). & \text{\axiom{R6}}
\end{tflalign*}
Each non-prefixed unguarded occurrence of $X$ within some $G_i$ or $H$ is eliminated as in the previous case. Note that the maximum nesting-depth of $\tau$ prefixes around $X$ in all $G_i$ has decreased, so we inductively continue this transformation until the $G_i$ and $H$ are free of unguarded $X$ occurrences. Ultimately, it remains to apply \axiom{R5} to get rid of the $\tau\prefix \psingle X$ term; we conclude by taking $F'' \synid \tau\prefix (\sum_{i \in I} G_i \nchoice H)$.
\end{proof}

\subsection{Equation systems} \label{sec:completeness:sec:equation-systems}

\noindent
As already mentioned, \emph{equation systems} (ESs) are an important ingredient of our completeness proof. We prefer reasoning about ESs over plain expressions, because the former naturally express recursions in a standardised form. 

Formally, for variables $\vec X = (X_1, \ldots, X_n)$ and expressions $\vec S = (S_1, \ldots, S_n)$, let $\mathcal S : \vec X = \vec S$ denote an equation system with \emph{formal variables} $\vec X$. Sometimes we write $\tilde X$ to address the corresponding set $\{X_1, \ldots, X_n\}$. All non-formal variables occurring in some expression $S_i$ are called \emph{free} and we denote $\Var(\mathcal S)$ as the set of all free variables in $\mathcal S$. We say an ES $\mathcal S : \vec X = \vec S$ is \emph{satisfied} by some expression $E$, if there exist expressions $\vec E = (E_1, \ldots, E_n)$ with $E \synid E_1$, such that $\vec E = \vec S \subst{\vec E}{\vec X}$ according to our axioms.

Before we continue clarifying the intuition behind ESs, let us first introduce an important property, which we almost always presuppose:

\begin{definition}[Standardness] \label{def:es-standardness}\rm
An equation system $\mathcal S: \vec X = \vec S$ with $\vec X = (X_1, \ldots, X_n)$ and $\vec S = (S_1, \ldots, S_n)$ is called \emph{standard} if every expression $S_i$ has the shape
\[
	S_i \synid \sum_j \alpha_{i, j}\prefix \psum_{k=1}^n {p_{i, j, k}} X_k \nchoice \sum_j V_{i, j},
\]
where $V_{i, j} \in \Var(\mathcal S)$.
\end{definition}

\noindent
Intuitively, just like expressions, we continue visualising standard equation systems (SESs) as PA: The formal variables may be interpreted as states and the transition relation ${\trans{}_{\mathcal S}} \subseteq \tilde X \times \Act \times \Distr{\tilde X}$ is characterised by\vspace{-1ex}
\[
	X_i \trans\alpha_{\mathcal S} \mu \quad\text{iff}\quad S_i \trans\alpha \mu,
\]
according to our operational semantics. Note that this definition crucially depends on the standardness property; otherwise $\mu$ might support expressions not contained in $\tilde X$.

\begin{example}\rm
Consider the expressions $E$ and $F$ from \autoref{exm:completeness}, after being transformed into guarded expressions. The following SES $\mathcal S$ (resp. $\mathcal T$) is satisfied by $E$ (resp. $F$):
\begin{align*}
	\mathcal S : X_1 &= \tau\prefix X_2, & \mathcal T : Y_1 &= b\prefix Y_4 \nchoice \tau\prefix Y_2, \\
	X_2 &= \tau\prefix X_3 \nchoice b\prefix X_4, & Y_2 &= \tau\prefix (\psingle{Y_2} \pchoice{2/3} \psingle{Y_3}) \nchoice b\prefix Y_4, \\
	X_3 &= a\prefix X_4, & Y_3 &= a\prefix Y_4, \\
	X_4 &= \nil, & Y_4 &= \nil.
\end{align*}
Let us prove that $F$ satisfies $\mathcal T$, exemplary. Choose expressions $\vec F = (F_1, \ldots, F_4)$ with $F \synid F_1 \synid b\prefix\nil \nchoice \tau\prefix \rec Y (\tau\prefix (\psingle Y \pchoice{2/3} \psingle{a\prefix\nil}) \nchoice b\prefix\nil), F_2 \synid \rec Y (\tau\prefix (\psingle Y \pchoice{2/3} \psingle{a\prefix\nil}) \nchoice b\prefix\nil), F_3 \synid a\prefix\nil$ and $F_4 \synid \nil$. Now plug these values into the equations of $\mathcal T$ by replacing each $Y_i$ by $F_i$. The axiom \axiom{R1} is required to prove the second equation, the others are trivial.

The next section generalises what we have achieved here for this particular example: All guarded expressions satisfy some (guarded) SES.
\end{example}

\noindent
There are some other important properties that we rely on later in this proof. We have already defined a strong transition relation $\trans\alpha_{\mathcal S}$ on SESs, let weak transitions $\wtrans{}_{\mathcal S}$ and \raisebox{0pt}[0pt][0pt]{$\wtrans\alpha_{\mathcal S}$} be defined as before using derivations.

\begin{definition}[Guardedness] \label{def:es-guardedness}\rm
A standard equation system $\mathcal S\!: \vec X = \vec S$ is called \emph{(probabilistically) guarded}, if there exists no $i$, such that \raisebox{0pt}[0pt][0pt]{$\dirac{X_i} \wtrans\tau_{\mathcal S} \dirac{X_i}$}.
\end{definition}

\begin{definition}[Saturatedness] \label{def:es-saturatedness}\rm
A standard equation system $\mathcal S: \vec X = \vec S$ is called \emph{saturated}, if the following conditions hold:
\begin{enumerate}[label=(\roman*)]
\item whenever \raisebox{0pt}[0pt][0pt]{$\dirac{X_i} \wtrans\alpha_{\mathcal S} \mu$}, then also $\dirac{X_i} \trans\alpha_{\mathcal S} \mu$, \label{def:es-saturatedness:itm:trans}
\item whenever $\dirac{X_i} \wtrans{}_{\mathcal S} \mu$ and $V \in \Var(\mathcal S)$ occurs in $S_j$ for all $j$ with $\mu(X_j) > 0$, then $V$ occurs in $S_i$.\label{def:es-saturatedness:itm:var}
\end{enumerate}
\end{definition}

\noindent
Note that both definitions essentially coincide with Milner's versions in the nonprobabilistic setting. However, asserting these properties when dealing with infinitary probabilistic weak transitions results in a substantial overhead in many of the following steps.

Hereinafter, it proves useful to introduce some handy notation: When considering some probability distribution $\mu$ over $\tilde X$, we sometimes write $\mu[\vec X]$ to clarify the domain of $\mu$. Furthermore, we may write $\mu[\vec S]$ to cast $\mu$ to a distribution over the expressions in $\vec S$, i.e.\ $\mu[\vec S] = \distr{S_j \mapsto \mu(X_j)}$.

\subsection{Equational characterisation} \label{sec:completeness:sec:equational-characterisation}

\noindent
The completeness proof continues by transforming a given (guarded) expression $E$ into an SES $\mathcal S$ satisfied by $E$. In fact, we construct $\mathcal S$ to be guarded and maintain this property until the very end of the completeness proof, where it becomes highly significant.

Although the general idea à la Milner carries over easily, the details of proving $\mathcal S$ being guarded turn out to be quite complex.

\begin{restatable}[Equational characterisation]{theorem}{equationalcharacterisation} \label{thm:equational-characterisation}
Given some guarded expression $E$, there is a guarded SES $\mathcal S$ with free variables in $\Var(E)$ satisfied by $E$.
\end{restatable}
\begin{proof}
See \autoref{apx:equational-characterisation}.
\end{proof}

\subsection{Saturation} \label{sec:completeness:sec:saturation}

\noindent
The next step of the completeness proof transforms the guarded SESs obtained by the previous section into saturated, guarded SESs.

Milner's saturation proof proceeds essentially as follows: We simply shortcut along every weak transition of the ES until we reach saturatedness. Note that there are two implicit assumptions: There is only a finite number of different weak transitions and each weak transition is of finite length. Unfortunately, both assumptions fail in our context! We use infinitary probabilistic weak transitions defined by derivations, i.e.\ infinite sequences. Furthermore, the number of weak transitions is not guaranteed to be bounded: Consider the PA in \autoref{fig:infinitary}. There exist transitions \raisebox{0pt}[0pt][0pt]{$s \wtrans\tau \distr{s \mapsto \frac 1 2, t \mapsto \frac 1 2}$}, \raisebox{0pt}[0pt][0pt]{$s \wtrans\tau \distr{s \mapsto \frac 1 4, t \mapsto \frac 3 4}, \ldots$}, more precisely: there exists a transition \raisebox{0pt}[0pt][0pt]{$s \wtrans\tau \distr{s \mapsto (\frac 1 2)^k, t \mapsto 1 - (\frac 1 2)^k}$} for all $k \geq 1$. Even worse: We already observed that weak transitions are convex, thus the amount of different weak transitions is uncountable.

We deal with these issues by strongly exploiting an important result of \cite{Deng2009}, originating from Markov Decision Theory \cite{Puterman1994}:

\begin{proposition}[\cite{Deng2009}] \label{prp:pa-finite-generability-alpha}
In a finite-state PA with finitely many transitions\footnote{The authors of
    \cite{Deng2009} refer to PA with a finite amount of states and transitions as
    \emph{finitary}.}, for a fixed $\nu$ and any action $\alpha$, there is a finite set of distributions $\{\mu_1, \ldots, \mu_m\}$, such that $\nu \wtrans\alpha \mu_i$ for all $i$ and whenever $\nu \wtrans\alpha \mu$, then $\mu$ is a convex combination of $\mu_1, \ldots, \mu_m$.
\end{proposition}

\noindent
Using this fact, the first infinity is solved: In every ES $\mathcal S: \vec X = \vec S$, there is only a finite number of `states' $X_i$, and for each one \autoref{prp:pa-finite-generability-alpha} yields a finite number of interesting transitions. So all in all, there is only a finite number of weak transitions left that needs to be translated into strong transitions. We continue with another related result of \cite{Deng2009}:

\begin{proposition}[\cite{Deng2009}] \label{prp:pa-finite-generability-stationary}
In a finite-state PA with finitely many transitions, for a fixed $\nu$, there is a finite set of stationary policies $\{\Theta_1, \ldots, \Theta_m\}$ deriving transitions $\nu \wtrans{}^{\Theta_i} \mu_i$, such that whenever $\nu \wtrans{} \mu$, then $\mu$ is a convex combination of $\mu_1, \ldots, \mu_m$.
\end{proposition}

\noindent
Here, a stationary policy $\Theta : S \to \SubDistr S$ resolves the nondeterministic choice in each state of the PA by scheduling a target subdistribution, i.e.\ whenever $\Theta(s) = \mu$, then $|\mu| \cdot \dirac s \trans\tau \mu$. The non-assigned part of the subdistribution corresponds to stopping in the respective state. A weak transition $\nu \wtrans{} \mu$ is derivable by some stationary policy $\Theta$, denoted by $\nu \wtrans{}^\Theta \mu$, if\footnote{The authors of \cite{Deng2009} introduced (stationary) policies acting on derivations. In our context, we omit the additional definitions for the sake of convenience and settle with the given easily-derivable product formula.}
\[
	\mu = \sum_{s_0 \cdots s_i \in S^+} \nu(s_0) \cdot ({\textstyle\prod_{j=0}^{i-1} \Theta(s_j)(s_{j+1})}) \cdot (1 - |\Theta(s_i)|) \cdot \dirac{s_i}.
\]

\subsubsection{Transition trees} \label{sec:completeness:sec:saturation:sec:transition-trees}

\noindent
So now each transition to be considered is derivable by some stationary policy. The derivation might still be infinite. However, a certain degree of regularity is introduced, which we are going to exploit. To be properly prepared, we introduce a concept equivalent to derivations, namely \emph{transition trees}. A derivation only yields information about the linear course of a transition, meaning that after every step, all target distributions are combined into one single distribution. A transition tree also keeps track of the branching structure, i.e.\ it assigns probabilities to all possible runs $s_0 \cdots s_i \in S^+$.

\begin{definition}[Transition Tree] \label{def:tt}\rm
A \emph{transition tree} (TT) is a mapping $\vartheta : S^* \to \SubDistr S$ satisfying the following conditions, for all $s_0 \cdots s_i \in S^+$:
\begin{enumerate}[label=(\roman*)]
\item $|\vartheta(s_0 \cdots s_i)| \leq \vartheta(s_0 \cdots s_{i-1})(s_i)$, and \label{def:tt:itm:consistent}
\item $|\vartheta(s_0 \cdots s_i)| \cdot \dirac{s_i} \trans\tau \vartheta(s_0 \cdots s_i)$. \label{def:tt:itm:transition}
\end{enumerate}
\end{definition}

\noindent
We will argue precisely about the correspondence between derivations and transition trees in the next proof (even though we only show one direction), before, we need some definitions: We refer to a \emph{node} as a sequence $s_0 \cdots s_i \in S^+$. $\vartheta(s_0 \cdots s_i)(s_{i+1})$ is interpreted as the probability of following a path with prefix $s_0 \cdots s_i s_{i+1}$ in the weak transition and the probability of stopping right after reaching $s_i$ is given by $\vartheta(s_0 \cdots s_{i-1})(s_i) - |\vartheta(s_0 \cdots s_i)|$. This motivates the next term: The \emph{volume} of some node $s_0 \cdots s_i$ is defined as
\[
	\Vol_\vartheta(s_0 \cdots s_i) = \sum_{s_{i+1} \cdots s_j \in S^*} (\vartheta(s_0 \cdots s_{j-1})(s_j) - |\vartheta(s_0 \cdots s_j)|) \cdot \dirac{s_j},
\]
which expresses the target probability distribution of the transition induced by $\vartheta$ after following the sequence $s_0 \cdots s_i$ through the PA. We say that $\vartheta$ \emph{starts from} $\vartheta(\varepsilon)$ and \emph{leads to} $\sum_{s_0 \in S} \Vol_\vartheta(s_0)$, which we sometimes denote by $\vartheta(\varepsilon) \wtrans{}^\vartheta \sum_{s_0 \in S} \Vol_\vartheta(s_0)$. As already mentioned, transition trees in general are equipotent to derivations, so we are focusing on a restricted subclass, namely \emph{stationary} transition trees:

\begin{definition} \label{def:stt}\rm
A transition tree $\vartheta$ is called \emph{stationary} (STT) if for all nodes $s_0 \cdots s_i$ and $t_0 \cdots t_j$ with $\vartheta(s_0 \cdots s_{i-1})(s_i) > 0$, $\vartheta(t_0 \cdots t_{j-1})(t_j) > 0$ and $s_i = t_j$,
\[
	\frac{\Vol_\vartheta(s_0 \cdots s_i)}{\vartheta(s_0 \cdots s_{i-1})(s_i)} = \frac{\Vol_\vartheta(t_0 \cdots t_j)}{\vartheta(t_0 \cdots t_{j-1})(t_j)}.
\]
\end{definition}

\noindent
So in a stationary transition tree $\vartheta$, the volume of any node $s_0 \cdots s_i$ only depends on $s_i$, up to some scalar. This property now allows a finite representation of transition trees, as explained in the following. We call a node $s_0 \cdots s_i$ \emph{pioneer} if there is no state occurring more than once in the sequence $s_0 \cdots s_i$. Furthermore, we call $s_0 \cdots s_i$ \emph{stopping} if $\vartheta(s_0 \cdots s_i) = \emptydistr$ and \emph{non-stopping} otherwise.

As an example, consider the automaton depicted in \autoref{fig:stt:sub:pa}; think of a weak transition that stops in both $s$ and $t$ with probability $\frac 1 2$ and halts in $u$ certainly. Otherwise, in $t$, the combined transition leading to both $s$ and $u$ with probability $\frac 1 2$, respectively, is scheduled. That weak transition is induced exactly by the following stationary policy $\Theta$:
\begin{align*}
	\Theta(s) &= \distr{t \mapsto \tfrac 1 3, u \mapsto \tfrac 1 6}, &
	\Theta(t) &= \distr{s \mapsto \tfrac 1 4, u \mapsto \tfrac 1 4}, &
	\Theta(u) &= \emptydistr.
\end{align*}
This transition may as well be drawn as an STT; \autoref{fig:stt:sub:original} shows every pioneer, non-stopping node (here each node is labelled only with its last state). The dashed connection leading from $t$ back to $s$ symbolises a \emph{recursion}. Formally, the tree continues here infinitely long, however, the stationary property allows us to think of a recursive loop instead.

It is easy to see that this correspondence between stationary policies and STTs holds in general:

\begin{figure}
\centering
\tikzset{
	transition diagram,
	state/.append style={on grid},
	node distance=1.5cm and .8cm,
	baseline=(s)
}
\begin{subfigure}[t]{.225\textwidth}
\centering
\begin{tikzpicture}
\node[initial, initial where=above, state] (s) {$s$};
\node[state, below=of s] (t) {$t$};
\node[state, below=of t] (u) {$u$};

\path (t) edge[transition, "$\tau$"] (s);
\path (t) edge[transition, "$\tau$", auto=right] (u);

\node[splitter] (*) at ($(s)!0.7!(t) + (.7cm, 0)$) {};
\path (s) edge["$\tau$", bend left] (*);
\path (*) edge[transition, "$\frac 2 3$", auto=right, bend left] (t);
\path (*) edge[transition, "$\frac 1 3$", bend left] (u);
\end{tikzpicture}
\subcaption{The PA.} \label{fig:stt:sub:pa}
\end{subfigure}
\hfill
\begin{subfigure}[t]{.225\textwidth}
\centering
\begin{tikzpicture}
\node[initial, initial where=above, state] (s) {$s$};
\node[state, below right=of s] (su) {$u$};
\node[state, below left=of s] (st) {$t$};
\node[state, below=of st] (stu) {$u$};

\path (s) edge[transition, "$\frac 1 3$"] (st);
\path (s) edge[transition, "$\frac 1 6$"] (su);
\path (st) edge[transition, "$\frac 1 4$"] (stu);
\path (st) edge[transition, dashed, "$\frac 1 4$", bend left] (s);
\end{tikzpicture}
\subcaption{The STT $\vartheta$ induced by the stationary policy $\Theta$.} \label{fig:stt:sub:original}
\end{subfigure}
\hfill
\begin{subfigure}[t]{.225\textwidth}
\centering
\begin{tikzpicture}
\node[initial, initial where=above, state] (s) {$s$};
\node[state, below right=of s] (su) {$u$};
\node[state, below left=of s] (st) {$t$};

\path (s) edge[transition, "$\frac 1 6$"] (st);
\path (s) edge[transition, "$\frac 1 4$"] (su);
\path (s) edge[transition, dashed, "$\frac 1 {12}$", loop left] (s);
\end{tikzpicture}
\subcaption{Shortcutting $t$.} \label{fig:stt:sub:shortcutting}
\end{subfigure}
\hfill
\begin{subfigure}[t]{.225\textwidth}
\centering
\begin{tikzpicture}
\node[initial, initial where=above, state] (s) {$s$};
\node[state, below right=of s] (su) {$u$};
\node[state, below left=of s] (st) {$t$};

\path (s) edge[transition, "$\frac 2 {11}$", auto=right] (st);
\path (s) edge[transition, "$\frac 3 {11}$"] (su);
\end{tikzpicture}
\subcaption{Linearising $s$.} \label{fig:stt:sub:linearising}
\end{subfigure}
\caption{A weak transition depicted as stationary transition trees.} \label{fig:stt}
\end{figure}
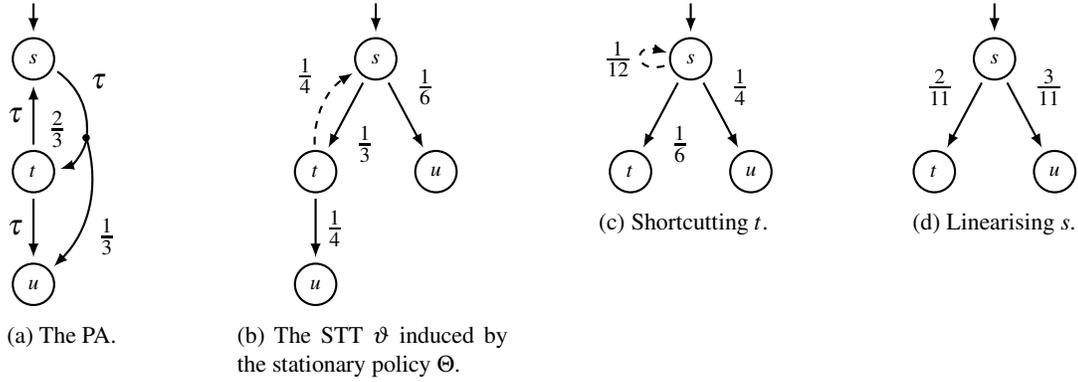

\begin{lemma} \label{lem:stt-existence}
Whenever $\nu \wtrans{} \mu$, there is some transition tree $\vartheta$, such that $\nu \wtrans{}^\vartheta \mu$. Furthermore, if $\nu \wtrans{}^\Theta \mu$ is derivable by some stationary policy $\Theta$, then $\vartheta$ is stationary.
\end{lemma}

\noindent
We proceed by presenting two operations on STTs: \emph{shortcutting} and \emph{linearising}. Roughly speaking, shortcutting removes a node and connects its incoming and outgoing transitions appropriately, whereas linearising means removing a node's self-loop. After repeatedly applying these operations, we lastly achieve a TT of depth $1$, which is (nearly) equivalent to a strong transition.

The authors of \cite{Eisentraut2013} have developed an algorithm to compute the minimal quotient PA with respect to $\wbis$, which conceptually uses both procedures reversely: Instead of expanding the PA by adding transitions, the respective transitions are removed.

Technically, we perform an induction on the size of the transition tree, which is yet to be defined: On the one hand, whenever we shortcut a node, then the number of non-stopping pioneer nodes decreases. On the other hand, whenever a loop is linearised, the number of recursions is guaranteed to decrease. Therefore, the size $|\vartheta|$ of some stationary transition tree $\vartheta$ is defined as the sum of all non-stopping, pioneer states and all recursions. Formally, a pioneer node $s_0 \cdots s_i$ recurs on some node $s_0 \cdots s_j$ with $j < i$, if $\vartheta(s_0 \cdots s_i)(s_j) > 0$.

Recall that we are concerned with weak transitions induced by ESs in this section. So whenever we perform one of the two operations, the underlying ES is transformed (only using the axioms), such that the modified STT remains derivable. Hence, given an SES $\mathcal S : \vec X = \vec S$, we make sure that both operations return an SES $\mathcal T : \vec X = \vec T$, which is \emph{augmented} with respect to $\mathcal S$, that is, $\mathcal T$ is satisfied by exactly all expressions satisfying $\mathcal S$ and meets the syntactical requirement $T_i \synid S_i \nchoice T_i'$ for some expressions $T_i'$. The last condition ensures that every strong transition in $\mathcal S$ occurs in $\mathcal T$, too.

\subsubsection{Shortcutting} \label{sec:completeness:sec:saturation:sec:shortcutting}

\noindent
This operation essentially extends the nonprobabilistic case; the infinitary property of weak transitions does not play a role, here. Nevertheless, we still give a detailed proof, as the technical skeleton introduced in this section requires careful treatment.

\begin{lemma}[Shortcutting] \label{lem:shortcutting}
Let $\mathcal S: \vec X = \vec S$ be a guarded SES and let $\mu \wtrans{}_{\mathcal S}^\vartheta \nu$ for some STT $\vartheta$. If there is a non-stopping, pioneer node $\cdots X_i X_j$, such that there exists no node recurring in $\cdots X_i X_j$, then there is some augmented guarded SES $\mathcal T: \vec X = \vec T$ with $\mu \wtrans{}_{\mathcal T}^{\vartheta'} \nu$ for some STT $\vartheta'$ with $|\vartheta'| < |\vartheta|$.
\end{lemma}
\begin{proof}
The key idea is to bypass the node $\cdots X_i X_j$ whenever it is possible. In detail, if there is a transition leaving $\cdots X_i X_j$, we might as well combine this transition with the incoming transition to $\cdots X_i X_j$, such that $\cdots X_i X_j$ is not visited in between. It remains possible that $\cdots X_i X_j$ is reachable with positive probability, however, then the TT stops here inevitably. The crucial modification of the underlying ES is justified by the axiom \axiom{T3} (or \axiom{T4}, in fact both axioms are equivalent in this context, since we are only dealing with $\tau$ actions).

By $\vartheta$ being a transition tree, we infer that $|\vartheta(\cdots X_i)| > 0$ and $|\vartheta(\cdots X_i X_j)| > 0$. Moreover, there are transitions\vspace{-1ex}
\begin{align}
	\dirac{X_i} &\trans\tau \frac{1}{|\vartheta(\cdots X_i)|} \cdot \vartheta(\cdots X_i), \label{lem:shortcutting:eqn:transition-xi} \\
	\dirac{X_j} &\trans\tau \frac{1}{|\vartheta(\cdots X_i X_j)|} \cdot \vartheta(\cdots X_i X_j). \label{lem:shortcutting:eqn:transition-xj}
\end{align}
Therefore, we conclude that the expressions $S_i$ and $S_j$ may be transformed to
\begin{align}
	S_i &= S_i \nchoice \tau\prefix \psum_{k=1}^n {\frac{\vartheta(\cdots X_i)}{|\vartheta(\cdots X_i)|} (X_k)} X_k, \label{lem:shortcutting:eqn:rewrite-si} \\
	S_j &= S_j \nchoice \tau\prefix \psum_{l=1}^n {\frac{\vartheta(\cdots X_i X_j)}{|\vartheta(\cdots X_i X_j)|} (X_l)} X_l. \label{lem:shortcutting:eqn:rewrite-sj}
\end{align}
Note that \eqref{lem:shortcutting:eqn:transition-xi} and \eqref{lem:shortcutting:eqn:transition-xj} only imply combined transitions. Thus, beside the usual $\nchoice$ axioms, it might have been necessary to apply the convexity axiom \axiom{C}.

We choose $\mathcal T: \vec X = \vec T$ as follows:
\[
	T_k \synid
	\begin{cases}
		S_k &\text{if $k \neq i$,} \\
		S_i \nchoice \tau\prefix \psum_{l=1}^n {\frac{\vartheta(\cdots X_i) - |\vartheta(\cdots X_i X_j)| \cdot \dirac{X_j} + \vartheta(\cdots X_i X_j)}{|\vartheta(\cdots X_i)|} (X_l)} X_l &\text{if $k = i$.}
	\end{cases}
\]
This assignment of $T_i$ yields a transition
\begin{equation} \label{lem:shortcutting:eqn:transition-shortcut}
	|\vartheta(\cdots X_i)| \cdot \dirac{X_i} \trans\tau_{\mathcal T} \vartheta(\cdots X_i) - |\vartheta(\cdots X_i X_j)| \cdot \dirac{X_j} + \vartheta(\cdots X_i X_j),
\end{equation}
which is required in the later part of the proof. First, we assert that any expression $E$ satisfies $\mathcal S$ if and only if it satisfies $\mathcal T$. So let $E$ satisfy $\mathcal S$, i.e.\ we gain expressions $\vec E$, such that $\vec E = \vec S\subst{\vec E}{\vec X}$. For any $k \neq i$, we clearly have $E_k = S_k\subst{\vec E}{\vec X} \synid T_k\subst{\vec E}{\vec X}$. So consider the interesting case:
\begin{tflalign*}
	& E_i & \\
	={} & S_i\subst{\vec E}{\vec X} & \\
	={} & S_i\subst{\vec E}{\vec X} \nchoice \tau\prefix \psum_{k=1}^n {\frac{\vartheta(\cdots X_i)}{|\vartheta(\cdots X_i)|} (X_k)} E_k & \text{\eqref{lem:shortcutting:eqn:rewrite-si}} \\
\intertext{Recall that $\vartheta$ is a transition tree, thus it is guaranteed that $\vartheta(\cdots X_i)(X_j) \geq |\vartheta(\cdots X_i X_j)|$. This justifies the following expansion:}
	={} & S_i\subst{\vec E}{\vec X} \nchoice \tau\prefix \psum_{k=1}^n {\frac{\vartheta(\cdots X_i) - |\vartheta(\cdots X_i X_j)| \cdot \dirac{X_j} + |\vartheta(\cdots X_i X_j)| \cdot \dirac{X_j}}{|\vartheta(\cdots X_i)|} (X_k)} E_k & \\
	={} & S_i\subst{\vec E}{\vec X} \nchoice \tau\prefix \Big( \psingle{E_j} \pchoice{\frac{|\vartheta(\cdots X_i X_j)|}{|\vartheta(\cdots X_i)|}} \psum_{k=1}^n {\frac{\vartheta(\cdots X_i) - |\vartheta(\cdots X_i X_j)| \cdot \dirac{X_j}}{|\vartheta(\cdots X_i)| - |\vartheta(\cdots X_i X_j)|} (X_k)} E_k \Big) & \\
	={} & S_i\subst{\vec E}{\vec X} \nchoice \tau\prefix \Big( \psingle{S_j\subst{\vec E}{\vec X} \nchoice \tau\prefix \psum_{l=1}^n {\frac{\vartheta(\cdots X_i X_j)}{|\vartheta(\cdots X_i X_j)|} (X_l)} E_l} & \text{\eqref{lem:shortcutting:eqn:rewrite-sj}} \\
	& \pchoice{\frac{|\vartheta(\cdots X_i X_j)|}{|\vartheta(\cdots X_i)|}} \psum_{k=1}^n {\frac{\vartheta(\cdots X_i) - |\vartheta(\cdots X_i X_j)| \cdot \dirac{X_j}}{|\vartheta(\cdots X_i)| - |\vartheta(\cdots X_i X_j)|} (X_k)} E_k \Big) & \\
	={} & S_i\subst{\vec E}{\vec X} \nchoice \tau\prefix \Big( \psingle{S_j\subst{\vec E}{\vec X} \nchoice \tau\prefix \psum_{l=1}^n {\frac{\vartheta(\cdots X_i X_j)}{|\vartheta(\cdots X_i X_j)|} (X_l)} E_l} & \text{\axiom{T1}} \\
	& \pchoice{\frac{|\vartheta(\cdots X_i X_j)|}{|\vartheta(\cdots X_i)|}} \psum_{k=1}^n {\frac{\vartheta(\cdots X_i) - |\vartheta(\cdots X_i X_j)| \cdot \dirac{X_j}}{|\vartheta(\cdots X_i)| - |\vartheta(\cdots X_i X_j)|} (X_k)} (\nil \nchoice \tau\prefix \psingle{E_k}) \Big) & \\
	={} & S_i\subst{\vec E}{\vec X} \nchoice \tau\prefix \Big( \Big( \psum_{l=1}^n {\frac{\vartheta(\cdots X_i X_j)}{|\vartheta(\cdots X_i X_j)|} (X_l)} E_l \Big) \pchoice{\frac{|\vartheta(\cdots X_i X_j)|}{|\vartheta(\cdots X_i)|}} \Big(\psum_{k=1}^n {\frac{\vartheta(\cdots X_i) - |\vartheta(\cdots X_i X_j)| \cdot \dirac{X_j}}{|\vartheta(\cdots X_i)| - |\vartheta(\cdots X_i X_j)|} (X_k)} E_k \Big) \Big) & \text{\axiom{T3}} \\
	={} & S_i\subst{\vec E}{\vec X} \nchoice \tau\prefix \psum_{k=1}^n {\frac{\vartheta(\cdots X_i) - |\vartheta(\cdots X_i X_j)| \cdot \dirac{X_j} + \vartheta(\cdots X_i X_j)}{|\vartheta(\cdots X_i)|} (X_k)} E_k & \\
	\synid{} & T_i\subst{\vec E}{\vec X}. &
\end{tflalign*}
So $\mathcal T$ is satisfied by some expression $E$ if and only if $\mathcal S$ is satisfied by $E$. Moreover, the syntactical condition is clearly met, so $\mathcal T$ is in fact augmented with respect to $\mathcal S$. Note that the only strong transition derivable in $\mathcal T$, but not in $\mathcal S$, \eqref{lem:shortcutting:eqn:transition-shortcut}, exists as a weak transition in $\mathcal S$. So by some elementary reasoning about weak transitions, we find that the guardedness property holds on $\mathcal T$ as well.

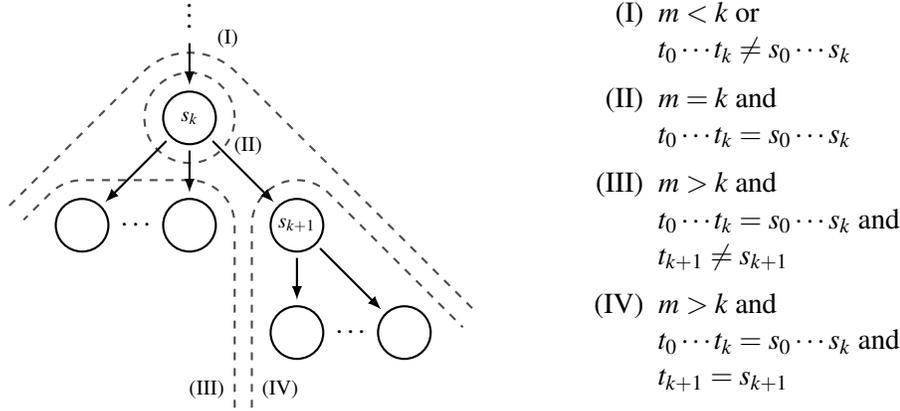
\begin{figure}
\begin{subfigure}{.55\textwidth}
\begin{center}
\begin{tikzpicture}[
	transition diagram,
	state/.append style={font=\scriptsize, minimum size=.7cm, inner sep=.01cm},
	node distance=.7cm,
	region/.style={thick, dashed, black!70}
]
\node[] (0) {${\vdots}$};
\node[state, below=of 0] (1) {$s_k$};
\node[state, below=of 1] (3) {};
\node[state, left=of 3] (2) {};
\node[state, right=of 3] (4) {$s_{k+1}$};
\node[state, below=of 4] (5) {};
\node[state, right=of 5] (6) {};

\node[] at ($(2)!.5!(3)$) {${\cdots}$};
\node[] at ($(5)!.5!(6)$) {${\cdots}$};

\path (0) edge[transition] (1);
\path (1) edge[transition] (2);
\path (1) edge[transition] (3);
\path (1) edge[transition] (4);
\path (4) edge[transition] (5);
\path (4) edge[transition] (6);

\begin{scope}[on background layer]
	\draw[region] let \p1 = ($(0) + (0, -.6cm) - (1)$) in ($(2) + (135:\y1) + (-135:.5cm)$) -- ($(1) + (135:\y1)$) arc (135:45:\y1) -- ($(6) + (45:\y1) + (-45:.5cm)$);
	\draw[region] ($(1) + (.6cm, 0)$) arc (0:360:.6cm);
	\draw[region] ($(3 |- 5) + (.6cm, -1cm)$) -- ($(3) + (.6cm, 0)$) arc (0:90:.6cm) -- ($(2) + (0, .6cm)$) arc (90:135:.6cm) -- ($(2) + (135:.6cm) + (-135:.5cm)$);
	\draw[region] ($(5) + (-.6cm, -1cm)$) -- ($(4) + (-.6cm, 0)$) arc (180:45:.6cm) -- ($(6) + (45:.6cm) + (-45:.5cm)$);
\end{scope}
\begin{scope}[on background layer, font=\scriptsize]
	\node[below right=-.25cm and .05cm of 0] {\ref{fig:shortcutting-regions:itm:1}};
	\node[below right=-.15cm and .2cm of 1] {\ref{fig:shortcutting-regions:itm:2}};
	\node[left] at ($(3 |- 5) + (.6cm, -.75cm)$) {\ref{fig:shortcutting-regions:itm:3}};
	\node[right] at ($(5) + (-.6cm, -.75cm)$) {\ref{fig:shortcutting-regions:itm:4}};
\end{scope}
\end{tikzpicture}
\end{center}
\end{subfigure}
\hfill
\begin{subfigure}{.44\textwidth}
\begin{enumerate}[label=(\Roman*), rightmargin=0cm]
\item $m < k$ or\\$t_0 \cdots t_k \neq s_0 \cdots s_k$ \label{fig:shortcutting-regions:itm:1}
\item $m = k$ and\\$t_0 \cdots t_k = s_0 \cdots s_k$ \label{fig:shortcutting-regions:itm:2}
\item $m > k$ and\\$t_0 \cdots t_k = s_0 \cdots s_k$ and\\$t_{k+1} \neq s_{k+1}$ \label{fig:shortcutting-regions:itm:3}
\item $m > k$ and\\$t_0 \cdots t_k = s_0 \cdots s_k$ and\\$t_{k+1} = s_{k+1}$ \label{fig:shortcutting-regions:itm:4}
\end{enumerate}
\end{subfigure}
\caption{Each sequence $t_0 \cdots t_m \in \tilde X^*$ is contained in a specially treated region of $\vartheta$.} \label{fig:shortcutting-regions}
\end{figure}

It remains to show, that there is some STT $\vartheta'$ smaller than $\vartheta$ and deriving $\mu \wtrans{}_{\mathcal T}^{\vartheta'} \nu$. Let $s_0 \cdots s_k s_{k+1}$ denote the designated sequence from before, i.e.\ $s_k = X_i$ and $s_{k+1} = X_j$. We assign $\vartheta'(t_0 \cdots t_m)$ depending on the region in which $t_0 \cdots t_m$ is contained in $\vartheta$. \autoref{fig:shortcutting-regions} distinguishes the four cases:
\[
	\vartheta'(t_0 \cdots t_m) =
	\begin{cases}
		\vartheta(t_0 \cdots t_m) &\text{if \ref{fig:shortcutting-regions:itm:1},} \\
		\vartheta(s_0 \cdots s_k) - |\vartheta(s_0 \cdots s_{k+1})| \cdot \dirac{s_{k+1}} + \vartheta(s_0 \cdots s_{k+1}) &\text{if \ref{fig:shortcutting-regions:itm:2},} \\
		\vartheta(s_0 \cdots s_{k+1} t_{k+1} \cdots t_m) + \vartheta(t_0 \cdots t_m) &\text{if \ref{fig:shortcutting-regions:itm:3},} \\
		\emptydistr &\text{if \ref{fig:shortcutting-regions:itm:4}.}
	\end{cases}
\]

To visualise this construction, recall the example from before considering \autoref{fig:stt}. After shortcutting the node $t$ in \autoref{fig:stt:sub:original}, we obtain the STT depicted in \autoref{fig:stt:sub:shortcutting}. $t$ is originally reached with probability $\frac 1 3$ and both $s$ and $u$ are reachable from $t$ with probability $\frac 1 4$. Now, the shortcutting connects these transitions, such that both $s$ and $u$ become immediate successors of $s$ with probability $\frac 1 3 \cdot \frac 1 4 = \frac 1 {12}$, respectively; $u$ is already reachable with probability $\frac 1 6$, so we end up with $\frac 1 6 + \frac 1 {12} = \frac 1 4$. It is important to observe that $t$ remains a successor of $s$ with a positive probability of $\frac 1 6$. However, $t$ itself has no successor any longer, which makes $t$ stopping.

We continue by asserting that $\vartheta'$ is in fact a transition tree. So let $t_0 \cdots t_m \in \tilde X^+$ be some sequence and distinguish as before: The cases \ref{fig:shortcutting-regions:itm:1} and \ref{fig:shortcutting-regions:itm:4} are trivial. If \ref{fig:shortcutting-regions:itm:2} arises, we make use of \eqref{lem:shortcutting:eqn:transition-shortcut} to prove \ref{def:tt:itm:transition}. Moreover, it is easy to see that $|\vartheta'(s_0 \cdots s_k)| = |\vartheta(s_0 \cdots s_k)|$, thus \ref{def:tt:itm:consistent} holds as well. So consider the case \ref{fig:shortcutting-regions:itm:3}. \ref{def:tt:itm:transition} is straightforward by $\vartheta$ being a transition tree. In order to show \ref{def:tt:itm:consistent}, we need to distinguish further between $m = k + 1$ and $m > k + 1$. The latter case is easy; the former case is solved by:
\begin{tflalign*}
	& |\vartheta'(t_0 \cdots t_{k+1})| & \\
	={} & |\vartheta(s_0 \cdots s_{k+1} t_{k+1})| + |\vartheta(t_0 \cdots t_{k+1})| & \\
	\leq{} & \vartheta(s_0 \cdots s_{k+1})(t_{k+1}) + \vartheta(t_0 \cdots t_k)(t_{k+1}) & \\
	={} & \vartheta'(t_0 \cdots t_k)(t_{k+1}). &
\end{tflalign*}

It turns out to be exhausting, yet technically easy to prove that $\vartheta'$ is indeed stationary; we omit the details here. Moreover, we then find that $\vartheta'$ leads to $\nu$, too.

We continue by showing that $|\vartheta'| < |\vartheta|$. It is easy to check that any recursion in $\vartheta'$ results from at least one recursion in $\vartheta$, so the number of recursions is not increasing. However, there are fewer non-stopping pioneer nodes in $\vartheta'$ than in $\vartheta$: For any non-stopping pioneer node $t_0 \cdots t_m$ in $\vartheta'$, we find (at least) one corresponding non-stopping node in $\vartheta$. If $m \leq k$ or $t_0 \cdots t_k \neq s_0 \cdots s_k$, then we choose $t_0 \cdots t_m$. Otherwise, if $m > k$ and $t_0 \cdots t_k = s_0 \cdots s_k$ is non-stopping and pioneer in $\vartheta'$, then so is $s_0 \cdots s_{k+1} t_{k+1} \cdots t_m$ or $t_0 \cdots t_m$ in $\vartheta$. Here comes into play that there exists no node recurring on $s_0 \cdots s_{k+1}$ in $\vartheta$; this assumption is necessary, since otherwise $s_0 \cdots s_{k+1} t_{k+1} \cdots t_m$ might be non-stopping, but not pioneer. Note that this correspondence is injective and $s_0 \cdots s_{k+1}$ in $\vartheta$ finds no non-stopping partner in $\vartheta'$. Hence, the number of non-stopping pioneer nodes in $\vartheta'$ is strictly smaller in comparison to $\vartheta$.
\end{proof}

\subsubsection{Linearising} \label{sec:completeness:sec:saturation:sec:linearising}

\noindent
The next operation seems far more interesting in our context: Intuitively, we unroll transitions that partially loop back to their origin state until reaching their `limit transitions'; clearly, this problem emerges not before considering probabilistic finite-state systems. The infinitary nature of weak transitions, captured by \axiom{R3}, does the trick here.

\begin{lemma}[Linearising] \label{lem:linearising}
Let $\mathcal S: \vec X = \vec S$ be a guarded SES and let $\mu \wtrans{}_{\mathcal S}^\vartheta \nu$ for some STT $\vartheta$. If there is a pioneer node $\cdots X_i$ with $\vartheta(\cdots X_i)(X_i) > 0$, then there is some augmented guarded SES $\mathcal T: \vec X = \vec T$ with $\mu \wtrans{}_{\mathcal T}^{\vartheta'} \nu$ for some STT $\vartheta'$ with $|\vartheta'| < |\vartheta|$.
\end{lemma}
\begin{proof}
There are two possible cases: Either there is another variable $X_j \neq X_i$ with $\vartheta(\cdots X_i)(X_j) > 0$, or there is not. The latter case represents $X_i$ stopping with a certain probability and self-looping, otherwise. We easily find an STT $\vartheta'$ that immediately stops once reaching $X_i$. So let us focus on the first case. $\vartheta$ is a TT, so we have
\begin{equation} \label{lem:linearising:eqn:transition-xi} 
	\dirac{X_i} \trans\tau_{\mathcal S} \frac{1}{|\vartheta(\cdots X_i)|} \cdot \vartheta(\cdots X_i).
\end{equation}
Importantly, this implies $\vartheta(\cdots X_i)(X_i) < |\vartheta(\cdots X_i)| \leq \vartheta(\cdots)(X_i)$, since there would exist a weak transition $\dirac{X_i} \wtrans\tau_{\mathcal S} \dirac{X_i}$ otherwise, but $\mathcal S$ is guarded. We continue transforming the underlying ES, such that
\begin{align}
	S_i
	&= S_i \nchoice \tau\prefix \psum_{k=1}^n {\frac{\vartheta(\cdots X_i)}{|\vartheta(\cdots X_i)|} (X_k)} X_k \notag \\
	&= S_i \nchoice \tau\prefix \bigg(\psingle{X_i} \pchoice{\frac{\vartheta(\cdots X_i)(X_i)}{|\vartheta(\cdots X_i)|}} \psum_{k=1}^n {\frac{\vartheta(\cdots X_i) - \vartheta(\cdots X_i)(X_i) \cdot \dirac{X_i}}{|\vartheta(\cdots X_i)| - \vartheta(\cdots X_i)(X_i)} (X_k)} X_k \bigg) \label{lem:linearising:eqn:rewrite-si} ,
\end{align}
possibly using \axiom{C}.

Now assign $\vec T$ as follows:
\[
	T_k \synid
	\begin{cases}
		S_k &\text{if $k \neq i$,} \\
		S_i \nchoice \tau\prefix \psum_{l=1}^n {\frac{\vartheta(\cdots X_i) - \vartheta(\cdots X_i)(X_i) \cdot \dirac{X_i}}{|\vartheta(\cdots X_i)| - \vartheta(\cdots X_i)(X_i)} (X_l)} X_l &\text{if $k = i$.}
	\end{cases}
\]
Note that $\mathcal T$ derives a strong transition
\begin{equation} \label{lem:linearising:eqn:transition-linearised}
	(|\vartheta(\cdots X_i)| - \vartheta(\cdots X_i)(X_i)) \cdot \dirac{X_i} \trans\tau_{\mathcal T} \vartheta(\cdots X_i) - \vartheta(\cdots X_i)(X_i) \cdot \dirac{X_i},
\end{equation}
which is going to be important when reducing the STT. We show that $\mathcal S$ and $\mathcal T$ are satisfied by the same expressions. So let $\vec E$ satisfy $\mathcal S$, i.e.\ $\vec E = \vec S \subst{\vec E}{\vec X}$. For all $k \neq i$, clearly $E_k = S_k \subst{\vec E}{\vec X} \synid T_k\subst{\vec E}{\vec X}$. Moreover:
\begin{tflalign*}
	& E_i & \\
	={} & S_i \subst{\vec E}{\vec X} & \\
	={} & S_i \subst{\vec E}{\vec X} \nchoice \tau\prefix \Big(\psingle{E_i} \pchoice{\frac{\vartheta(\cdots X_i)(X_i)}{|\vartheta(\cdots X_i)|}} \psum_{k=1}^n {\frac{\vartheta(\cdots X_i) - \vartheta(\cdots X_i)(X_i) \cdot \dirac{X_i}}{|\vartheta(\cdots X_i)| - \vartheta(\cdots X_i)(X_i)} (X_k)} E_k \Big) & \text{\eqref{lem:linearising:eqn:rewrite-si}}
\intertext{Notice that $E_i$ satisfies the equation
\[
	\textstyle S_i \subst{E_k}{X_k \mid k \neq i} \nchoice \tau\prefix \Big(\psingle{X_i} \pchoice{\frac{\vartheta(\cdots X_i)(X_i)}{|\vartheta(\cdots X_i)|}} \psum_{k=1}^n {\frac{\vartheta(\cdots X_i) - \vartheta(\cdots X_i)(X_i) \cdot \dirac{X_i}}{|\vartheta(\cdots X_i)| - \vartheta(\cdots X_i)(X_i)} (X_k)} E_k \Big)
\]
with variable $X_i$. We are certain that $X_i$ occurs guarded in $S_i \subst{E_k}{X_k \mid k \neq i}$, since $\mathcal S$ is guarded, and obviously, the right-hand side of the $\nchoice$ operator guards $X_i$ as well. Hence, by \axiom{R2}, infer:}
	={} & \rec{X_i} \Big(S_i \subst{E_k}{X_k \mid k \neq i} \nchoice{} \tau\prefix \Big(\psingle{X_i} \pchoice{\frac{\vartheta(\cdots X_i)(X_i)}{|\vartheta(\cdots X_i)|}} \psum_{k=1}^n {\frac{\vartheta(\cdots X_i) - \vartheta(\cdots X_i)(X_i) \cdot \dirac{X_i}}{|\vartheta(\cdots X_i)| - \vartheta(\cdots X_i)(X_i)} (X_k)} E_k \Big) \Big) & \\
	={} & \rec{X_i} \Big( S_i \subst{E_k}{X_k \mid k \neq i} \nchoice \tau\prefix \psum_{k=1}^n {\frac{\vartheta(\cdots X_i) - \vartheta(\cdots X_i)(X_i) \cdot \dirac{X_i}}{|\vartheta(\cdots X_i)| - \vartheta(\cdots X_i)(X_i)} (X_k)} E_k \Big) & \text{\axiom{R3}} \\
	={} & S_i \subst{\vec X}{\vec E} \nchoice \tau\prefix \psum_{k=1}^n {\frac{\vartheta(\cdots X_i) - \vartheta(\cdots X_i)(X_i) \cdot \dirac{X_i}}{|\vartheta(\cdots X_i)| - \vartheta(\cdots X_i)(X_i)} (X_k)} E_k & \\
	\synid{} & T_i \subst{\vec X}{\vec E}. &
\end{tflalign*}
Evidently, $\mathcal T$ is augmented with respect to $\mathcal S$. Similar to the proof of \autoref{lem:shortcutting}, the ES $\mathcal T$ is guarded; there is only one strong transition existing in $\mathcal T$ exclusively, \eqref{lem:linearising:eqn:transition-linearised}, which exists in $\mathcal S$ as a weak transition, nevertheless.

It remains to construct $\vartheta'$ as desired. Let $s_0 \cdots s_k$ denote the designated sequence from before, such that $s_k = X_i$ and let $s_{k+1} = X_i$, too. Consider \autoref{fig:shortcutting-regions} again, which illustrates the cases we distinguish; $\vartheta'$ is designed following the key idea, namely not revisiting $X_i$, after $X_i$ is reached once:
\[
	\vartheta'(t_0 \cdots t_m) =
	\begin{cases}
		\vartheta(t_0 \cdots t_m) &\text{if \ref{fig:shortcutting-regions:itm:1},} \\
		\lambda \cdot (\vartheta(s_0 \cdots s_k) - \vartheta(s_0 \cdots s_k)(s_k) \cdot \dirac{s_k}) &\text{if \ref{fig:shortcutting-regions:itm:2},} \\
		\lambda \cdot \vartheta(t_0 \cdots t_m) &\text{if \ref{fig:shortcutting-regions:itm:3},} \\
		\emptydistr &\text{if \ref{fig:shortcutting-regions:itm:4},} \\
	\end{cases}
\]
where
\[
	\lambda = \frac{\vartheta(s_0 \cdots s_{k-1})(s_k)}{\vartheta(s_0 \cdots s_{k-1})(s_k) - \vartheta(s_0 \cdots s_k)(s_k)} > 1.
\]
Again, the construction is illustrated by \autoref{fig:stt}: After linearising $s$, as depicted in \autoref{fig:stt:sub:linearising}, the self-loop has been removed, i.e.\ the probability of reaching $s$ multiple times is set to $0$. The probabilities of all branches descending from $s$ are scaled by $\lambda = \frac{1}{1 - 1/12} = \frac{12}{11}$.

It is easy to prove that $\vartheta'$ is a TT. The cases \ref{fig:shortcutting-regions:itm:1} and \ref{fig:shortcutting-regions:itm:4} are trivial. In the case \ref{fig:shortcutting-regions:itm:2}, \ref{def:tt:itm:consistent} follows after observing that $|\vartheta'(s_0 \cdots s_k)| = |\vartheta(s_0 \cdots s_k)|$. \ref{def:tt:itm:transition} exploits the transition freshly introduced by $\mathcal T$, \eqref{lem:linearising:eqn:transition-linearised}. So focus on the remaining case \ref{fig:shortcutting-regions:itm:3}: most properties are immediate by $\vartheta$ satisfying the properties of a TT. The only interesting case obtains if $m = k + 1$, and we would like to prove \ref{def:tt:itm:consistent}:
\[
	|\vartheta'(t_0 \cdots t_{k+1})| = \lambda \cdot |\vartheta(t_0 \cdots t_{k+1})| \leq \lambda \cdot \vartheta(t_0 \cdots t_k)(t_{k+1}) = \vartheta'(t_0 \cdots t_k)(t_{k+1}).
\]
Besides, it is not hard to prove that $\vartheta'$ is stationary.

There is one last part missing: We are still required to show that indeed $|\vartheta'| < |\vartheta|$. Clearly, any node is non-stopping and pioneer in $\vartheta'$ if and only if it is in $\vartheta$. The number of recursions, however, has decreased: The recursion leading from $s_0 \cdots s_k$ to itself does not exist in $\vartheta'$ anymore. The existence of any other recursion remains untouched (although the actual probabilities may have been scaled).
\end{proof}

\noindent
We finally gathered all necessary preparations; all parts are assembled in the proof of the next theorem:

\begin{theorem}[Saturation] \label{thm:saturation}
Let $\mathcal S$ be a guarded SES satisfied by some expression $E$. Then there is a guarded, saturated SES $\mathcal T$ satisfied by $E$.
\end{theorem}
\begin{proof}
We begin proving the first condition of \autoref{def:es-saturatedness}. Let $\mathcal S: \vec X = \vec S$; repeat the following transformations for every $X_k$ and for every action $\alpha$ occurring as a prefix in $\mathcal S$. By \autoref{prp:pa-finite-generability-alpha}, we only need to consider a finite number of weak transitions $\dirac{X_k} = \mu \wtrans{} \gamma \trans\alpha \eta \wtrans{} \nu$, and, by \autoref{prp:pa-finite-generability-stationary}, $\mu \wtrans{} \gamma$ (and $\eta \wtrans{} \nu$ analogously) is a convex combination of weak transitions $\mu \wtrans{}^{\Theta_i} \gamma_i$ generated by stationary schedulers $\Theta_i$. Fix some $i$, \autoref{lem:stt-existence} translates the respective transition into a STT $\mu \wtrans{}^\vartheta \gamma$. Now perform an induction on the size of $\vartheta$:

If $|\vartheta| > 1$, pick an arbitrary pioneer node $s_0 \cdots s_j$, such that any non-stopping continuation of $s_0 \cdots s_j$ is no longer pioneer; one might think of $s_0 \cdots s_j$ as a leaf in the finite tree representation. Now distinguish: First, assume that there is a node recurring on $s_0 \cdots s_j$. Then, by the choice of $s_0 \cdots s_j$, that node must be $s_0 \cdots s_j s_j$, so we can linearise. Formally, apply \autoref{lem:linearising} to obtain an augmented ES and a smaller STT, and conclude by induction. Otherwise, if there is no node recurring on $s_0 \cdots s_j$, observe that $j \geq 1$ since $|\vartheta| > 1$, so we exactly meet the requirements of \autoref{lem:shortcutting} and shortcut instead.

If $|\vartheta| \leq 1$, then the only non-stopping node in $\vartheta$ (if there is any) is the root. So simply by the definition of a TT, there are transitions
\begin{align*}
	\mu - \mu' &\trans\tau_{\mathcal S} \gamma - \mu', &
	\eta - \eta' &\trans\tau_{\mathcal S} \nu - \eta',
\end{align*}
for $\mu', \eta'$ with $\mu' \leq \mu$, $\mu' \leq \gamma$, $\eta' \leq \eta$ and $\eta' \leq \nu$. Additionally, recall that $\gamma \trans\alpha \eta$, so by a very similar argumentation as in \autoref{lem:shortcutting}, we can shortcut these transitions using both laws \axiom{T3} and \axiom{T4} once (here, the full generality of \axiom{T4} is exploited). The result is a guarded SES $\mathcal S'$, which is augmented with respect to $\mathcal S$ and yields $\mu \trans\alpha_{\mathcal S'} \nu$.

It remains to establish \autoref{def:es-saturatedness} \ref{def:es-saturatedness:itm:var}, so suppose there was some transition $\dirac{X_i} \wtrans{} \mu$, where $S_j \unguarded V$ for all $j$ with and $\mu(X_j) > 0$, but $S_i \not\unguarded V$. Now apply the same procedure as before: $\dirac{X_i} \wtrans{} \mu$ is decomposed into several weak transitions that result into STTs; then shortcut and/or linearise until $S_i \unguarded V$. Although the linearising method remains as it is, using \axiom{R3}, the shortcutting is modified to rely on \axiom{T2} instead of \axiom{T3}.
\end{proof}

\noindent
As the proof of \autoref{thm:saturation} might seem hard to visualise, we demonstrate the transformations exemplary:

\begin{figure}[t]
\begin{center}
\begin{tikzpicture}[transition diagram, state/.append style={node distance=1.5cm and .8cm, on grid}]
\node[state] (1) {$Y_2$};
\node[initial, state, above=of 1] (0) {$Y_1$};
\node[state, below left=of 1] (2) {$Y_3$};
\node[state, below right=of 1] (3) {$Y_4$};

\path (0) edge[transition, "$\tau$"] (1);
\path (0) edge[transition, bend left, "$b$" pos=.2] (3);

\node[splitter] (*) at ($(1)!0.5!(2) + ({atan(1.5/.8) + 90}:.5cm)$) {};
\path (1) edge[bend right, "$\tau$", auto=right] (*);
\path (*) edge[transition, bend right, "$\frac 1 3$", auto=right] (2);
\path (*) edge[transition, bend right, "$\frac 2 3$", auto=right] (1);
\path (2) edge[transition, "$a$"] (3);
\path (1) edge[transition, "$b$"] (3);

\path (1) ++(1.5cm, 0) node {$=$} node[above=.25ex] {\scriptsize\axiom{R3}};

\node[state] (1) at ($(1) + (3cm, 0)$) {$Y_2$};
\node[initial, state, above=of 1] (0) {$Y_1$};
\node[state, below left=of 1] (2) {$Y_3$};
\node[state, below right=of 1] (3) {$Y_4$};

\path (0) edge[transition, "$\tau$"] (1);
\path (0) edge[transition, bend left, "$b$" pos=.2] (3);

\path (1) edge[transition, bend right=35, "$\tau$", auto=right] (2);
\node[splitter] (*) at ($(1)!0.5!(2) + ({atan(1.5/.8) + 90}:.-.22cm)$) {};
\path (1) edge[bend left, "$\tau$" pos=.6] (*);
\path (*) edge[transition, bend left=22, "$\frac 1 3$" {pos=.3, xshift=-.07cm, yshift=.19cm}] (2);
\path (*) edge[transition, bend left=22, "$\frac 2 3$" {pos=.4, xshift=.05, yshift=-.03cm}] (1);
\path (2) edge[transition, "$a$", auto=right] (3);
\path (1) edge[transition, "$b$"] (3);

\path (1) ++(1.5cm, 0) node {$=$} node[above=.25ex] {\scriptsize\axiom{T3}};

\node[state] (1) at ($(1) + (3cm, 0)$) {$Y_2$};
\node[initial, state, above=of 1] (0) {$Y_1$};
\node[state, below left=of 1] (2) {$Y_3$};
\node[state, below right=of 1] (3) {$Y_4$};

\path (0) edge[transition, bend right=41, "$\tau$" pos=.2, auto=right] (2);
\path (0) edge[transition, "$\tau$"] (1);
\path (0) edge[transition, bend left=41, "$b$" pos=.2] (3);

\path (1) edge[transition, bend right=35, "$\tau$", auto=right] (2);
\node[splitter] (*) at ($(1)!0.5!(2) + ({atan(1.5/.8) + 90}:.-.22cm)$) {};
\path (1) edge[bend left, "$\tau$" pos=.6] (*);
\path (*) edge[transition, bend left=22, "$\frac 1 3$" {pos=.3, xshift=-.07cm, yshift=.19cm}] (2);
\path (*) edge[transition, bend left=22, "$\frac 2 3$" {pos=.4, xshift=.05, yshift=-.05cm}] (1);
\path (2) edge[transition, "$a$", auto=right] (3);
\path (1) edge[transition, "$b$"] (3);

\path (1) ++(1.5cm, 0) node {$=$} node[above=.25ex] {\scriptsize\axiom{T3}};

\node[state] (1) at ($(1) + (3cm, 0)$) {$Y_2$};
\node[initial, state, above=of 1] (0) {$Y_1$};
\node[state, below left=of 1] (2) {$Y_3$};
\node[state, below right=of 1] (3) {$Y_4$};

\path (0) edge[transition, bend right=41, "$\tau$" pos=.2, auto=right] (2);
\path (0) edge[transition, "$\tau$"] (1);
\path (0) edge[transition, bend left=41, "${a,b}$" pos=.2] (3);

\path (1) edge[transition, bend right=35, "$\tau$", auto=right] (2);
\node[splitter] (*) at ($(1)!0.5!(2) + ({atan(1.5/.8) + 90}:.-.22cm)$) {};
\path (1) edge[bend left, "$\tau$" pos=.6] (*);
\path (*) edge[transition, bend left=22, "$\frac 1 3$" {pos=.3, xshift=-.07cm, yshift=.19cm}] (2);
\path (*) edge[transition, bend left=22, "$\frac 2 3$" {pos=.4, xshift=.05, yshift=-.05cm}] (1);
\path (2) edge[transition, "$a$", auto=right] (3);
\path (1) edge[transition, "${a,b}$"] (3);
\end{tikzpicture}
\end{center}
\caption{The SES $\mathcal T$ becomes saturated step-by-step.} \label{fig:saturation}
\end{figure}
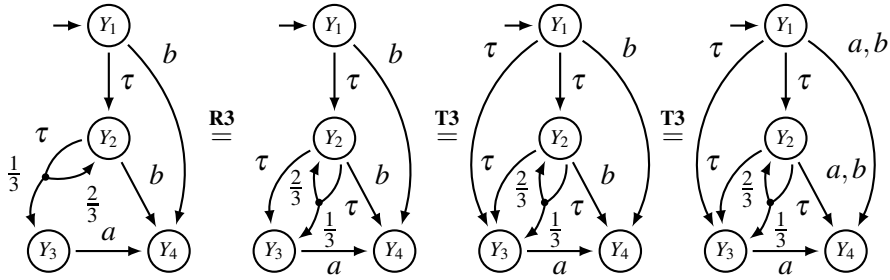

\begin{example}\rm
Recall the expression $F$ from the previous examples, translated into the guarded SES $\mathcal T : \vec Y = \vec T$. The saturation proceeds as follows: First, the infinitary transition $\dirac{Y_2} \wtrans\tau \dirac{Y_3}$ is linearised to obtain $\dirac{Y_2} \trans\tau \dirac{Y_3}$. Then, we add the missing $\tau$ and $\alpha$ transitions by shortcutting. (See \autoref{fig:saturation}.)
\end{example}

\subsection{Joining equation systems} \label{sec:completeness:sec:join}

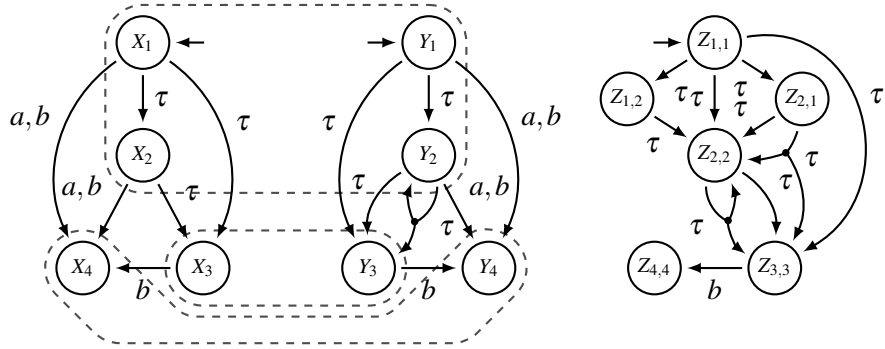
\begin{figure}[t]
\begin{center}
\tikzset{
	state/.append style={minimum size=.7cm, inner sep=0, node distance=1.5cm and .8cm, on grid}
}
\begin{tikzpicture}[
	transition diagram,
	region/.style={thick, dashed, black!70}
]
\node[initial, initial where=right, state] (X1) {$X_1$};
\node[state, below=of X1] (X2) {$X_2$};
\node[state, below right=of X2] (X3) {$X_3$};
\node[state, below left=of X2] (X4) {$X_4$};

\path (X1) edge[transition, "$\tau$"] (X2);
\path (X1) edge[transition, bend left=40, "$\tau$"] (X3);
\path (X1) edge[transition, bend right=40, "{$a, b$}", auto=right] (X4);
\path (X2) edge[transition, "$\tau$"] (X3);
\path (X2) edge[transition, "{$a, b$}", auto=right] (X4);
\path (X3) edge[transition, "$b$"] (X4);

\node[initial, state] (Y1) at ($(X1) + (3.8cm, 0)$) {$Y_1$};
\node[state, below=of Y1] (Y2) {$Y_2$};
\node[state, below left=of Y2] (Y3) {$Y_3$};
\node[state, below right=of Y2] (Y4) {$Y_4$};

\path (Y1) edge[transition, "$\tau$"] (Y2);
\path (Y1) edge[transition, bend right=40, "$\tau$", auto=right] (Y3);
\path (Y1) edge[transition, bend left=40, "{$a, b$}"] (Y4);
\path (Y2) edge[transition, "{$a, b$}"] (Y4);
\path (Y3) edge[transition, "$b$", auto=right] (Y4);

\path (Y2) edge[transition, bend right, "$\tau$", auto=right] (Y3);
\node[splitter] (*) at ($(Y2)!0.5!(Y3) + ({atan(1.6/1) + 90}:.-.25cm)$) {};
\path (Y2) edge[bend left, "$\tau$" pos=.6] (*);
\path (*) edge[transition, bend left=22] (Y3);
\path (*) edge[transition, bend left=22] (Y2);

\begin{scope}[on background layer]
	\def\r{.5cm}
	\def\D{.5cm}
	\def\d{1cm}
	\draw[region] ($(Y1) + (0, \r)$) arc (90:0:\r) -- ($(Y2) + (\r, 0)$) arc (0:-90:\r) -- ($(X2) + (0, -\r)$) arc (-90:-180:\r) -- ($(X1) + (-\r, 0)$) arc (-180:-270:\r) -- cycle;
	\draw[region] ($(Y3) + (0, \r)$) arc (90:-90:\r) -- ($(X3) + (0, -\r)$) arc (-90:-270:\r) -- cycle;
	\draw[region] ($(Y4) + (135:\r)$) arc (135:-45:\r) -- ($(Y4) + (-\D, -\D) + (-45:\r)$) arc (-45:-90:\r) -- ($(X4) + (\D, -\D) + (-90:\r)$) arc(-90:-135:\r) -- ($(X4) + (-135:\r)$) arc (-135:-315:\r) [rounded corners]-- ($(X4) + (45:\r) + (\d, -\d)$) [rounded corners]-- ($(Y4) + (135:\r) + (-\d, -\d)$) -- cycle;
\end{scope}

\node[initial, state] (Z11) at ($(Y1) + (3.8cm, 0)$) {$Z_{1, 1}$};
\node[state, below=of Z11] (Z22) {$Z_{2, 2}$};
\node[state, left=of $(Z11)!0.5!(Z22)$] (Z12) {$Z_{1, 2}$};
\node[state, right=of $(Z11)!0.5!(Z22)$] (Z21) {$Z_{2, 1}$};
\node[state, below right=of Z22] (Z33) {$Z_{3, 3}$};
\node[state, below left=of Z22] (Z44) {$Z_{4, 4}$};

\path (Z11) edge[transition, "$\tau$", auto=right] (Z22);
\path (Z11) edge[transition, "$\tau$" pos=.7] (Z12);
\path (Z12) edge[transition, "$\tau$", auto=right] (Z22);
\path (Z11) edge[transition, "$\tau$", auto=right] (Z21);
\path (Z21) edge[transition, "$\tau$", auto=right] (Z22);

\path (Z11) edge[transition, "$\tau$", out=10, in=30, looseness=1.45] (Z33);

\node[splitter] (*) at ($(Z21)!.5!(Z22) + ({atan(.8/.75) + 90}:-.5cm)$) {};
\path (Z21) edge[bend left=10, "$\tau$"] (*);
\path (*) edge[transition, bend left=10] (Z22);
\path (*) edge[transition, bend left] (Z33);

\path (Z22) edge[transition, bend left, "$\tau$"] (Z33);
\node[splitter] (*) at ($(Z22)!.5!(Z33) + ({atan(.8/1.5)}:.-.25cm)$) {};
\path (Z22) edge[bend right, "$\tau$" pos=.6, auto=right] (*);
\path (*) edge[transition, bend right=22] (Z33);
\path (*) edge[transition, bend right=22] (Z22);

\path (Z33) edge[transition, "$b$"] (Z44);
\end{tikzpicture}
\end{center}
\caption{Depicts the SESs $\mathcal S$ and $\mathcal T$ after being saturated on the left; the dashed regions designate the equivalence classes induced by $\wbis$. The right PA shows the SES $\mathcal U$ constructed as in \autoref{thm:joining}. For the sake of readability, we omit all quantities and all transitions leading from $Z_{1, 1}, Z_{1, 2}, Z_{2, 1}$ and $Z_{2, 2}$ to $Z_{4, 4}$, labelled with $a$ or $b$.} \label{fig:example-joining}
\end{figure}

\noindent
We achieve the next milestone by creating an SES $\mathcal U$ satisfied by both $E$ and $F$. To this end, we finally take into account that $E \wcong F$. However, as we are not dealing with expressions anymore, we are required to find a similar correspondence on the level of SESs. Therefore, in the original proof, Milner assumed (a nonprobabilistic version of) the following proposition: ``Let $\vec E = (E_1, \ldots, E_n)$ satisfy $\mathcal S : \vec X = \vec S$. Whenever $E_i \wtrans\alpha \mu$, then $X_i \wtrans\alpha_{\mathcal S} \nu[\vec X]$ and $\mu \wbis \nu[\vec S]$''. A formal proof may for instance be found in \cite{Lohrey2002}. Certainly, this statement holds in our probabilistic setting as well. Still, it turns out to be hard to show. So instead, let us simplify: We assume an even stronger statement: Whenever $\dirac{E_i} \wtrans\alpha \mu$, then $\Supp\mu \subseteq \{E_1, \ldots, E_n\}$ and
\[
	\dirac{E_i} \wtrans\alpha \mu[\vec E] \quad\text{iff}\quad \dirac{X_i} \wtrans\alpha_{\mathcal S} \mu[\vec X].
\]
Obviously, this statement does not hold for arbitrary expressions $\vec E$, so we call a solution $E$ \emph{perfect} if expressions $\vec E$ exist as desired. Intuitively, the PA induced by $\mathcal S$ and by a perfect solution $E$ of $\mathcal S$ are isomorphic, i.e.\ identical up to renaming of states. \autoref{thm:joining}, illustrated in \autoref{fig:example-joining}, assumes perfect solutions $E$ and $F$, whose existence will be justified by the next section.

\begin{theorem} \label{thm:joining}
Let $\mathcal S$ (resp.\ $\mathcal T$) be a guarded, saturated SES with perfect solution $E$ (resp. $F$) and let $E \wcong F$. Then there is a guarded SES $\mathcal U$ satisfied by both $E$ and $F$.
\end{theorem}
\begin{proof}
Let $\mathcal S: \vec X = \vec S$ and $\mathcal T: \vec Y = \vec T$. Since $E$ (resp.\ $F$) satisfies $\mathcal S$ (resp.\ $\mathcal T$), there are expressions \raisebox{0pt}[0pt][0pt]{$\vec E = (E_1, \ldots, E_n)$} and \raisebox{0pt}[0pt][0pt]{$\vec F = (F_1, \ldots, F_m)$} with $E \synid E_1$ and $F \synid F_1$, such that \raisebox{0pt}[0pt][0pt]{$\vec E \mathbin= \vec S \subst{\vec E}{\vec X}$} and \raisebox{0pt}[0pt][0pt]{$\vec F \mathbin= \vec T \subst{\vec F}{\vec Y}$}. We begin by proving the following implications for all $E_i \wbis F_j$:
\begin{enumerate}[label=(\arabic*)]
\item Whenever $X_i \trans\alpha_{\mathcal S} \mu[\vec X]$, then one of the following cases holds: \label{thm:joining:itm:x}
	\begin{enumerate}[label=(\alph*)]
	\item $\alpha = \tau$ and $(1 \mathord- p) \cdot \dirac{Y_j} \trans\tau_{\mathcal T} \nu[\vec Y]$, such that $\mu[\vec E] \wbis p \cdot \dirac{F_j} + \nu[\vec F]$, for some $0 < p \leq 1$, \label{thm:joining:itm:x:itm:idling}
	\item $\dirac{Y_j} \trans\alpha_{\mathcal T} \nu[\vec Y]$ and $\mu[\vec E] \wbis \nu[\vec F]$, \label{thm:joining:itm:x:itm:synch}
	\end{enumerate}
\item Whenever $Y_j \trans\alpha_{\mathcal T} \nu[\vec Y]$, then one of the following cases holds: \label{thm:joining:itm:y}
	\begin{enumerate}[label=(\alph*)]
	\item $\alpha = \tau$ and $(1 \mathord- p) \cdot \dirac{X_i} \trans\tau_{\mathcal S} \mu[\vec X]$, such that $\nu[\vec F] \wbis p \cdot \dirac{E_i} + \mu[\vec E]$, for some $0 < p \leq 1$, \label{thm:joining:itm:y:itm:idling}
	\item $\dirac{X_i} \trans\alpha_{\mathcal S} \mu[\vec X]$ and $\nu[\vec F] \wbis \mu[\vec E]$, \label{thm:joining:itm:y:itm:synch}
	\end{enumerate}
\item $S_i \unguarded Z$ if and only if $T_j \unguarded Z$, for all free variables $Z$. \label{thm:joining:itm:variables}
\end{enumerate}
Focus on \ref{thm:joining:itm:x} first, \ref{thm:joining:itm:y} can be shown analogously. Suppose that $X_i \trans\alpha_{\mathcal S} \mu[\vec X]$, then $S_i \subst{\vec E}{\vec X} \trans\alpha \mu[\vec E]$. By $S_i \subst{\vec E}{\vec X} \wcong E_i$ due to soundness and $E_i \wbis F_j$, we conclude that \raisebox{0pt}[0pt][0pt]{$F_j \wtrans{\hat\alpha} \nu$ with $\nu \wbis \mu[\vec E]$}. Since $F$ is a perfect solution of $\mathcal T$, we know that $\nu$ is a distribution only over expressions in $\tilde F$, write $\nu[\vec F]$. Moreover, there is a transition \raisebox{0pt}[0pt][0pt]{$Y_j \wtrans{\hat\alpha} \nu[\vec Y]$}. Finally, distinguish two cases: If $\alpha \neq \tau$ or there is a transition \raisebox{0pt}[0pt][0pt]{$Y_j \wtrans\tau \nu[\vec Y]$}, then we have shown \ref{thm:joining:itm:x}\ref{thm:joining:itm:x:itm:synch}, by $\mathcal T$ being saturated. So suppose that $\alpha = \tau$ and there is no transition \raisebox{0pt}[0pt][0pt]{$Y_j \wtrans\tau \nu[\vec Y]$}. Then, by \autoref{prp:wtrans-decompose-tau}, $Y_j$ idles with some probability $p > 0$ and there is a transition \raisebox{0pt}[0pt][0pt]{$(1 \mathord- p) \cdot \dirac{Y_j} \wtrans\tau \nu'[\vec Y]$}, such that $\nu[\vec Y] \synid p \cdot \dirac{Y_j} + \nu'[\vec Y]$. Again, after exploiting the saturatedness of $\mathcal T$, we end up with \ref{thm:joining:itm:x}\ref{thm:joining:itm:x:itm:idling}.

\ref{thm:joining:itm:variables} is straightforward by $\mathcal S$ and $\mathcal T$ being standard and $E_i \wbis F_j$.

If we restrict $E_i$ and $F_j$ further to be weakly congruent, $E_i \wcong F_j$, then we find stricter implications following an analogous argumentation: Both cases \ref{thm:joining:itm:x}\ref{thm:joining:itm:x:itm:idling} and \ref{thm:joining:itm:y}\ref{thm:joining:itm:y:itm:idling} must not occur, since weak congruence is more restrictive by requiring $\tau$ transitions to not be simulated by idling.

Now let \raisebox{0pt}[0pt][0pt]{$I = \{(i, j) \mid E_i \wbis F_j\}$} and define, for any two distributions $\mu, \nu$ with $\mu[\vec E] \wbis \nu[\vec F]$, a distribution $\mu \times \nu \in \Distr I$ where
\[
	(\mu \times \nu)(i, j) := \frac{\mu(X_i) \cdot \nu(Y_j)}{\sum_{i' : E_i \wbis E_{i'}} \mu(X_{i'})} = \frac{\mu(X_i) \cdot \nu(Y_j)}{\sum_{j' : F_j \wbis F_{j'}} \nu(Y_{j'})}.
\]
By construction \raisebox{0pt}[0pt][0pt]{$\mu(X_i) \mathbin= \sum_{j : (i, j) \in I} (\mu \times \nu)(i, j)$} and symmetrically \raisebox{0pt}[0pt][0pt]{$\nu(Y_j) \mathbin= \sum_{i : (i, j) \in I} (\mu \times \nu)(i, j)$}. For each entry $(i, j) \in I$, let $Z_{i, j}$ be a fresh variable and let \raisebox{0pt}[0pt][0pt]{$\vec Z = (Z_{i, j})_{(i, j) \in I}$}. To construct the SES $\mathcal U: \vec Z = \vec U$ (with distinguished variable $Z_{1, 1}$), we now assign $\vec U = (U_{i, j})_{(i, j) \in I}$ where $U_{i, j}$ sums over:
\begin{itemize}
\item \raisebox{0pt}[0pt][0pt]{$\alpha\prefix \psum_{(k, \ell) \in I} {(\mu \times \nu)(k, \ell)} Z_{k, \ell}$}, whenever $X_i \trans\alpha_{\mathcal S} \mu[\vec X]$ and, according to \ref{thm:joining:itm:x}\ref{thm:joining:itm:x:itm:synch}, $\dirac{Y_j} \trans\alpha_{\mathcal T} \nu[\vec Y]$ with $\mu[\vec E] \wbis \nu[\vec F]$,
\item \raisebox{0pt}[0pt][0pt]{$\tau\prefix \psum_{(k, \ell) \in I} {(\mu \times (p \cdot \dirac{Y_j} + \nu))(k, \ell)} Z_{k, \ell}$}, whenever $X_i \trans\tau_{\mathcal S} \mu[\vec X]$ and, according to \ref{thm:joining:itm:x}\ref{thm:joining:itm:x:itm:idling},\\ $(1 \mathord- p) \cdot \dirac{Y_j} \trans\tau_{\mathcal T} \nu[\vec Y]$ with $\mu[\vec E] \wbis p \cdot \dirac{F_j} + \nu[\vec F]$,
\item \raisebox{0pt}[0pt][0pt]{$\alpha\prefix \psum_{(k, \ell) \in I} {(\mu \times \nu)(k, \ell)} Z_{k, \ell}$, whenever $Y_j \mathbin{\trans\alpha_{\mathcal T}} \nu[\vec Y]$} and, according to \ref{thm:joining:itm:y}\ref{thm:joining:itm:y:itm:synch}, $\dirac{X_i} \trans\alpha_{\mathcal S} \mu[\vec X]$ with $\mu[\vec E] \wbis \nu[\vec F]$,
\item \raisebox{0pt}[0pt][0pt]{$\tau\prefix \psum_{(k, \ell) \in I} {((p \cdot \dirac{X_i} + \mu) \times \nu)(k, \ell)} Z_{k, \ell}$}, whenever $Y_j \trans\tau_{\mathcal T} \nu[\vec Y]$ and, according to \ref{thm:joining:itm:y}\ref{thm:joining:itm:y:itm:idling},\\ $(1 \mathord- p) \cdot \dirac{X_i} \trans\tau_{\mathcal S} \mu[\vec X]$ with $p \cdot \dirac{E_i} + \mu[\vec E] \wbis \nu[\vec F]$,
\item $Z$, for any free variable $Z$ with $S_i \unguarded Z$, or equivalently as claimed by \ref{thm:joining:itm:variables}, $T_j \unguarded Z$.
\end{itemize}

It suffices to show that $E$ satisfies $\mathcal U$; the same can be shown analogously for $F$. For $(i, j) \in I$, we distinguish the following two (exhaustive and mutually exclusive) cases:
\begin{enumerate}[label=(\roman*)]
\item whenever $Y_j \trans\tau_{\mathcal T} \nu[\vec Y]$, then always case \ref{thm:joining:itm:y}\ref{thm:joining:itm:y:itm:synch} holds, \label{thm:joining:itm:always-synch}
\item there is some transition $Y_j \trans\alpha_{\mathcal T} \nu[\vec Y]$, such that not \ref{thm:joining:itm:y}\ref{thm:joining:itm:y:itm:synch}. \label{thm:joining:itm:exists-idling}
\end{enumerate}
Then define expressions $\vec G = (G_{i, j})_{(i, j) \in I}$ with
\[
	G_{i, j} \synid
	\begin{cases}
		E_i &\text{if \ref{thm:joining:itm:always-synch} for $(i, j)$,}\\
		\rec X (E_i \nchoice H_{i, j}) &\text{if \ref{thm:joining:itm:exists-idling} for $(i, j)$,}
	\end{cases}
\]
where $X$ is a fresh variable and $H_{i, j}$ contains a summand
\[
	\tau\prefix \bigg(\psingle X \pchoice p \psum_{k=1}^n {\frac{\mu}{1 - p}(X_k)} E_k\bigg)
\]
for any transition $Y_j \trans\alpha_{\mathcal T} \nu[\vec Y]$ that cannot be matched by $X_i$ as described in \ref{thm:joining:itm:y}\ref{thm:joining:itm:y:itm:synch}, but only as in \ref{thm:joining:itm:y}\ref{thm:joining:itm:y:itm:idling}. Every case $p = 1$ may be treated as an expression $\tau\prefix \psingle X$ here, nevertheless, we omit these corner cases for the sake of simplicity. Next, we assert that
\begin{equation} \label{thm:joining:eqn:inner-equ}
	\alpha\prefix (\psingle{G_{i, j}} \pchoice q P) = \alpha\prefix (\psingle{E_i} \pchoice q P),
\end{equation}
for all $q$ and $P$. The statement is trivial if \ref{thm:joining:itm:always-synch} holds for $(i, j)$, so assume \ref{thm:joining:itm:exists-idling}. Consider the equation posed by $\tau\prefix \psingle{E_i \nchoice H_{i, j}}$ in which $X$ is guarded. We show that $\tau\prefix \psingle{\rec X (E_i \nchoice H_{i, j})}$ yields a solution:
\begin{tflalign*}
	& \tau\prefix \psingle{(E_i \nchoice H_{i, j}) \subst{\tau\prefix \psingle{\rec X (E_i \nchoice H_{i, j})}}{X}} & \\
	={} & \tau\prefix \psingle{(E_i \nchoice H_{i, j}) \subst{\rec X (E_i \nchoice H_{i, j})}{X}} & \text{\axiom{T1}} \\
	={} & \tau\prefix \psingle{\rec X (E_i \nchoice H_{i, j})}. & \text{\axiom{R1}}
\end{tflalign*}
Now \axiom{R2} entails $\rec X \tau\prefix \psingle{E_i \nchoice H_{i, j}} = \tau\prefix \psingle{\rec X (E_i \nchoice H_{i, j})}$. So continue by:
\begin{tflalign*}
	& \tau\prefix \psingle{\rec X (E_i \nchoice H_{i, j})} & \\
	={} & \rec X \tau\prefix \psingle{E_i \nchoice H_{i, j}} & \\
	={} & \rec X (E_i \nchoice \tau\prefix \psingle X \nchoice H_{i, j}) & \text{\axiom{R5}} \\
	\synid{} & \rec X (E_i \nchoice \tau\prefix \psingle X \nchoice \sum \{\tau\prefix (\psingle X \pchoice p \psum*_{k=1}^n {\frac{\mu}{1 - p}(X_k)} E_k)\}) & \\
	={} & \rec X (E_i \nchoice \tau\prefix \psingle X \nchoice \sum \{\tau\prefix (\psingle X \pchoice p \psum*_{k=1}^n {\frac{\mu}{1 - p}(X_k)} E_k)\} \nchoice \sum \{\tau\prefix \psum*_{k=1}^n {\frac{\mu}{1 - p}(X_k)} E_k\}) & \text{\axiom{R3}} \\
	={} & \rec X (E_i \nchoice \tau\prefix \psingle X \nchoice \sum \{\tau\prefix \psum*_{k=1}^n {\frac{\mu}{1 - p}(X_k)} E_k\}) & \text{\axiom{C}} \\
	={} & \rec X \tau\prefix \psingle*{E_i \nchoice \sum \{\tau\prefix \psum*_{k=1}^n {\frac{\mu}{1 - p}(X_k)} E_k\}} & \text{\axiom{R5}} \\
	={} & \tau\prefix \psingle*{E_i \nchoice \sum \{\tau\prefix \psum*_{k=1}^n {\frac{\mu}{1 - p}(X_k)} E_k\}}. & \text{\axiom{R1}}
\intertext{Recall that each distribution $\mu$ is the target of some transition $(1 - p) \cdot \dirac{X_i} \trans\tau_{\mathcal S} \mu$, so by convexity, idempotence of $\nchoice$ and $E_i = S_i \subst{\vec E}{\vec S}$:}
	={} & \tau\prefix \psingle{E_i}.
\end{tflalign*}
Finally, we obtain the desired equality:
\begin{tflalign*}
	& \alpha\prefix (\psingle{G_{i, j}} \pchoice q P) & \\
	\synid{} & \alpha\prefix (\psingle{\rec X (E_i \nchoice H_{i, j})} \pchoice q P) & \\
	={} & \alpha\prefix (\psingle{\tau\prefix \psingle{\rec X (E_i \nchoice H_{i, j})}} \pchoice q P) & \text{\axiom{T1}} \\
	={} & \alpha\prefix (\psingle{\tau\prefix \psingle{E_i}} \pchoice q P) & \\
	={} & \alpha\prefix (\psingle{E_i} \pchoice q P). & \text{\axiom{T1}}
\end{tflalign*}
\autoref{fig:partial-idling} visualises abstractly how to understand the transformations above.

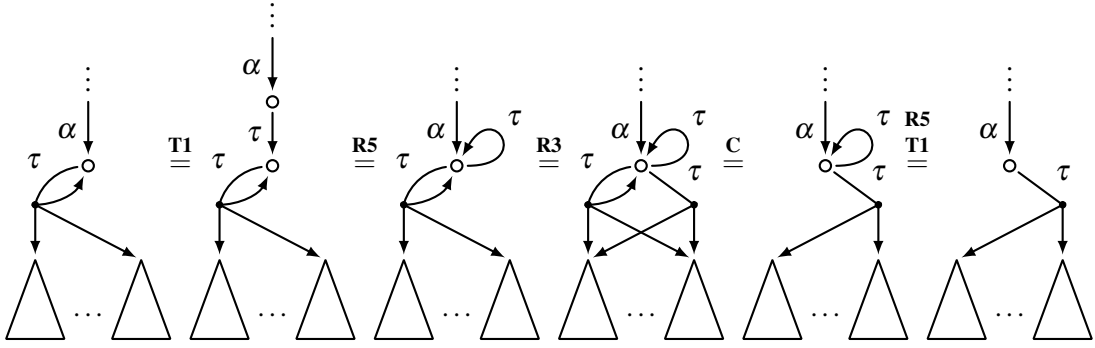
\begin{figure}
\begin{center}
\begin{tikzpicture}[process diagram, node distance=1.2cm and .65cm]
\def\pwidth{.775cm}
\def\pheight{1.05cm}
\def\d{2.45cm}

\node[state, lonely] (1) {};
\path (1) ++(0, .85cm) node[anchor=south] (0) {${\vdots}$};
\node[state, below left=of 1] (p1) {};
\node[state, below right=of 1] (pn) {};

\pic at (p1) {process={0, \pwidth, \pheight}};
\pic at (pn) {process={0, \pwidth, \pheight}};

\node at ($(p1)!0.5!(pn) + (0, -.7 * \pheight)$) {${\cdots}$};

\path (0) edge[transition, "$\alpha$", auto=right] (1);

\node[splitter] (*l) at ($(p1) + (0, .7 * \pheight)$) {};
\path (1) edge[space <, "$\tau$", bend right=32, auto=right] (*l);
\path (*l) edge[transition, bend right=32] (1);
\path (*l) edge[transition] (p1);
\path (*l) edge[transition] (pn);

\path (1) ++(.5 * \d, 0) node {$=$} node[above=.25ex] {\scriptsize\axiom{T1}};

\node[state, lonely] (1) at ($(1) + (\d, 0)$) {};
\path (1) ++(0, .85cm) node[state, lonely] (i) {} ++(0, .85cm) node[anchor=south] (0) {${\vdots}$};
\node[state, below left=of 1] (p1) {};
\node[state, below right=of 1] (pn) {};

\pic at (p1) {process={0, \pwidth, \pheight}};
\pic at (pn) {process={0, \pwidth, \pheight}};

\node at ($(p1)!0.5!(pn) + (0, -.7 * \pheight)$) {${\cdots}$};

\path (0) edge[transition, "$\alpha$", auto=right] (i);
\path (i) edge[transition, space <, "$\tau$", auto=right] (1);

\node[splitter] (*l) at ($(p1) + (0, .7 * \pheight)$) {};
\path (1) edge[space <, "$\tau$", bend right=32, auto=right] (*l);
\path (*l) edge[transition, bend right=32] (1);
\path (*l) edge[transition] (p1);
\path (*l) edge[transition] (pn);

\path (1) ++(.5 * \d, 0) node {$=$} node[above=.25ex] {\scriptsize\axiom{R5}};

\node[state, lonely] (1) at ($(1) + (\d, 0)$) {};
\path (1) ++(0, .85cm) node[anchor=south] (0) {${\vdots}$};
\node[state, below left=of 1] (p1) {};
\node[state, below right=of 1] (pn) {};

\pic at (p1) {process={0, \pwidth, \pheight}};
\pic at (pn) {process={0, \pwidth, \pheight}};

\node at ($(p1)!0.5!(pn) + (0, -.7 * \pheight)$) {${\cdots}$};

\path (0) edge[transition, "$\alpha$", auto=right] (1);
\path (1) edge[transition, space <, out=0, in=60, looseness=32, "$\tau$" pos=.6, auto=right] (1);

\node[splitter] (*l) at ($(p1) + (0, .7 * \pheight)$) {};
\path (1) edge[space <, "$\tau$", bend right=32, auto=right] (*l);
\path (*l) edge[transition, bend right=32] (1);
\path (*l) edge[transition] (p1);
\path (*l) edge[transition] (pn);

\path (1) ++(.5 * \d, 0) node {$=$} node[above=.25ex] {\scriptsize\axiom{R3}};

\node[state, lonely] (1) at ($(1) + (\d, 0)$) {};
\path (1) ++(0, .85cm) node[anchor=south] (0) {${\vdots}$};
\node[state, below left=of 1] (p1) {};
\node[state, below right=of 1] (pn) {};

\pic at (p1) {process={0, \pwidth, \pheight}};
\pic at (pn) {process={0, \pwidth, \pheight}};

\node at ($(p1)!0.5!(pn) + (0, -.7 * \pheight)$) {${\cdots}$};

\path (0) edge[transition, "$\alpha$", auto=right] (1);
\path (1) edge[transition, space <, out=0, in=60, looseness=32, "$\tau$" pos=.6, auto=right] (1);

\node[splitter] (*l) at ($(p1) + (0, .7 * \pheight)$) {};
\path (1) edge[space <, "$\tau$", bend right=32, auto=right] (*l);
\path (*l) edge[transition, bend right=32] (1);
\path (*l) edge[transition] (p1);
\path (*l) edge[transition] (pn);

\node[splitter] (*r) at ($(pn) + (0, .7 * \pheight)$) {};
\path (1) edge[space <, "$\tau$" pos=.6] (*r);
\path (*r) edge[transition] (p1);
\path (*r) edge[transition] (pn);

\path (1) ++(.5 * \d, 0) node {$=$} node[above=.25ex] {\scriptsize\axiom{C}};

\node[state, lonely] (1) at ($(1) + (\d, 0)$) {};
\path (1) ++(0, .85cm) node[anchor=south] (0) {${\vdots}$};
\node[state, below left=of 1] (p1) {};
\node[state, below right=of 1] (pn) {};

\pic at (p1) {process={0, \pwidth, \pheight}};
\pic at (pn) {process={0, \pwidth, \pheight}};

\node at ($(p1)!0.5!(pn) + (0, -.7 * \pheight)$) {${\cdots}$};

\path (0) edge[transition, "$\alpha$", auto=right] (1);
\path (1) edge[transition, space <, out=0, in=60, looseness=32, "$\tau$" pos=.6, auto=right] (1);

\node[splitter] (*r) at ($(pn) + (0, .7 * \pheight)$) {};
\path (1) edge[space <, "$\tau$" pos=.6] (*r);
\path (*r) edge[transition] (p1);
\path (*r) edge[transition] (pn);

\path (1) ++(.5 * \d, 0) node {$=$} node[above=.25ex, align=center] {\scriptsize\axiom{R5}\\[-.7ex]\scriptsize\axiom{T1}};

\node[state, lonely] (1) at ($(1) + (\d, 0)$) {};
\path (1) ++(0, .85cm) node[anchor=south] (0) {${\vdots}$};
\node[state, below left=of 1] (p1) {};
\node[state, below right=of 1] (pn) {};

\pic at (p1) {process={0, \pwidth, \pheight}};
\pic at (pn) {process={0, \pwidth, \pheight}};

\node at ($(p1)!0.5!(pn) + (0, -.7 * \pheight)$) {${\cdots}$};

\path (0) edge[transition, "$\alpha$", auto=right] (1);

\node[splitter] (*r) at ($(pn) + (0, .7 * \pheight)$) {};
\path (1) edge[space <, "$\tau$" pos=.6] (*r);
\path (*r) edge[transition] (p1);
\path (*r) edge[transition] (pn);
\end{tikzpicture}
\end{center}
\caption{Conceptual procedure of transforming a transition to not target its origin state anymore.} \label{fig:partial-idling}
\end{figure}

In order to show that $E$ satisfies $U$, it suffices to show that $E \synid E_1 \synid G_{1, 1}$ and \raisebox{0pt}[0pt][0pt]{$\vec G = \vec U \subst{\vec G}{\vec Z}$}. The former requirement is easy: Recall that $E_1 \wcong F_1$ implies \ref{thm:joining:itm:y}\ref{thm:joining:itm:y:itm:synch}, so we have \ref{thm:joining:itm:always-synch} and by definition $E_1 \synid G_{1, 1}$. Next, fix some pair $(i, j) \in I$; it remains to check that indeed \raisebox{0pt}[0pt][0pt]{$G_{i, j} = U_{i, j} \subst{\vec G}{\vec Z}$}.

We begin by assuming \ref{thm:joining:itm:always-synch} for $(i, j)$, then the shape of $U_{i, j}$ simplifies to
\begin{tflalign*}
	& U_{i, j} \subst{\vec G}{\vec Z} & \\
	\synid{} & \sum \raisebox{0pt}[0pt][0pt]{$\{\alpha\prefix \psum_{(k, \ell) \in I} {(\mu \times \nu)(k, \ell)} G_{k, \ell} \mid \text{$X_i \trans\alpha_{\mathcal S} \mu[\vec X]$ and \ref{thm:joining:itm:x}\ref{thm:joining:itm:x:itm:synch}}\}$} \nchoice{} & \\
	& \sum \raisebox{0pt}[0pt][0pt]{$\{\tau\prefix \psum_{(k, \ell) \in I} {(\mu \times (p \cdot \dirac{Y_j} + \nu))(k, \ell)} G_{k, \ell} \mid \text{$X_i \trans\tau_{\mathcal S} \mu[\vec X]$ and \ref{thm:joining:itm:x}\ref{thm:joining:itm:x:itm:idling}}\}$} \nchoice{} & \\
	& \sum \raisebox{0pt}[0pt][0pt]{$\{\alpha\prefix \psum_{(k, \ell) \in I} {(\mu \times \nu)(k, \ell)} Z_{k, \ell} \mid \text{$Y_j \trans\alpha_{\mathcal T} \nu[\vec Y]$ and \ref{thm:joining:itm:y}\ref{thm:joining:itm:y:itm:synch}}\}$} \nchoice{} & \\
	& \sum \raisebox{0pt}[0pt][0pt]{$\{\text{$Z$ a free variable} \mid \text{$S_i \unguarded Z$ and $T_j \unguarded Z$}\}$} & \\
	={} & \sum \raisebox{0pt}[0pt][0pt]{$\{\alpha\prefix \psum_{k=1}^n {\mu(X_k)} E_k \mid \text{$X_i \trans\alpha_{\mathcal S} \mu[\vec X]$ and \ref{thm:joining:itm:x}\ref{thm:joining:itm:x:itm:synch}}\}$} \nchoice{} & \text{\eqref{thm:joining:eqn:inner-equ}} \\
	& \sum \raisebox{0pt}[0pt][0pt]{$\{\tau\prefix \psum_{k=1}^n {\mu(X_k)} E_k \mid \text{$X_i \trans\tau_{\mathcal S} \mu[\vec X]$ and \ref{thm:joining:itm:x}\ref{thm:joining:itm:x:itm:idling}}\}$} \nchoice{} & \text{\eqref{thm:joining:eqn:inner-equ}} \\
	& \sum \raisebox{0pt}[0pt][0pt]{$\{\alpha\prefix \psum_{k=1}^n {\mu(X_k)} E_k \mid \text{$Y_j \trans\alpha_{\mathcal T} \nu[\vec Y]$ and \ref{thm:joining:itm:y}\ref{thm:joining:itm:y:itm:synch}}\}$} & \text{\eqref{thm:joining:eqn:inner-equ}} \\
	& \sum \raisebox{0pt}[0pt][0pt]{$\{\text{$Z$ a free variable} \mid \text{$S_i \unguarded Z$}\}$} & \text{\ref{thm:joining:itm:variables}}
\intertext{Note that the first, second and last row characterise exactly $S_i \subst{\vec E}{\vec X}$. The third row is eliminated by convexity and idempotence, again.}
	={} & S_i \subst{\vec E}{\vec X} & \\
	={} & E_i & \\
	\synid{} & G_{i, j}. &
\end{tflalign*}

Now consider the case \ref{thm:joining:itm:exists-idling}. In contrast to before, $U_{i, j}$ features summands resulting from some transition \raisebox{0pt}[0pt][0pt]{$Y_j \trans\tau_{\mathcal T} \nu[\vec Y]$}, for which \ref{thm:joining:itm:y}\ref{thm:joining:itm:y:itm:idling} holds, but \ref{thm:joining:itm:y}\ref{thm:joining:itm:y:itm:synch} does not. After applying the same transformations as before:
\begin{tflalign*}
	& U_{i, j} \subst{\vec G}{\vec Z} & \\
	={} & E_i \nchoice \sum \{\tau\prefix \psum_{(k, \ell) \in I} {((p \cdot \dirac{X_i} + \mu) \times \nu)(k, \ell)} Z_{k, \ell} \mid \text{$Y_j \trans\tau_{\mathcal T} \nu[\vec Y]$ and only \ref{thm:joining:itm:y}\ref{thm:joining:itm:y:itm:idling}}\} & \\
	={} & E_i \nchoice \sum \{\tau\prefix (\psingle{G_{i, j}} \pchoice p \psum*_{k=1}^n {\frac{\mu}{1 - p}(X_k)} E_k) \mid \text{$Y_j \trans\tau_{\mathcal T} \nu[\vec Y]$ and only \ref{thm:joining:itm:y}\ref{thm:joining:itm:y:itm:idling}}\} & \text{\eqref{thm:joining:eqn:inner-equ}} \\
	={} & (E_i \nchoice H_{i, j}) \subst{G_{i, j}}{X} & \\
	\synid{} & (E_i \nchoice H_{i, j}) \subst{\rec X (E_i \nchoice H_{i, j})}{X} & \\
	={} & \rec X (E_i \nchoice H_{i, j}) & \text{\axiom{R1}} \\
	\synid{} & G_{i, j}.
\end{tflalign*}

So both $E$ and $F$ satisfy the SES $\mathcal U$. Clearly $\mathcal U$ is standard, however, it takes some effort to prove that $\mathcal U$ is guarded. For a subdistribution $\mu[\vec Z]$, let $\mu[\vec X]$ denote the projection of $\mu$ on $\vec X$, i.e.\ \raisebox{0pt}[0pt][0pt]{$\mu[\vec X] \synid \distr{X_k \mapsto \sum_{\ell : (k, \ell) \in I} \mu(Z_{k, \ell})}$} and analogously for $\mu[\vec Y]$. It is now easy to check that each non-combined transition $Z_{i, j} \trans\tau_{\mathcal U} \mu[\vec Z]$ implies either $X_i \trans\tau_{\mathcal S} \mu[\vec X]$ or there is some $p > 0$, such that $\mu[\vec X] \synid p \cdot \dirac{X_i} + \mu'[\vec X]$, where $(1 \mathord- p) \cdot \dirac{X_i} \trans\tau_{\mathcal S} \mu'[\vec X]$ and $\dirac{Y_j} \trans\tau_{\mathcal T} \mu[\vec Y]$. Suppose that $\mathcal U$ is not guarded, thus, we are given a transition $\dirac{Z_{i, j}} \wtrans{}_{\mathcal U} \mu[\vec Z] \trans\tau_{\mathcal U} \nu[\vec Z] \wtrans{}_{\mathcal U} \dirac{Z_{i, j}}$. It is easy to translate the $\wtrans{}$ parts to
\begin{align}
	\dirac{X_i} &\wtrans{}_{\mathcal S} \mu[\vec X], & \dirac{Y_j} &\wtrans{}_{\mathcal T} \mu[\vec Y], \label{thm:joining:eqn:xi-to-mu} \\
	\nu[\vec X] &\wtrans{}_{\mathcal S} \dirac{X_i}, & \nu[\vec Y] &\wtrans{}_{\mathcal T} \dirac{Y_j}. \label{thm:joining:eqn:nu-to-xi}
\end{align}
Now we destructure $\mu[\vec Z] \trans\tau_{\mathcal U} \nu[\vec Z]$ into a part that is derivable in $\mathcal S$ and a part that idles, i.e.\ let $\gamma[\vec X] \leq \mu[\vec X]$ and $\eta[\vec X] \leq \nu[\vec X]$, such that $\gamma[\vec X] \trans\tau_{\mathcal S} \eta[\vec X]$ and $\mu[\vec X] - \gamma[\vec X] = \nu[\vec X] - \eta[\vec X]$. W.l.o.g.\ we assume that $\gamma[\vec X] \not\synid \emptydistr$, otherwise proceed with transitions in $\mathcal T$ instead. Next, we aim to construct a weak transition
\begin{equation} \label{thm:joining:eqn:xi-to-gamma}
	\dirac{X_i} \wtrans{}_{\mathcal S} \frac{1}{|\gamma[\vec X]|} \cdot \gamma[\vec X].
\end{equation}
The sequence $(\gamma_j^\proc, \gamma_j^\halt)_{j \in \Nat}$ is constructed as:
\begin{align*}
	\gamma_0^\proc &\synid \dirac{X_i}, & \gamma_0^\halt &\synid \emptydistr, \\
	\gamma_{j+1}^\proc &\synid (1 - |\gamma[\vec X]|)^j \cdot (\mu[\vec X] - \gamma[\vec X]) & \gamma_{j+1}^\halt, &\synid (1 - |\gamma[\vec X]|)^j \cdot \gamma[\vec X].
\end{align*}
Notice that $\dirac{X_i} \synid \gamma_0 \synid \gamma_0^\proc + \gamma_0^\halt$ and that
\[
	\gamma[\vec X] \synid \sum_{j \in \Nat} \gamma_j^\halt \synid \sum_{j \in \Nat} (1 - |\gamma[\vec X]|)^j \cdot \gamma[\vec X] \synid \frac{1}{|\gamma[\vec X]|} \cdot \gamma[\vec X].
\]
Now, by \autoref{prp:wtrans-weak-construction} it suffices to assert \raisebox{0pt}[0pt][0pt]{$\gamma_j^\proc \wtrans{}_{\mathcal S} \gamma_{j+1}$} for all $j$. The base case $j = 0$ is solved by \eqref{thm:joining:eqn:xi-to-mu}. Observing that $\nu[\vec X] \synid \eta[\vec X] + (\nu[\vec X] - \eta[\vec X])$, we decompose \eqref{thm:joining:eqn:nu-to-xi} to find
\begin{align}
	\eta[\vec X] &\wtrans{}_{\mathcal S} |\eta[\vec X]| \cdot \dirac{X_i}, & \nu[\vec X] - \eta[\vec X] &\wtrans{}_{\mathcal S} (1 - |\eta[\vec X]|) \cdot \dirac{X_i}, \label{thm:joining:eqn:eta-to-xi}
\end{align}
so we have
\begin{tflalign*}
	& \gamma_{j+1}^\proc & \\
	\synid{} & (1 - |\gamma[\vec X]|)^j \cdot (\mu[\vec X] - \gamma[\vec X]) & \\
	\synid{} & (1 - |\gamma[\vec X]|)^j \cdot (\nu[\vec X] - \eta[\vec X]) & \mu[\vec X] - \gamma[\vec X] \synid \nu[\vec X] - \eta[\vec X] \\
	\wtrans{}_{\mathcal S}{} & (1 - |\gamma[\vec X]|)^{j+1} \cdot \dirac{X_i} & \text{\eqref{thm:joining:eqn:eta-to-xi}} \\
	\wtrans{}_{\mathcal S}{} & (1 - |\gamma[\vec X]|)^{j+1} \cdot \mu[\vec X] & \text{\eqref{thm:joining:eqn:xi-to-mu}} \\
	\synid{} & \gamma_{j+2}. &
\end{tflalign*}
Finally, given \eqref{thm:joining:eqn:xi-to-gamma} and \eqref{thm:joining:eqn:eta-to-xi}, there is a transition:
\[
	\dirac{X_i} \wtrans{}_{\mathcal S} \frac{1}{|\gamma[\vec X]|} \cdot \gamma[\vec X] \trans\tau_{\mathcal S} \frac{1}{|\eta[\vec X]|} \cdot \eta[\vec X] \wtrans{}_{\mathcal S} \dirac{X_i},
\]
which contradicts $\mathcal S$ being guarded.
\end{proof}

\noindent
There is one additional remark worth mentioning: The proof of \autoref{thm:joining} applies the axiom \axiom{R5}, which originally served only the purpose of reducing from arbitrary to probabilistically guarded expressions! Consequently, our guardedness notion turns out to be too weak in this context; in fact, one could wish for an additional condition preventing certain non-initial $\tau$ transitions. For the sake of convenience, we have not introduced such a condition, as it doubtlessly leads to a considerable overhead of checking additional properties in each step of the completeness proof.

Nevertheless, note that probabilistic guardedness still proves itself necessary altogether, as it suffices to guarantee a unique solution of equations up to $\wcong$ and results in a simple version of \axiom{R2}.

\subsection{Unique solution of equations} \label{sec:completeness:sec:unique-solution}

\noindent
The last step of the completeness proof asserts the existence and uniqueness of equation's solutions: For every SES $\mathcal S$, there exists a solution $E$ of $\mathcal S$, and, provided that $\mathcal S$ is guarded, any expression $F$ satisfying $\mathcal S$, too, is provably equal to $E$: $E = F$.

In the case of a single equation, both properties are guaranteed by the laws \axiom{R1} and \axiom{R2}, respectively; the general case follows by an easy inductive argument. The additional requirement that $E$ be a perfect solution of $\mathcal S$ only seemingly leads to complications: In fact, the original proof \cite{Milner1989b} already constructs a perfect solution, without explicitly mentioning it.

\begin{restatable}[Unique solution of equations up to $=$]{theorem}{uniquesolutionequ} \label{thm:unique-solution-equ}
Let $\mathcal S$ be a guarded SES. There is a perfect solution $E$ of $\mathcal S$ with free variables in $\Var(\mathcal S)$. Furthermore, if $F$ satisfies $\mathcal S$, then $E = F$.
\end{restatable}
\begin{proof}
See \autoref{apx:unique-solution}.
\end{proof}

\subsection{Assembling all parts} \label{sec:completeness:sec:assembling}

\noindent
We are finally ready to obtain the completeness theorem! The proof follows the preceding sections exactly, without bringing up any surprises.

\begin{theorem}[Completeness] \label{thm:completeness}
If $E \wcong F$, then $E = F$.
\end{theorem}
\begin{proof}
By \autoref{thm:guardedness}, assume that both $E$ and $F$ are guarded. Next, by \autoref{thm:equational-characterisation} and \autoref{thm:saturation}, we obtain guarded, saturated SESs $\mathcal S$ and $\mathcal T$, satisfied by $E$ and $F$, respectively. \autoref{thm:unique-solution-equ} finds a perfect solution $E' = E$ (resp.\ $F' = F$) of $\mathcal S$ (resp.\ $\mathcal T$), which satisfies the joint guarded SES $\mathcal U$, given by \autoref{thm:joining}. Finally, \autoref{thm:unique-solution-equ} yields $E = E' = F' = F$.
\end{proof}

\section{Conclusion and further work} \label{sec:conclusion}

\noindent
In this paper, we established a sound and complete axiomatisation of infinitary probabilistic weak bisimilarity --- a conservative extension of the nonprobabilistic setting. It turned out that allowing infinitary transitions is not only desirable from a modelling perspective, as randomised and deterministic implementations of the same problem should usually be indistinguishable, but also enables sound and complete axioms even (or especially) in the presence of unguarded recursion.

The main difficulty was obviously given by the completeness proof. Although most transformations followed Milner's steps, asserting properties like guardedness of ESs led to a tremendous overhead, since weak transitions $\wtrans{}$ were here defined as infinite sequences.

We are confident that this approach carries over nicely to the other points in the branching time spectrum \cite{vanGlabbeek1993a}. Further work might address for instance the development of an axiomatisation of branching bisimilarity as an extension of \cite{vanGlabbeek1993b} or focus on the interesting connection of infinitary and divergence-sensitive semantics \cite{Lohrey2002}.

Moreover, we only considered finite-state systems, so follow-up work could study the consequences of adding parallel composition and restriction operators.

\appendix
\section{Soundness of \axiom{R2}} \label{apx:soundness-r2}

\noindent
Recall that \axiom{R2} guarantees that, for expressions $E$ and $G$, whenever $G = E \subst G X$, in which case we say that $G$ is a solution of the \emph{equation} $E$ in variable $X$, then $G = \rec X E$, provided that $X$ is probabilistically guarded in $E$.
For $H$ a potentially different solution of $E$, we must have that $G = \rec X E = H$, therefore, \axiom{R2} expresses an interesting property of $=$ and $\wcong$ consequently, namely that any equation $E$ has a unique solution up to $\wcong$.

However, note that the premise that $X$ be probabilistically guarded in $E$ cannot be omitted. It might be helpful to understand the following counterexample:
\[
	E \synid \tau\prefix (\psingle X \pchoice{1/4} \psingle{a\prefix \nil \nchoice \tau\prefix X}),
\]
here $E \unguarded X$. Both $G$ and $H$, where
\begin{align*}
	G &\synid \rec X \tau\prefix (\psingle X \pchoice{1/4} \psingle{a\prefix \nil \nchoice \tau\prefix X}), \\
	H &\synid b\prefix \nil \nchoice \rec X \tau\prefix (\psingle{b\prefix \nil \nchoice X} \pchoice{1/4} \psingle{a\prefix \nil \nchoice \tau\prefix (b\prefix \nil \nchoice X)}),
\end{align*}
turn out to be solutions of $E$. Yet, clearly $G \not\wcong H$, since $H$ contains the action $b$, which does neither occur in $G$, nor in $E$. So how is this possible? Intuitively, $E$ not guarding $X$ yields a weak transition reaching $X$ with full probability $1$. Hence, the alien action $b$ sneaking into the left-hand side of the equation may now be simulated by some transition \raisebox{0pt}[0pt][0pt]{$\wtrans b$} of the right-hand side, although $b$ does not occur in $E$, originally. Fortunately, any expression $E$ guarding $X$ prevents this effect.

Before we prove that every guarded equation has a unique solution, we need to gather some additional results related to weak transitions and guardedness.

\begin{table}
\caption{Inference rules for strong unguardedness.} \label{tab:strongunguardedness}
\hsep
\begin{center}
\hfill
	\AxiomC{\vphantom{$E \blacktriangleright V$}}
	\UnaryInfC{$X \blacktriangleright X$}
	\DisplayProof
\hfill
	\AxiomC{$E \blacktriangleright X$}
	\RightLabel{$Y \neq X$}
	\UnaryInfC{$\rec Y E \blacktriangleright X$}
	\DisplayProof
\hfill
	\AxiomC{$E \blacktriangleright X$}
	\UnaryInfC{$E \nchoice F \blacktriangleright X$}
	\DisplayProof
\hfill
	\AxiomC{$F \blacktriangleright X$}
	\UnaryInfC{$E \nchoice F \blacktriangleright X$}
	\DisplayProof
\hspace*{\fill}
\end{center}
\hsep
\end{table}

\begin{definition}[Strong guardedness]\label{def:strongunguardedness}\rm
A free variable $X$ is said to occur \emph{strongly unguarded} in $E$ if $E \blacktriangleright X$ according to \autoref{tab:strongunguardedness}.
\end{definition}

\begin{lemma}\label{strongly unguarded transitions}
If $E \blacktriangleright X$ and $G \trans{\alpha} \mu$ then $E\subst G X \trans{\alpha} \mu$.
\end{lemma}
\begin{proof}
A straightforward induction on the derivation of $E \blacktriangleright X$ from the rules of
\autoref{tab:strongunguardedness}, using the rules of \autoref{tab:semantics}.
\end{proof}

\begin{corollary}\label{strongly unguarded weak transition E}
If $E \blacktriangleright X$ and $G \wtrans{\alpha} \mu$ then $E\subst G X \wtrans{\alpha} \mu$.
\end{corollary}

\noindent
Write $\pi \blacktriangleright X$ iff $E \blacktriangleright X$ for all $E \in\Supp\pi$.

\begin{corollary}\label{strongly unguarded weak transitions}
If $\pi \blacktriangleright X$ and $|\pi|\cdot \dirac{G} \wtrans{\alpha} \mu$ then $\pi\subst G X \wtrans{\alpha}\mu$.
\end{corollary}

\noindent
Write $\pi \blacktriangleright V$ if (1) for each $X\in V$ there is an $E\in\Supp\pi$ with $E\blacktriangleright X$
and (2) for each $E\in\Supp\pi$ there is an $X\in V$ with $E\blacktriangleright X$.
A term $F$ is a \emph{$+$-resolution} of a term $E$ if $F$ is obtained from $E$ by recursively
replacing each subterm $E_1+E_2$ of $E$ by either $E_1$ or $E_2$. Thus, a $+$-resolution is $+$-free
and hence has at most one outgoing transition (to a $+$-free distribution).

\begin{lemma}\label{lem:characterisation unguardedness}
  $E \unguarded V$ iff $F \wtrans{} \pi$ for some $+$-resolution $F$ of $E$ and a distribution
  $\pi\in\Distr{\Exp}$ with $\pi\blacktriangleright V$.
\end{lemma}
\begin{proof}
A straightforward induction on $E$.
\end{proof}

\begin{corollary}\label{cor:characterisation unguardedness resolution}
  $E \unguarded X$ iff $F \wtrans{} \pi$ for some $+$-resolution $F$ of $E$ and a distribution
  $\pi\in\Distr{\Exp}$ with $\pi\blacktriangleright X$.
\end{corollary}

\begin{corollary}\label{cor:characterisation unguardedness}
  $E \unguarded X$ iff $E \wtrans{} \pi$ for a distribution
  $\pi\in\Distr{\Exp}$ with $\pi\blacktriangleright X$.
\end{corollary}

\begin{lemma} \label{degree of guardedness}
For each $E \in \Exp$ with $E \not\unguarded X$ there exists a $p_E > 0$ such that if $E \wtrans{} p \cdot \pi_1 + (1\mathord-p) \cdot \pi_2$ with $\pi_i \in \Distr\Exp$ and $\pi_2 \blacktriangleright X$ then $p \geq p_E$.
\end{lemma}
\begin{proof}
Let $E\in\Exp$ be such that there is no such $p_E>0$. As there are only finitely many states reachable
from $E$, the distributions reachable from $E$ can be seen as points in a finite-dimensional
Euclidean space. According to  \cite{Deng2009} the set $\Pi:= \{\pi\in\Distr{\Exp} \mid E
\wtrans{}\pi\}$ is compact (i.e.\ bounded and topologically closed).
Now consider an infinite sequence $(\mu_i)_{i\in\Nat}$ in $\Pi$ where $\mu_i$ has the form 
$p \cdot \pi_1 + (1\mathord-p) \cdot \pi_2$ with $\pi_2 \blacktriangleright X$ and $p<\frac{1}{i}$.
Since each infinite sequence in a compact space has a convergent subsequence, which has a limit
within that space, there must be a distribution $\mu\in\Pi$ with $\mu\blacktriangleright X$.
This contradicts \autoref{cor:characterisation unguardedness}.
\end{proof}

\noindent
In the following statements $E,F,G,H \in\Exp$ and $\mu,\nu,\pi,\gamma\in\SubDistr{\Exp}$.
Moreover, we assume $G, H, \mu, \nu$ to be closed, while $E$, $F$, $\pi$ and $\gamma$ are allowed to
contain only the free variable $X$.

The next lemma says that a transition from $\pi\subst GX$ can always
be decomposed in a part that stems from $\pi$ and a part originating
from $G$.

\begin{lemma}\label{transition decomposition}
$\pi \subst G X \trans{\alpha} \mu$ iff, for some $\pi_1$, $\pi_2$, $\pi'_1$ and $\mu'$,
$\pi \synid \pi_1 + \pi_2$,
$\pi_1 \trans{\alpha} \pi'_1$, $\pi_2 \blacktriangleright X$, $|\pi_2|\cdot \dirac G \trans{\alpha}\mu'$
and $\mu \synid \pi'_1\subst G X + \mu'$.
\end{lemma}
\begin{proof}
Obviously, $\pi_1 \trans{\alpha} \pi'_1$ implies  $\pi_1 \subst G X \trans{\alpha} \pi'_1 \subst G X$.
Using this, as well as \autoref{strongly unguarded transitions} (or \autoref{strongly unguarded weak transitions}
with $\trans{\alpha}$ instead of $\wtrans{\alpha}$),
``if'' (i.e.\ the implication from right to left) follows immediately from the
convex lifting of $\trans{}$ to a relation between subdistributions.

``Only if'': The transition $\pi \subst G X \mathbin{\trans{\alpha}} \mu$ can be decomposed into transitions
\mbox{$\pi_1 \subst G X \mathbin{\trans{\alpha}} \mu_1$} and $\pi_2 \subst G X \trans{\alpha} \mu_2$ with
$\pi \synid \pi_1+\pi_2$ and $\mu \synid \mu_1+\mu_2$, where the transitions contributing to 
$\pi_2 \subst G X \trans{\alpha} \mu_2$ are obtained through rules \textbf{rec},
\textbf{choice-l} and \textbf{choice-r} from a transition with source $G$,
and the ones in $\pi_1 \subst G X \trans{\alpha} \mu_1$ are not.
It then follows that $\pi_2 \blacktriangleright X$, $|\pi_2|\cdot \dirac{G} \trans{\alpha} \mu_2$ and
$\mu_1$ has the form $\pi_1'\subst G X$, where $\pi_1\trans{\alpha}\pi'_1$.
\end{proof}

\noindent
The following lemma adapts the idea from \autoref{transition decomposition} to weak transitions.
Any weak transition from $\pi\subst G X \wtrans{} \mu$ can be decomposed in a part stemming from
$\pi$ and a part originating from $G$. Although the lemma is intuitively obvious, its proof involves
a lot of bookkeeping. We do not believe this can be done much simpler.

\begin{lemma}\label{weak transition decomposition}
$\pi\subst G X \wtrans{} \mu$ iff, for some $\pi_1$, $\pi_2$ and $\mu'$, $\pi \wtrans{} \pi_1 + \pi_2$,
$\pi_2 \blacktriangleright X$, $|\pi_2|\cdot\dirac G \wtrans{\tau}\mu'$
and $\mu \synid \pi_1\subst G X + \mu'$.
\end{lemma}

\begin{proof}
Obviously, $\pi \wtrans{} \pi'$ implies  $\pi \subst G X \wtrans{} \pi' \subst G X$.
Using this, as well as \autoref{strongly unguarded weak transitions},
``if'' follows immediately from the linearity, reflexivity and transitivity of $\wtrans{}$
(Propositions~\ref{prp:linearity-decomposition} and \ref{prp:preorder}).

``Only if'': Let $\pi\subst G X \wtrans{} \mu$. Then there exists a derivation
$(\mu_i^\proc, \mu_i^\halt)_{i \in \Nat}$ such that
$\pi\subst G X \synid \mu_0 \synid \mu_0^\proc + \mu_0^\halt$ and $\mu \synid \sum_{i \in \Nat} \mu_i^\halt$. 
We now define inductively a derivation $(\gamma_i^\proc, \gamma_i^\halt)_{i \in \Nat}$,
such that $\gamma_i\subst G X \leq \mu_i$
for all $i\in\Nat$.

Let $\gamma_0 \synid \pi$, so that $\gamma_0\subst G X \synid \mu_0$.

Given $\gamma_i$ with $\gamma_i\subst G X \leq \mu_i$,
let $\gamma_i^?$ and $\gamma_i^{\halt!}$ be such that
\begin{align*}
	\gamma_i^{\halt!}\subst G X \synid \mu_i^\halt \cap \gamma_i\subst G X &&
	\text{and} &&
	\gamma_i \mathbin{\synid} \gamma_i^? + \gamma^{\halt!}.
\end{align*}
Here $\mu\cap\gamma\in\SubDistr{S}$ is defined by
$(\mu\mathop\cap\gamma)(s) \mathbin{:=} \min(\mu(s),\gamma(s))$ for all $s\mathbin\in S$.
Now
\begin{align*}
	\gamma_i^{\halt!}\subst G X \leq \mu_i^\halt &&
	\text{and} &&
	\gamma_i^?\subst G X \leq \mu_i^\proc.
\end{align*}
(Namely, for each $H\mathbin\in\CExp$, if $\mu_i^\halt (H) \geq \gamma_i\subst G X (H)$
 then $\gamma_i^?\subst G X (H)=\gamma_i\subst G X (H) - \gamma_i^{\halt!}\subst G X (H) = 0 \leq  \mu_i^\proc(H)$,
 and if $\mu_i^\halt (H) < \gamma_i\subst G X (H)$ then
 $\gamma_i^?\subst G X (H)= \gamma_i\subst G X(H) - \mu_i^\halt (H) \leq \mu_i(H) -  \mu_i^\halt (H) \linebreak[3]= \mu_i^\proc (H)$.)
In the special case that $k=0$ we have $\gamma_0^{\halt!}\subst G X \synid \mu_0^\halt$ and $\gamma_0^?\subst G X \synid \mu_0^\proc$.
By decomposition (cf.\ \autoref{prp:linearity-decomposition}), using that $\mu_i^\proc \trans{\tau} \mu_{i+1}$,
there is a $\mu^?_{i+1} \mathbin\leq \mu_{i+1}$ with \mbox{$\gamma_i^?\subst G X \trans{\tau} \mu^?_{i+1}$} and
$\mu_i^\proc - \gamma_i^?\subst G X \trans{\tau} \mu_{i+1}-\mu^?_{i+1}$.
In the special case that $k=0$ we have $\mu^?_{1} \synid \mu_{1}$.

By \autoref{transition decomposition}, using that $\gamma_i^?\subst G X \trans{\tau} \mu^?_{i+1}$,
there are  $\gamma_i^\proc$, $\gamma_i^-$, $\gamma_{i+1}$ and $\mu^-_{i+1}$,\linebreak[3] such that
$\gamma_i^? \synid \gamma_i^\proc + \gamma_i^-$, $\gamma_i^\proc \trans{\tau} \gamma_{i+1}$,
$\gamma_i^- \blacktriangleright X$, $|\gamma_i^-|\cdot\dirac G \trans{\tau} \mu^-_{i+1}$ and
$\mu^?_{i+1} \synid \gamma_{i+1}\subst G X + \mu^-_{i+1}$.
Let $\gamma_i^\halt \synid \gamma_i^- + \gamma_i^{\halt!}$. Then $\gamma_i \synid \gamma_i^\proc + \gamma_i^\halt$.
Moreover, $\gamma_{i+1}\subst G X \leq \mu_{i+1}^? \leq \mu_{i+1}$.
Thus $(\gamma_i^\proc, \gamma_i^\halt)_{i \in \Nat}$ indeed is a derivation with the required property.

It follows that $\pi \wtrans{} \pi':\synid \sum_{i \in \Nat} \gamma_i^\halt$. 
Let $\pi_1\mathbin{\synid} \sum_{i \in \Nat} \gamma_i^{\halt!}$ and $\pi_2\mathbin{\synid} \sum_{i \in \Nat} \gamma_i^-$.
Then $\pi' \synid \pi_1 + \pi_2$ and $\pi_2 \blacktriangleright X$.
Let $\mu^- \synid \sum_{i\geq 1} \mu_i^-$.
Then $|\pi_2|\cdot\dirac G \trans\tau \mu^-$.
It remains to find a $\mu'$ such that $\mu^- \wtrans{\tau}\mu'$
and $\mu \synid \pi_1\subst G X + \mu'$.

By induction on $k\geq 1$ we define $\eta_{ki}$,
$\eta_{ki}^\rightarrow$ and $\eta_{ki}^\halt$, for $1\leq i\leq k$, such that
\begin{align*}
	\eta_{k1} \synid \mu_k^-, &&
	\eta_{ki} \synid \eta_{ki}^\rightarrow + \eta_{ki}^\halt, &&
	\eta_{ki}^\rightarrow \trans{\tau} \eta_{(k+1)(i+1)},
\end{align*}
\vspace{-5ex}
\begin{align*}
	\sum_{i=1}^k \eta_{ki} + \gamma_k\subst G X &\synid \mu_k, &
	\text{and} &&
	\sum_{i=1}^k \eta_{ki}^\halt + \gamma_k^{\halt!}\subst G X &\synid \mu_k^\halt.
\end{align*}
\emph{Induction base:}
Let $\eta_{11} :\synid \mu_1^-$. 
Using that $\mu_1 \synid \mu_1^? \synid \gamma_1\subst G X + \mu_1^-$,
the two leftmost equations are satisfied for $k=1$.
All other statements will be dealt with fully by the induction step.
\\[1ex]
\emph{Induction step:} Suppose $\eta_{ki}$ for $1\leq i\leq k$ are
already known, and $\mu_k \synid \sum_{i=1}^k \eta_{ki} + \gamma_k\subst G X$.
With induction on $i>0$ we define $\eta_{ki}^\halt :\synid \eta_{ki} \cap (\mu_k^\halt - \nu_{ki})$
with $\nu_{ki} :\synid \gamma_k^{\halt!}\subst GX + \sum_{j=1}^{i-1}\eta_{kj}^\halt$ and establish that $\nu_{ki} \leq \mu_k^\halt$.
Namely,
surely $\nu_{k1} \synid \gamma_k^{\halt!}\subst GX \leq \mu_k^\halt$,
and when assuming that $\nu_{ki}\leq \mu_k^\halt$, for some
$1\mathbin\leq i < k$, and
defining $\eta_{ki}^\halt :\synid \eta_{ki} \cap (\mu_k^\halt-\nu_{ki})$
we obtain that $\nu_{k(i+1)} \synid \eta_{ki}^\halt + \nu_{ki} \leq
(\mu_k^\halt-\nu_{ki}) + \nu_{ki} \synid \mu_k^\halt$.

With induction on $i$ we now establish, for  $1\mathbin\leq i\mathbin\leq k$, that
$\sum_{j=i}^k\eta_{kj} + \nu_{ki} \mathbin\geq \mu_k^\halt$.
Here we use the trivial lemma that $\nu+(\eta \cap \mu) \geq \mu$ iff $\nu+\eta \geq \mu$.

\noindent
The induction base ($i\mathbin=1$): Since $\sum_{i=1}^k \eta_{ki} + \gamma_k\subst G X \synid \mu_k \geq \mu_k^\halt$
we also have
\[\sum_{i=1}^k \eta_{ki}  + \nu_{k1} \synid \sum_{j=1}^k \eta_{kj}  + \gamma_k^{\halt!}\subst G X
\synid \sum_{j=1}^k \eta_{kj}  + (\gamma_k\subst G X \cap \mu_k^\halt) \geq \mu^\halt_k.\]

\noindent
The induction step: Assume $\sum_{j=i}^k\eta_{kj} + \nu_{ki} \mathbin\geq \mu_k^\halt$ for some $1\leq i < k$.
Then 
$(\sum_{j=i}^k\eta_{kj} - \eta_{ki}) + \eta_{ki}
\synid \sum_{j=i}^k\eta_{kj} \geq \mu_k^\halt - \nu_{ki}$, so by the trivial lemma also
$\sum_{j=i+1}^k\eta_{kj} + \eta_{kj}^\halt \synid (\sum_{j=i}^k\eta_{kj} - \eta_{ki}) +
(\eta_{ki}\cap(\mu_k^\halt - \nu_{ki})) \geq \mu_k^\halt - \nu_{ki}$, and thus
$\sum_{j=i+1}^k\eta_{kj} + \nu_{k(i+1)}
\geq \mu_k^\halt$.

In particular (by taking $j\mathbin=k$) $\eta_{kk} + \nu_{kk} \mathbin\geq \mu_k^\halt$,
which implies $\eta_{kk}\geq \mu_k^\halt-\nu_{kk}$.
Therefore $\eta_{kk}^\halt \synid \mu_k^\halt-\nu_{kk}$.
Hence $\sum_{j=1}^{k}\eta_{kj}^\halt + \gamma_k^{\halt!}\subst GX \synid \eta_{kk}^\halt + \nu_{kk} \synid \mu_k^\halt$.

Now define $\eta_{ki}^\rightarrow :\synid \eta_{ki} - \eta_{ki}^\halt$, using that $\eta_{ki}^\halt \leq \eta_{ki}$.
This yields $\eta_{ki} \synid \eta_{ki}^\rightarrow + \eta_{ki}^\halt$, and thereby\vspace{-2ex}
\[
	\mu_k^\rightarrow \synid \mu_k - \mu_k^\halt \synid 
	(\sum_{i=1}^k \eta_{ki} + \gamma_k\subst G X) - (\sum_{i=1}^k \eta_{ki}^\halt + \gamma^{\halt!}_k\subst G X) \synid
	\sum_{i=1}^k \eta_{ki}^\rightarrow + \gamma^?_k\subst G X.\vspace{-1ex}
\]
Since $\sum_{i=1}^k \eta_{ki}^\rightarrow \synid
(\mu_k^\rightarrow - \gamma_k^?\subst G X) \trans{\tau}
(\mu_{k+1} - \mu^?_{k+1})$, by decomposition
$\mu_{k+1} - \mu^?_{k+1} \synid \sum_{i=1}^k\eta_{(k+1)(i+1)}$
for some subdistributions $\eta_{(k+1)(i+1)}$
such that $\eta_{ki}^\rightarrow \trans{\tau} \eta_{(k+1)(i+1)}$ for $i=1,\ldots,k$.
Furthermore, define $\eta_{(k+1)1} :\synid \mu_k^-$.
It follows that
\begin{tflalign*}
	& \mu_{k+1} & \\
	\synid{} & \sum_{i=1}^{k} \eta_{(k+1)(i+1)} + \mu^?_{k+1} & \\
	\synid{} & \sum_{i=2}^{k+1} \eta_{(k+1)i} + (\gamma_{k+1}\subst G X + \mu^-_{k+1}) & \\
	\synid{} & \sum_{i=1}^{k+1} \eta_{(k+1)i} + \gamma_{k+1}\subst GX. &
\end{tflalign*}
This ends the inductive definition and proof.

Now, for all $i\geq 1$, let $\theta_i :\synid \sum_{k=i}^\infty \eta_{ki}$,
 $\theta_i^\rightarrow :\synid \sum_{k=i}^\infty \eta_{ki}^\rightarrow$ and
 $\theta_i^\halt :\synid \sum_{k=i}^\infty \eta_{ki}^\halt$.\linebreak[2]
It follows that $\theta_1 \mathbin{\synid} \sum_{k=1}^\infty \eta_{k1} \mathbin{\synid}
\sum_{k=1}^\infty \mu^-_{k} \mathbin{\synid} \mu^-$,~
$\theta_i \mathbin{\synid} \theta_i^\rightarrow+\theta_i^+$, and
$\theta_i^\rightarrow \trans{\tau} \theta_{i+1}$.
Moreover, $\sum_{i=1}^\infty \theta_i^\halt \mathbin{\synid}
 \sum_{i=1}^\infty \sum_{k=i}^\infty \eta^\halt_{ki} \mathbin{\synid}
 \sum_{k=1}^\infty \sum_{i=1}^k \eta_{ki}^\halt \mathbin{\synid}
 \sum_{k=1}^\infty (\mu_k^\halt - \gamma_k^{\halt!}\subst G X)\linebreak[2]
 \synid \sum_{k=0}^\infty (\mu_k^\halt - \gamma_k^{\halt!}\subst G X)
 \synid \mu - \pi_1\subst G X \synid: \mu'$.
So $\mu^-\wtrans{}\mu'$. 
\end{proof}

\noindent
The next lemma implies that if $X$ is guarded in $E$, and $G$ is a solution of $E$ in the variable
  $X$, then, up to weak bisimilarity, the weak transitions that can be performed by $E\subst GX$ are
  entirely determined by $E$. This lemma plays a crucial role in the soundness proof of \axiom{R2}.

\begin{lemma}\label{weak transition decomposition guarded}
Let $E \mathbin{\not\unguarded} X$ and $G \wcong E \subst G X$. Then
  $E\subst G X \wtrans{}\wbis \mu \in \Distr{\Exp}$ iff, for some $\pi_i\in\Distr{\Exp}$ and $p\mathbin\in(0,1]$,
  $E \wtrans{} p\cdot \pi_1 + (1\mathord-p)\cdot\pi_2$,
  $\pi_2 \blacktriangleright X$
  and
  $\mu \wbis \pi_1\subst G X$.
\end{lemma}
\begin{proof}
  ``If'': We inductively construct a sequence $(\mu_i^\proc, \mu_i^\halt)_{i \in \Nat}$
  as occurs in \autoref{prp:wtrans-weak-construction}.
  Let $\mu_0:\synid\mu_0^\proc :\synid E\subst GX$ and $\mu_0^\halt\synid\emptyset$.
  Then $\mu_0^\proc \wtrans{} \mu_1 :\synid \mu_1^\proc + \mu_1^\halt$,
  where $\mu_1^\halt :\synid p\cdot \pi_1 \subst GX$ and
  $\mu_1^\proc:\synid(1\mathord-p)\cdot\pi_2\subst GX$.

  Now assume $\mu_n^\proc \wbis (1\mathord-p)^n\cdot\pi_2\subst G X$.
  Since $G \mathbin{\wcong} E \subst G X$, we have
  $G \wtrans{} p\cdot \eta^\halt + (1\mathord-p)\cdot\eta^\proc$
  where $\eta^\halt \wbis \pi_1\subst GX$ and $\eta^\proc\wbis\pi_2\subst GX$.
  \autoref{strongly unguarded weak transitions} yields
  $\pi_2\subst GX \wtrans{} p\cdot \eta^\halt + (1\mathord-p)\cdot\eta^\proc$.
  So $\mu_n^\proc \wtrans{} p(1\mathord-p)^n\cdot\gamma^\halt + (1\mathord-p)^{n+1}\cdot\gamma^\proc$.
  where $\gamma^\halt \wbis \pi_1\subst GX$ and $\gamma^\proc\wbis\pi_2\subst GX$.
  Take $\mu^\halt_{n+1}:\synid p(1\mathord-p)^n\cdot\gamma^\halt$ and
  $\mu^\proc_{n+1} :\synid (1\mathord-p)^{n+1}\cdot\gamma^\proc$.
  Then $\mu_n^\proc \wtrans{} \mu_{n+1} :\synid \mu_{n+1}^\proc + \mu_{n+1}^\halt$.

  So, by \autoref{prp:wtrans-weak-construction}, $E\subst GX \wtrans{} \sum_{i\in\Nat}\mu_i^\halt \wbis
  \sum_{i\in\Nat}p(1\mathord-p)^i\cdot\pi_1\subst GX \synid \pi_1\subst GX$.

  ``Only if'': Suppose $E\subst G X \wtrans{}\wbis \mu^0 \mathbin\in\Distr{\Exp}$.
  By \autoref{weak transition decomposition}, there are $\pi_i^0\mathbin\in\Distr{\Exp}$,
  $\mu^1\mathbin\in\SubDistr{\Exp}$ and $p,q$ with $p+q\leq 1$
  such that $E \wtrans{} p\cdot\pi_1^0 + q\cdot\pi_2^0$,
  $\pi_2^0 \blacktriangleright X$, $G \wtrans{\tau}\mu^1$
  and $\mu^0 \wbis p\cdot\pi_1\subst G X + q\cdot\mu^1$. Since $\mu^0 \mathbin\in\Distr{\Exp}$
  it follows that $\mu^1\mathbin\in\Distr{\Exp}$ and $q=1\mathord-p$.
  By \autoref{degree of guardedness}, $p\geq p_E$.

  As $G \wcong E \subst G X$, $E\subst GX \wtrans{\tau}\wbis\mu^1$.
  Continuing inductively, we find, for all $k\in\Nat$,
  $\pi_i^k,\mu^{k+1}\in\Distr{\Exp}$ and $p_k\geq p_E$
  such that $E \wtrans{} p_k\cdot\pi_1^k + (1\mathord-p_k)\cdot\pi_2^k$,
  $\pi_2^k \blacktriangleright X$, $G \wtrans{\tau}\mu^{k+1}$
  and $\mu^k \wbis p_k\cdot\pi_1^k\subst G X + (1\mathord-p_k)\cdot\mu^{k+1}$.
  Thus $\mu^0 \wbis \sum_{k\in\Nat} \prod_{i=0}^{k-1} (1\mathord-p_i) \cdot p_k \cdot \pi^k_1\subst GX$.
  So let $\pi_1:\synid  \sum_{k\in\Nat} \prod_{i=0}^{k-1} (1\mathord-p_i) \cdot p_k \cdot \pi^k$.
  It remains to find $\pi_2\in\Distr{\Exp}$ and $p \in (0,1]$ such that
  $E \wtrans{} p\cdot \pi_1 + (1\mathord-p)\cdot\pi_2$ and $\pi_2 \blacktriangleright X$.

  Let $c=\sum_{k\in\Nat} \prod_{i=0}^{k-1} (1\mathord-p_i)$.
  Then $c\cdot p_E \leq \sum_{k\in\Nat} \prod_{i=0}^{k-1} (1\mathord-p_i)\cdot p_k = 1$, so $c$ is finite.
  It follows that $\dirac E \synid  \frac{1}{c}\sum_{k\in\Nat} \prod_{i=0}^{k-1} (1\mathord-p_i)\cdot \dirac E$.
  Since $E \wtrans{} p_k\cdot\pi_1^k + (1\mathord-p_k)\cdot\pi_2^k$, by linearity
  (\autoref{prp:linearity-decomposition})\linebreak
  $E \wtrans{} \frac{1}{c}\sum_{k\in\Nat} (\prod_{i=0}^{k-1} (1\mathord-p_i))\cdot( p_k\cdot\pi_1^k +
  (1\mathord-p_k)\cdot\pi_2^k) \synid
  \frac{1}{c}\pi_1 + \frac{1}{c}\sum_{k\in\Nat} (\prod_{i=0}^{k} (1\mathord-p_i))\cdot\pi_2^k$.
  We now choose $p:=\frac{1}{c} \in (0,1]$ and
  $\pi_2 :\synid \frac{1}{c-1}\sum_{k\in\Nat} (\prod_{i=0}^{k} (1\mathord-p_i))\cdot\pi_2^k$,
  since $\sum_{k\in\Nat} (\prod_{i=0}^{k} (1\mathord-p_i)) = \sum_{k>0} (\prod_{i=0}^{k-1} (1\mathord-p_i)) = c-1$.
  Now $E \wtrans{} p\cdot \pi_1 + (1\mathord-p)\cdot\pi_2$, using that $1-\frac 1c =\frac{c\mathord-1}c$.
  Finally, $\pi_2 \blacktriangleright X$.
\end{proof}

\noindent
The following two lemmas are generalisation of Lemmas~\ref{weak transition decomposition}
and~\ref{weak transition decomposition guarded} from weak transitions $\wtrans{}$ to $\wtrans{\alpha}$.

\begin{lemma}\label{transition decomposition full}
$F\subst G X \wtrans{\alpha} \mu$ iff, for some $\pi_1$, $\pi_2$, $\pi'_1$, $\mu_1$ and $\mu_2$,
$F \wtrans{} \pi_1 + \pi_2$, $\pi_1 \trans{\alpha} \pi'_1$,
$\pi_2 \blacktriangleright X$, $\pi'_1 \subst GX \wtrans{} \mu_1$,
$|\pi_2|\cdot\dirac G \wtrans{\alpha}\mu_2$
and $\mu \synid \mu_1 + \mu_2$.
\end{lemma}
\begin{proof}
  ``If'' follows just as for Lemmas~\ref{transition decomposition} and~\ref{weak transition decomposition}.

  ``Only if'': Let $F\subst G X \wtrans{\alpha} \mu$. Then
  $F\subst G X \wtrans{}\mu^\dagger\trans{\alpha}\mu^\ddagger \wtrans{} \mu$.
  So by \autoref{weak transition decomposition},
  for some $\pi^\dagger_1$, $\pi^\dagger_2$ and $\mu^\S$, $F \wtrans{} \pi^\dagger_1 + \pi^\dagger_2$,
  $\pi^\dagger_2 \blacktriangleright X$, $|\pi^\dagger_2|\cdot\dirac G \wtrans{\tau}\mu^\S$
  and $\mu^ \dagger \synid \pi^\dagger_1\subst G X + \mu^\S$.
  By decomposition, there are $\mu_1^\ddagger$ and $\mu_2^\ddagger$ such that 
  $\pi^\dagger_1\subst G X \trans\alpha \mu_1^\ddagger$, $\mu^\S\trans\alpha\mu_2^\ddagger$ and $\mu^\ddagger=\mu_1^\ddagger+\mu_2^\ddagger$.
  By \autoref{transition decomposition}, for some $\eta_1$, $\eta_2$, $\eta'_1$ and $\mu''$,
  $\pi^\dagger_1 \synid \eta_1 + \eta_2$,
  $\eta_1 \trans{\alpha} \eta'_1$, $\eta_2 \blacktriangleright X$, $|\eta_2|\cdot \dirac G \trans{\alpha}\mu''$
  and $\mu_1^\ddagger \synid \eta'_1\subst G X + \mu''$.
  Now take $\pi_1:\synid \eta_1$, $\pi_2:\synid \eta_2+\pi^\dagger_2$, $\pi_1' :\synid \eta_1'$ and
  $\mu' :\synid \mu_2^\ddagger+\mu''$. Then
$F \wtrans{} \pi_1 + \pi_2$, $\pi_1 \trans{\alpha} \pi'_1$,
$\pi_2 \blacktriangleright X$, $|\pi_2|\cdot\dirac G \wtrans{}\trans{\alpha}\mu'$
  and $\mu^\ddagger \synid \pi'_1\subst G X + \mu'$.
  Finally, by decomposition (\autoref{prp:linearity-decomposition}), $\mu \synid \mu_1 + \mu_2$,
  $\pi'_1\subst G X\wtrans{}\mu_1$ and $\mu' \wtrans{}\mu_2$.
\end{proof}

\begin{lemma}\label{Transition decomposition guarded}
  Let $E \mathbin{\not\unguarded} X$ and $G \mathbin{\wcong} E \subst G X$.
  Then $E\subst G X \wtrans{\alpha} \wbis \mu \mathbin\in\Distr{\Exp}$ iff,
  for some $\pi_1, \pi_2, \pi'_1\in\Distr{\Exp}$ and $p\mathbin\in(0,1]$,
  $E \wtrans{} p\cdot \pi_1 + (1\mathord-p)\cdot\pi_2$,
  $\pi_1 \trans{\alpha} \pi'_1$, $\pi_2 \blacktriangleright X$,
  and
  $\pi'_1\subst G X \wtrans{}\wbis \mu$.
\end{lemma}

\begin{proof}
``If'': Suppose $E \wtrans{} p\cdot \pi_1 + (1\mathord-p)\cdot\pi_2$,
  $\pi_1 \trans{\alpha} \pi'_1$, $\pi_2 \blacktriangleright X$ and $\pi'_1\subst G X \wtrans{}\wbis \mu$
  for some $\pi_1, \pi_2, \pi'_1\in\Distr{\Exp}$ and $p\mathbin\in(0,1]$.
  By \autoref{weak transition decomposition guarded}, 
  $E\subst G X \wtrans{} \wbis \pi_1\subst GX$.
  Moreover, $\pi_1\subst GX \trans{\alpha} \pi'_1\subst GX \wtrans{}\wbis\mu$.
  Thus $E\subst G X \wtrans{\alpha} \wbis \mu$.

  ``Only if'': Suppose $E\subst G X \wtrans{\alpha}\wbis \mu \synid: \mu^0 \mathbin\in\Distr{\Exp}$.
  By \autoref{transition decomposition full}, there are $\pi_1^0,\pi_2^0,\pi_3^0\mathbin\in\Distr{\Exp}$,
  $\mu_1^0,\mu^1\mathbin\in\SubDistr{\Exp}$ and $p_0,q$ with $p_0+q\leq 1$
  such that $E \wtrans{} p_0\cdot\pi_1^0 + q\cdot\pi_2^0$, $\pi_1^0 \trans{\alpha} \pi_3^0$,
  $\pi_2^0 \blacktriangleright X$, $\pi^0_3 \subst GX \wtrans{} \mu^0_1$, $G \wtrans{\alpha}\mu^1$
  and $\mu^0 \wbis p_0\cdot\mu_1^0 + q\cdot\mu^1$. Since $\mu^0 \mathbin\in\Distr{\Exp}$
  it follows that $\mu^0_1,\mu^1\mathbin\in\Distr{\Exp}$ and $q=1\mathord-p_0$.
  By \autoref{degree of guardedness}, $p_0\geq p_E$.

  As $G \wcong E \subst G X$, $E\subst GX \wtrans{\alpha}\wbis\mu^1$ by \autoref{prp:wcong-wtrans}.
  Continuing inductively, we find, for all $k\mathbin\in\Nat$,
  $\pi_i^k,\mu_1^k,\mu^{k+1}\mathbin\in\Distr{\Exp}$ and $p_k\geq p_E$
  such that \mbox{$E \wtrans{} p_k\cdot\pi_1^k + (1\mathord-p_k)\cdot\pi_2^k$},
  \mbox{$\pi_1^k \mathbin{\trans{\alpha}}\pi_3^k$},
  \mbox{$\pi_2^k {\blacktriangleright} X$}, \mbox{$\pi^k_3 \subst GX \mathbin{\wtrans{}} \mu^k_1$},
  \mbox{$G \mathbin{\wtrans{\alpha}}\mu^{k+1}$}
  and \mbox{$\mu^k \wbis p_k{\cdot}\mu_1^k + (1\mathord-p_k){\cdot}\mu^{k+1}$}.
  Let $\pi_1:\synid  \sum_{k\in\Nat} \prod_{i=0}^{k-1} (1\mathord-p_i) \cdot p_k \cdot \pi_1^k$,
  $\pi'_1:\synid  \sum_{k\in\Nat} \prod_{i=0}^{k-1} (1\mathord-p_i) \cdot p_k \cdot \pi_3^k$
  and $\mu_1:\synid  \sum_{k\in\Nat} \prod_{i=0}^{k-1} (1\mathord-p_i) \cdot p_k \cdot \mu_1^k$.
  Then $\pi_1 \trans{\alpha} \pi'_1$ and $\pi'_1\subst GX \wtrans{} \mu_1 \wbis \mu$.
  It remains to find $\pi_2\in\Distr{\Exp}$ and $p \in (0,1]$ such that
  $E \wtrans{} p\cdot \pi_1 + (1\mathord-p)\cdot\pi_2$ and $\pi_2 \blacktriangleright X$.
  This goes exactly as in the proof of \autoref{weak transition decomposition guarded}.
\end{proof}

\noindent
Using these lemmas we can now obtain the main result of this appendix.

\begin{theorem}[Unique solution of equations up to ${\wcong}$] \label{thm:unique-solution-obs}
Let $E \not\unguarded X$. If $G \wcong E \subst G X$ and $H \wcong E \subst H X$, then $G \wcong H$.
\end{theorem}
\begin{proof}
We show that
\[
	\mathcal R = \{(F \subst G X, F \subst H X) \mid \Var(F) \subseteq \{X\}\}
\]
is a two-sided rooted weak bisimulation up to $\wbis$,
for then, by choosing $F \synid X$, we obtain the desired statement by \autoref{lem:wcong-up-to}.
By symmetry, it is enough to prove
\begin{enumerate}[label=($\dagger$)]
\item whenever $F \subst G X \wtrans\alpha \mu$, then there exists a $\nu$, such that $F \subst H X \wtrans\alpha \nu$ and $\mu \wbis\mathrel{\mathcal R}\wbis \nu$. \label{thm:unique-solution-obs:itm:simulate-alpha}
\end{enumerate}
Here, just as in \autoref{def:wcong-up-to}, we only consider convergent weak transitions, so we can safely assume $|\mu| = 1$.

Following Milner \cite{Milner1989a}, we first prove the following similar property:
\begin{enumerate}[label=($\ddagger$)]
\item whenever $\pi \subst G X \wtrans{} \mu$, then there exists a $\nu$, such that $\pi \subst H X \wtrans{} \nu$ and $\mu \wbis\mathrel{\mathcal R}\wbis \nu$. \label{thm:unique-solution-obs:itm:simulate-tau}
\end{enumerate}
We start with the case $\pi \synid \dirac{E}$. So let $E \subst G X \wtrans{} \mu$.
By \autoref{weak transition decomposition guarded}, for some $\pi_i$ and $p\mathbin\in(0,1]$,
  $E \wtrans{} p\cdot \pi_1 + (1\mathord-p)\cdot\pi_2$,
  $\pi_2 \blacktriangleright X$
  and
  $\mu \wbis \pi_1\subst G X$.
Again by \autoref{weak transition decomposition guarded},
$E \subst H X \wtrans{}\wbis \pi_1\subst H X$. Moreover,
$\mu \wbis\pi_1\subst G X \mathrel{\mathcal{R}}\pi_1\subst H X$.

Now consider the case $\pi \not\synid \dirac{E}$. So let $\pi \subst G X \wtrans{} \mu$.
By \autoref{weak transition decomposition}, for some $\pi_1$, $\pi_2$ and $\mu'$, $\pi \wtrans{} \pi_1 + \pi_2$,
$\pi_2 \blacktriangleright X$, $|\pi_2|\cdot\dirac G \wtrans{\tau}\mu'$
and $\mu \synid \pi_1\subst G X + \mu'$.
Since $G \wcong E \subst G X$ there is a $\mu''$ such that
\mbox{$|\pi_2|\cdot\dirac{E \subst G X} \wtrans{}\mu''$} and $\mu' \wbis \mu''$.
So by the case considered above there exists a $\nu''$, such that
$|\pi_2|\cdot\dirac{E \subst H X} \wtrans{} \nu''$ and
$\mu'' \wbis\mathrel{\mathcal R}\wbis \nu''$.
Since $H \wcong E \subst H X$ there is a $\nu'$ such that
$|\pi_2|\cdot\dirac H \wtrans{}\nu'$ and $\nu'' \wbis \nu'$.
By \autoref{weak transition decomposition} (from right to left)
$\pi\subst H X \wtrans{} \nu :\synid \pi_1\subst H X + \nu'$.
Moreover, $\mu \wbis\mathrel{\mathcal R}\wbis \nu$.

From \ref{thm:unique-solution-obs:itm:simulate-tau} we also obtain
\begin{enumerate}[label=($\S$)]
\item whenever $\mu_1 \wbis\mathrel{\mathcal R}\wbis \nu_1$ and $\mu_1 \wtrans{} \mu$, then there
  exists a $\nu$, such that $\nu_1 \wtrans{} \nu$ and $\mu \wbis\mathrel{\mathcal R}\wbis \nu$. \label{thm:unique-solution-obs:closure}
\end{enumerate}

Using this, we establish \ref{thm:unique-solution-obs:itm:simulate-alpha}.
We start with the case $F \synid E$. So let $E \subst G X \wtrans{\alpha} \mu$.
By \autoref{Transition decomposition guarded},
for some $\pi_1, \pi_2, \pi'_1\in\Distr{\Exp}$ and $p\mathbin\in(0,1]$,
  $E \wtrans{} p\cdot \pi_1 + (1\mathord-p)\cdot\pi_2$,
  $\pi_1 \trans{\alpha} \pi'_1$, $\pi_2 \blacktriangleright X$,
  and
  $\pi'_1\subst G X \wtrans{}\wbis \mu$.
  By the case considered above there exists a $\nu$, such that $\pi \subst H X \wtrans{} \nu$ and
  $\mu\wbis\mathrel{\mathcal R}\wbis \nu$.
Again by \autoref{Transition decomposition guarded},
$E \subst H X \wtrans{\alpha} \wbis \nu$.

Now consider the case $F \not\synid E$. So let $F \subst G X \wtrans{\alpha}\mu$.
By \autoref{transition decomposition full}, for some $\pi_1$, $\pi_2$, $\pi'_1$, $\mu_1$ and $\mu_2$,
$F \wtrans{} \pi_1 + \pi_2$, $\pi_1 \trans{\alpha} \pi'_1$,
$\pi_2 \blacktriangleright X$, \mbox{$\pi'_1 \subst GX \wtrans{} \mu_1$},
$|\pi_2|\cdot\dirac G \wtrans{\alpha}\mu_2$
and $\mu \synid \mu_1 + \mu_2$.
Since $G \wcong E \subst G X$ there is a $\mu'_2$ such that
$|\pi_2|\cdot\dirac{E \subst G X} \wtrans{\alpha}\mu'_2$ and $\mu_2 \wbis \mu'_2$.
So by the case considered above there exists a $\nu_1$ such that
$\pi'_1 \subst HX \wtrans{} \nu_1$ and $\mu_1 \wbis\mathrel{\mathcal R}\wbis \nu_1$,
as well as a $\nu'_2$, such that
$|\pi_2|\cdot\dirac{E \subst H X} \wtrans{} \nu'_2$ and
$\mu'_2 \wbis\mathrel{\mathcal R}\wbis \nu'_2$.
Since $H \wcong E \subst H X$ there is a $\nu_2$ such that
$|\pi_2|\cdot\dirac H \wtrans{\alpha}\nu_2$ and $\nu'_2 \wbis \nu_2$.
By \autoref{transition decomposition full} (from right to left)
$F\subst H X \wtrans{\alpha} \nu_1 + \nu_2$.
Moreover, $\mu_2 \wbis\mathrel{\mathcal R}\wbis \nu_2$ and thus
$\mu_1+\mu_2 \wbis\mathrel{\mathcal R}\wbis \nu_1+\nu_2$.
\end{proof}

\noindent
The soundness of \axiom{R2} now follows easily; essentially, \axiom{R2} finds the canonical solution $\rec X E$ to the equation $E$ in variable $X$.

\soundnessrtwo*
\begin{proof}
Without loss of generality we may assume that $\Var(E) \subseteq \{X\}$ and $\Var(F) = \emptyset$. Let $F \wcong E \subst F X$, where $E$ guards $X$. Now $\rec X E \wcong E \subst{\rec X E}{X}$, by the soundness of \axiom{R1}, hence, \autoref{thm:unique-solution-obs} yields $F \wcong \rec X E$.
\end{proof}

\section{Equational characterisation} \label{apx:equational-characterisation}

\noindent
To be prepared properly, we first establish some useful foundations:

\begin{definition}[Cut] \label{def:cut}\rm
Let $\mathcal S, \mathcal T$ be equation systems over the formal variables $\vec X$. A weak transition $\mu \wtrans{}_{\mathcal T} \pi + \gamma$ is called a \emph{cut through} $\mu \wtrans{}_{\mathcal S} \nu$, if the following conditions are satisfied:
\begin{enumerate}[label=(\roman*)]
\item $\pi \leq \nu$, \label{def:cut:itm:partial}
\item $\gamma = \sum_{i \in \Nat} \gamma_i$ and there is some $\gamma' = \sum_{i \in \Nat} \gamma_i'$, such that for each $i$, \mbox{$\gamma_i \trans\tau_{\mathcal S} \gamma_i'$}; we call $\gamma_i$ the \emph{cut points}, \label{def:cut:itm:cut-points-existance}
\item for each $i$, $\gamma_i \trans\tau_{\mathcal S} \gamma_i'$ is \emph{not (even partially) derivable} in $\mathcal T$, that is, there is no $\emptyset \neq \xi \leq \gamma_i$, such that $\xi \trans\tau_{\mathcal T} \xi' \leq \gamma_i'$, and \label{def:cut:itm:cut-points-non-derivable}
\item $\mu \wtrans{}_{\mathcal S} \pi + \gamma'$ is an initial segment of $\mu \wtrans{}_{\mathcal S} \nu$. \label{def:cut:itm:initial-segment}
\end{enumerate}
\end{definition}

\noindent
The formal definition might seem deterrent, however, the intuition behind cuts is easily explained: Let $\mu \wtrans{}_{\mathcal S} \nu$; now try to mimic this transition in $\mathcal T$. In general this is not possible, since we could reach points $\gamma_i$ where the original transition continues with $\gamma_i \trans\tau_{\mathcal S} \gamma_i'$, but $\gamma_i \trans\tau_{\mathcal T} \gamma_i'$ does not hold. At these points, we simply stop. The remaining transition is of the shape $\mu \wtrans{}_{\mathcal T} \pi + \gamma$, where $\pi$ is the remainder of $\nu$ that could actually be reached in $\mathcal T$ and $\gamma$ is the sum of all cut points.

Intuitively, it is clear that there is a cut through any transition $\mu \wtrans{}_{\mathcal S} \nu$. Nevertheless, a precise proof is not obvious, due to the demanding specification given in \autoref{def:cut}.

\begin{lemma} \label{lem:cut-existence}
Let $\mathcal S, \mathcal T$ be SESs over the formal variables $\vec X$ and let $\mu \wtrans{}_{\mathcal S} \nu$. There are subdistributions $\pi$ and $\gamma$, such that $\mu \wtrans{}_{\mathcal T} \pi + \gamma$ is a cut through $\mu \wtrans{}_{\mathcal S} \nu$. Furthermore, if $\mu \wtrans\tau_{\mathcal S} \nu$, then $\mu \wtrans\tau_{\mathcal T} \pi + \gamma$.
\end{lemma}
\begin{proof}
First, observe that the following property holds:
\begin{enumerate}[label=($*$)]
\item If $\mu \trans\alpha_{\mathcal S} \nu$, then there is a split $\mu \synid \mu' + \mu''$ and $\nu \synid \nu' + \nu''$, such that $\mu' \trans\alpha_{\mathcal T} \nu'$ holds, but $\mu'' \trans\alpha_{\mathcal S} \nu''$ is not (even partially) derivable in $\mathcal T$, i.e.\ there is no $\emptyset \neq \xi \leq \mu''$ with $\xi \trans\alpha_{\mathcal T} \xi' \leq \nu''$.\label{lem:cut-existence:itm:strong-split}
\end{enumerate}
Initially, let $\mu'' \synid \mu$, $\nu'' \synid \nu$ and $\mu' \synid \nu' \synid \emptydistr$. Now suppose that there is some $\emptyset \neq \xi \leq \mu''$ with $\xi \trans\alpha_{\mathcal T} \xi' \leq \nu''$, then take $\xi$ to be maximal and reduce $\mu''$ by $\xi$ and $\nu''$ by $\xi'$. Repeat this procedure until $\mu'' \trans\alpha_{\mathcal S} \nu''$ is no longer derivable in $\mathcal T$.

\hspace{-6pt}%
We continue with the proof of the actual statement. Given $\mu \mathbin{\wtrans{}_{\mathcal S}} \nu$, there is a derivation $(\mu_i^\proc\!, \mu_i^\halt)_{i \in \Nat}$, i.e.\vspace{-1ex}
\begin{equation}
	\mu_i^\proc \trans\tau_{\mathcal S} \mu_{i+1} \synid \mu_{i+1}^\proc + \mu_{i+1}^\halt, \label{lem:cut-existence:eqn:mu-derivation}
\end{equation}
such that $\mu \synid \mu_0 \synid \mu_0^\proc + \mu_0^\halt$ and $\sum_{i \in \Nat} \mu_i^\halt \synid \nu$. We construct a sequence $(\nu_i^\proc, \nu_i^\halt + \gamma_i)_{i \in \Nat}$ as follows: In each step, using \ref{lem:cut-existence:itm:strong-split}, we cut off a maximal branch $\gamma_i$ that is not derivable in $\mathcal T$; this branch stops in the new derivation. We are going to define $\nu_i^\proc, \nu_i^\halt, \gamma_i$ and the auxiliary values $\eta_i$ recursively, as shown in the following diagram; the dashed connections represent transitions that are not (even partially) derivable in $\mathcal T$.
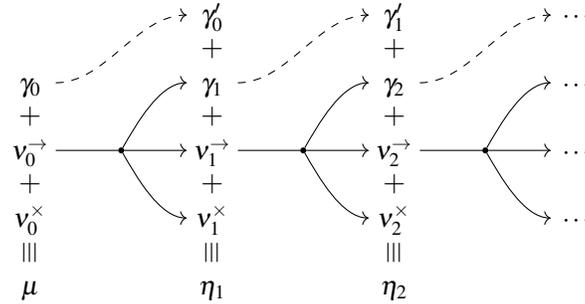
\begin{center}
\begin{tikzpicture}[relation diagram, every edge/.append style={shorten <=.34cm, shorten >=.34cm}, splitter/.append style={outer sep=-.34cm}]
\matrix[relation matrix, row sep={.9cm,between origins}, column sep={2.42cm,between origins}] (m) {
	& \gamma_0' & \gamma_1' & {\cdots} \\
	\gamma_0 & \gamma_1 & \gamma_2 & {\cdots} \\
	\nu_0^\proc & \nu_1^\proc & \nu_2^\proc & {\cdots} \\
	\nu_0^\halt & \nu_1^\halt & \nu_2^\halt & {\cdots} \\
	\mu & \eta_1 & \eta_2 & \\
};

\foreach \x [evaluate={\X=int(1+\x);}] in {1, 2, 3}{
	\path (m-2-\x.center) edge[transition, out=0, in=180, dashed] (m-1-\X.center);
	
	\node[splitter] (*1) at ($(m-3-\x)!.5!(m-3-\X)$) {};
	\path (m-3-\x.center) edge[] (*1);
	\path (*1) edge[transition, out=60, in=180] (m-2-\X.center);
	\path (*1) edge[transition] (m-3-\X.center);
	\path (*1) edge[transition, out=-60, in=-180] (m-4-\X.center);
}

\foreach \y [evaluate={\Y=int(1+\y);}] in {2, 3}
\node at ($(m-\y-1.south)!.5!(m-\Y-1.north)$) {${+}$};

\foreach \x in {2, 3}
\foreach \y [evaluate={\Y=int(1+\y);}] in {1, 2, 3}
\node at ($(m-\y-\x.south)!.5!(m-\Y-\x.north)$) {${+}$};

\foreach \x in {1, 2, 3}
\node[rotate=90] at ($(m-4-\x.south)!.5!(m-5-\x.north)$) {${\synid}$};
\end{tikzpicture}
\end{center}

For the base case, let $\eta_1 \synid \mu_1$. We instantiate \eqref{lem:cut-existence:eqn:mu-derivation} to obtain
\begin{equation}
	\mu_0^\proc \trans\tau_{\mathcal S} \eta_1. \label{lem:cut-existence:eqn:mu-derivation-0}
\end{equation}
Then, we define
\begin{align*}
	\nu_0^\halt &\synid \mu_0^\halt, &
	\nu_0^\proc + \gamma_0 &\synid \mu_0^\proc,
\end{align*}
where we split \eqref{lem:cut-existence:eqn:mu-derivation-0} by \ref{lem:cut-existence:itm:strong-split} into
\begin{align}
	\nu_0^\proc &\trans\tau_{\mathcal T} \eta_1 - \gamma_0' &
	\gamma_0 &\trans\tau_{\mathcal S} \gamma_0', \label{lem:cut-existence:eqn:split-0}
\end{align}
where the second transition is not (even partially) derivable in $\mathcal T$.

So let $i > 0$ and assume that all values are defined up to $i - 1$. Observe that $\gamma_{i-1}' \leq \eta_i \leq \mu_i$, thus, by some basic reasoning about subprobability distributions, there is a split $\gamma_{i-1}' \synid \gamma_{i-1}'^\proc + \gamma_{i-1}'^\halt$ and $\eta_i \synid \eta_i^\proc + \eta_i^\halt$, such that $\gamma_{i-1}'^\proc \leq \eta_i^\proc \leq \mu_i^\proc$ and $\gamma_{i-1}'^\halt \leq \eta_i^\halt \leq \mu_i^\halt$. Now we obtain $\eta_i^\proc - \gamma_{i-1}'^\proc \leq \eta_i^\proc \leq \mu_i^\proc$, so we can rewrite $\mu_i^\proc \synid (\mu_i^\proc - (\eta_i^\proc - \gamma_{i-1}^\proc)) + (\eta_i^\proc - \gamma_{i-1}'^\proc)$. We may decompose \eqref{lem:cut-existence:eqn:mu-derivation} to find some $\eta_{i+1} \leq \mu_{i+1}$, such that
\begin{equation}
	\eta_i^\proc - \gamma_{i-1}^\proc \trans\tau_{\mathcal S} \eta_{i+1}, \label{lem:cut-existence:eqn:mu-derivation-decomp-eta}
\end{equation}
and
\begin{equation}
	\mu_i^\proc - (\eta_i^\proc - \gamma_{i-1}'^\proc) \trans\tau_{\mathcal S} \mu_{i+1} - \eta_{i+1}. \label{lem:cut-existence:eqn:mu-derivation-decomp-rem}
\end{equation}
Next, the $i$-th values are defined as:
\begin{align*}
	\nu_i^\halt &\synid \eta_i^\halt - \gamma_{i-1}'^\halt, &
	\nu_i^\proc + \gamma_i &\synid \eta_i^\proc - \gamma_{i-1}'^\proc,
\end{align*}
where we split along \eqref{lem:cut-existence:eqn:mu-derivation-decomp-eta}, such that
\begin{align}
	\nu_i^\proc &\trans\tau_{\mathcal T} \eta_{i+1} - \gamma_i', &
	\gamma_i &\trans\tau_{\mathcal S} \gamma_i'. \label{lem:cut-existence:eqn:split-i}
\end{align}
Again, the second transition is not (even partially) derivable in $\mathcal T$. The construction is finished.

It remains to show that
\begin{equation}
	\mu \wtrans{}_{\mathcal T} \pi + \gamma \label{lem:cut-existence:eqn:cut}
\end{equation}
with $\pi \synid \sum_{i \in \Nat} \nu_i^\halt$ and $\gamma \synid \sum_{i \in \Nat} \gamma_i$ and that \eqref{lem:cut-existence:eqn:cut} is a cut through $\mu \wtrans{}_{\mathcal S} \nu$. Note that
\[
	\mu \synid \mu_0^\proc + \mu_0^\halt \synid \nu_0^\proc + \nu_0^\halt + \gamma_0,
\]
hence, in order to prove \eqref{lem:cut-existence:eqn:cut}, it suffices to show that $(\nu_i^\proc, \nu_i^\halt + \gamma_i)_{i \in \Nat}$ is a derivation. This result is straightforward: For all $i$, we obtain
\begin{equation}
	\nu_i^\proc \trans\tau_{\mathcal T} \eta_{i+1} - \gamma_i' \synid (\eta_{i+1}^\proc - \gamma_i'^\proc) + (\eta_{i+1}^\halt - \gamma_i'^\halt) \synid \nu_{i+1}^\proc + \nu_{i+1}^\halt + \gamma_{i+1}, \label{lem:cut-existence:eqn:derivation-nu}
\end{equation}
by \eqref{lem:cut-existence:eqn:split-0} and \eqref{lem:cut-existence:eqn:split-i}.

We continue by proving that \eqref{lem:cut-existence:eqn:cut} is indeed a cut; each desired property of \autoref{def:cut} is considered separately: \ref{def:cut:itm:partial} is easily shown by
\[
	\pi \synid \sum_{i \in \Nat} \nu_i^\halt \synid \mu_0^\halt + \sum_{i \geq 1} (\eta_i^\halt - \gamma_{i - 1}'^\halt) \leq \mu_0^\halt + \sum_{i \geq 1} \eta_i^\halt \leq \mu_0^\halt + \sum_{i \geq 1} \mu_i^\halt \synid \nu,
\]
both \ref{def:cut:itm:cut-points-existance} and \ref{def:cut:itm:cut-points-non-derivable} are immediate by \eqref{lem:cut-existence:eqn:split-0} and \eqref{lem:cut-existence:eqn:split-i}. Proving \ref{def:cut:itm:initial-segment} requires a little more effort. First, we need to assert
\begin{equation}
	\mu \wtrans{}_{\mathcal S} \pi + \gamma'. \label{lem:cut-existence:eqn:initial-segment}
\end{equation}
It suffices to check that $(\nu_i^\proc + \gamma_i, \nu_i^\halt + \gamma_{i-1}')_{i \in \Nat}$ is a derivation; let the out-of-index distribution $\gamma_{-1}'$ be $\emptydistr$. As a combination of \eqref{lem:cut-existence:eqn:derivation-nu} and \eqref{lem:cut-existence:eqn:split-i}, we have
\[
	\nu_i^\proc + \gamma_i \trans\tau_{\mathcal T} \nu_{i+1}^\proc + \gamma_{i+1} + \nu_{i+1}^\halt + \gamma_i'.
\]
Additionally, \eqref{lem:cut-existence:eqn:initial-segment} is an initial segment of $\mu \wtrans{} \nu$. We choose $(\nu_i^\proc + \gamma_i, \nu_i^\halt + \gamma_{i-1}')_{i \in \Nat}$ as a derivation for \eqref{lem:cut-existence:eqn:initial-segment} and $(\mu_i^\proc, \mu_i^\halt)_{i \in \Nat}$ as a derivation for $\mu \wtrans{} \nu$. For $i = 0$, we find $\nu_0^\proc + \gamma_0 \synid \mu_0^\proc$ and $\nu_0^\proc + \gamma_0 + \nu_0^\halt + \gamma_{-1}' \synid \mu_0$, immediately. So let $i > 0$, then
\[
	\nu_i^\proc + \gamma_i \synid \eta_i^\proc - \gamma_{i-1}'^\proc \leq \eta_i^\proc \leq \mu_i^\proc
\]
and
\[
	\nu_i^\proc + \gamma_i + \nu_i^\halt + \gamma_{i-1}' \synid (\eta_i^\proc - \gamma_{i-1}'^\proc) + (\eta_i^\halt - \gamma_{i-1}'^\halt) + \gamma_{i-1}' \synid \eta_i \leq \mu_i.
\]
We are nearly finished: The last condition holds for $i = 0$ by
\[
	\mu_0^\proc - (\nu_0^\proc + \gamma_0) \synid \emptydistr \trans\tau_{\mathcal S} \emptydistr \synid \mu_1 - \mu_1 \synid \mu_1 - \eta_1 \synid \mu_1 - (\nu_1^\proc + \gamma_1 + \nu_1^\halt + \gamma_0'),
\]
and for $i > 0$, we exploit \eqref{lem:cut-existence:eqn:mu-derivation-decomp-rem}:
\[
	\mu_i^\proc - (\nu_i^\proc + \gamma_i) \synid \mu_i^\proc - (\eta_i^\proc - \gamma_{i-1}'^\proc) \trans\tau_{\mathcal S} \mu_{i+1} - \eta_{i+1} \synid \mu_{i+1} - (\nu_{i+1}^\proc + \gamma_{i+1} + \nu_{i+1}^\halt + \gamma_i').
\]

There is one small part missing: Let $\mu \wtrans\tau_{\mathcal S} \nu$, then we are obliged to show that $\mu \wtrans\tau_{\mathcal T} \pi + \gamma$. By \autoref{prp:wtrans-tau}, we may equivalently assume that $\mu_0^\halt \synid \emptydistr$. Clearly, $\nu_0^\halt + \gamma_{-1}' \synid \emptydistr$, too, thus, we are finished after applying \autoref{prp:wtrans-tau} once more.
\end{proof}

\equationalcharacterisation*
\begin{proof}
We strengthen the statement by additionally asserting the following property for all $V \subset \Var$:
\begin{enumerate}[label=(§)]
\item If $E \not\unguarded V$, then there exists no $\mu$, such that $X_1 \wtrans{}_{\mathcal S} \mu$ and each $X \in V$ occurs in some expression supported by $\mu[\vec S]$. \label{thm:equational-characterisation:itm:prop}
\end{enumerate}
Here, $\mathcal S: \vec X = \vec S$. The statement is proven by an induction on the structure of $E$. Sometimes, we write $\mathcal S \union \mathcal T$ to denote the union of two ESs $\mathcal S, \mathcal T$; the result is an ES containing every equation either occurring in $\mathcal S$ or in $\mathcal T$.

$E \synid \nil$ or $E \synid X$: Let $\mathcal S$ contain the single equation $X_1 = E$.

$E \synid \alpha\prefix \psum_{i=1}^n {p_i} F_i$: We inductively obtain SESs $\mathcal S_i$ with leading equations $X_{i, 1} = S_{i, 1}$, satisfied by $F_i$, respectively. Choose $\mathcal S = \{X_1 = \alpha\prefix \psum_{i=1}^n {p_i} X_{i, 1}\} \union \bigcup_{i=1}^n \mathcal S_i$ with distinguished variable $X_1$. $\mathcal S$ is obviously standard, both the guardedness property and \ref{thm:equational-characterisation:itm:prop} follow by induction.

$E \synid F \nchoice G$: We obtain SESs $\mathcal T$ and $\mathcal U$ with leading equations $Y_1 = T_1$ and $Z_1 = U_1$, satisfied by $F$ and $G$, respectively. We construct $\mathcal S = \{X_1 = T_1 \nchoice U_1\} \union \mathcal T \union \mathcal U$ with distinguished variable $X_1$; again, all properties follow easily by induction.

$E \synid \rec X F$: We find an SES $\mathcal T: \vec X = \vec T$ satisfied by $F$ by induction. Since $E$ is guarded, we infer that $X$ is guarded in $F$. The variable $X$ cannot occur in $T_1$, due to \ref{thm:equational-characterisation:itm:prop}. We now construct $\mathcal S : \vec X = \vec S$, where $\vec S = \vec T \subst{T_1}{X}$; note that $X$ does not occur in $\mathcal S$ anymore. By \axiom{R1}, we conclude that $E$ provably satisfies $\mathcal S$. Moreover, it is easy to see that $\mathcal S$ is standard; the other properties require more effort.

First, we show by contradiction that $\mathcal S$ is guarded. Thus, suppose there is some weak transition
\begin{equation}
	\dirac{X_k} \wtrans\tau_{\mathcal S} \dirac{X_k}. \label{thm:equational-characterisation:eqn:xk-to-xk}
\end{equation}
By \autoref{lem:cut-existence}, there is a cut
\begin{equation}
	\dirac{X_k} \wtrans\tau_{\mathcal T} \pi + \gamma \label{thm:equational-characterisation:eqn:gamma-pi-cut}
\end{equation}
with $\gamma \synid \sum_{i \in \Nat} \gamma_i$ through \eqref{thm:equational-characterisation:eqn:xk-to-xk}. Since $\pi \leq \dirac{X_k}$ and $|\pi| + |\gamma| = 1$, we may rewrite \eqref{thm:equational-characterisation:eqn:gamma-pi-cut} into:
\begin{equation}
	\dirac{X_k} \wtrans\tau_{\mathcal T} (1 - |\gamma|) \cdot \dirac{X_k} + \gamma. \label{thm:equational-characterisation:eqn:gamma-cut}
\end{equation}
We know that $\mathcal T$ is probabilistically guarded by induction, so there is no transition $\dirac{X_k} \wtrans\tau_{\mathcal T} \dirac{X_k}$. Hence, $\gamma \not\synid \emptydistr$. From now on, it is no longer necessary to distinguish between $\wtrans{}$ and $\wtrans\tau$ transitions. By \eqref{thm:equational-characterisation:eqn:gamma-cut} being a cut, we know that there are subdistributions $\gamma_i'$, such that $\gamma_i \trans\tau \gamma_i'$ is derivable in $\mathcal S$, but not (even partially) in $\mathcal T$. Note that the only transitions derivable by $\mathcal S$, but not by $\mathcal T$, arise from the substitution $\subst{T_1}{X}$. Thus, for all $i$ and for each $X_j \in \Supp{\gamma_i}$, we conclude that $X$ occurs in $T_j$. Moreover, the cut \eqref{thm:equational-characterisation:eqn:gamma-cut} asserts that
\begin{equation}
	\dirac{X_k} \wtrans{}_{\mathcal S} (1 - |\gamma|) \cdot \dirac{X_k} + \gamma' \label{thm:equational-characterisation:eqn:gamma-initial-segment}
\end{equation}
is an initial segment of \eqref{thm:equational-characterisation:eqn:xk-to-xk}, where $\gamma' \synid \sum_{i \in \Nat} \gamma_i'$. By \autoref{prp:initial-segment}, there is some continuation of \eqref{thm:equational-characterisation:eqn:gamma-initial-segment}, such that
\begin{equation}
	(1 - |\gamma|) \cdot \dirac{X_k} + \gamma' \wtrans{}_{\mathcal S} \dirac{X_k}. \label{thm:equational-characterisation:eqn:gamma-continuation}
\end{equation}
Now, by decomposition and linearity, we obtain
\begin{equation}
	\frac{1}{|\gamma'|} \cdot \gamma' \wtrans{}_{\mathcal S} \dirac{X_k}. \label{thm:equational-characterisation:eqn:gamma-to-xk}
\end{equation}
We may again apply \autoref{lem:cut-existence} in order to obtain a cut
\begin{equation}
	\frac{1}{|\gamma'|} \cdot \gamma' \wtrans{}_{\mathcal T} (1 - |\nu|) \cdot \dirac{X_k} + \nu \label{thm:equational-characterisation:eqn:nu-cut}
\end{equation}
through \eqref{thm:equational-characterisation:eqn:gamma-to-xk} with cut points $\nu_i$, such that $\nu = \sum_{i \in \Nat} \nu_i$; again, we could immediately rewrite the remaining part of $\dirac{X_k}$ as $(1 - |\nu|) \cdot \dirac{X_k}$. Following the same argumentation as above, we observe that whenever $X_j \in \Supp{\nu_i}$ for some $i$, then $X$ occurs in $T_j$.

There is one last piece missing: Recall that each transition $\gamma_i \trans\tau \gamma_i'$ is derived by expressions $T_1$ substituting the variable $X$. Thus, there is also a transition
\begin{equation}
	\dirac{X_1} \trans\tau_{\mathcal T} \frac{1}{|\gamma'|} \cdot \gamma'. \label{thm:equational-characterisation:eqn:x1-to-gamma}
\end{equation}

We are now ready to construct the weak transition leading to a contradiction:
\begin{equation}
	\dirac{X_1} \wtrans{}_{\mathcal T} \nu + \frac{1 - |\nu|}{|\gamma|} \cdot \gamma. \label{thm:equational-characterisation:eqn:x1-to-nu-gamma}
\end{equation}
The sequence $(\mu_i^\proc, \mu_i^\halt)_{i \in \Nat}$ is constructed with
\begin{align*}
	\mu_0^\proc &\synid \dirac{X_1}, & \mu_0^\halt &\synid \emptydistr, \\
	\mu_1^\proc &\synid \tfrac{1}{|\gamma'|} \cdot \gamma', & \mu_1^\halt &\synid \emptydistr, \\
	\mu_2^\proc &\synid (1 - |\nu|) \cdot \dirac{X_k}, & \mu_2^\halt &\synid \nu, \\
	\mu_{i+3}^\proc &\synid (1 - |\nu|) \cdot (1 - |\gamma|)^{i + 1} \cdot \dirac{X_k}, & \mu_{i+3}^\halt &\synid (1 - |\nu|) \cdot (1 - |\gamma|)^i \cdot \gamma.
\end{align*}
Note that $\mu_0 \synid \mu_0^\proc + \mu_0^\halt \synid \dirac{X_1}$ and
\[
	\sum_{i \in \Nat} \mu_i^\halt \synid \nu + \sum_{i \in \Nat} (1 - |\nu|) \cdot (1 - |\gamma|)^i \cdot \gamma \synid \nu + \frac{1 - |\nu|}{|\gamma|} \cdot \gamma,
\]
since $|\gamma| > 0$. By \autoref{prp:wtrans-weak-construction}, it suffices to show that $\mu_i^\proc \wtrans{} \mu_{i + 1} \synid \mu_{i + 1}^\proc + \mu_{i + 1}^\halt$ for all $i$ in order to prove \eqref{thm:equational-characterisation:eqn:x1-to-nu-gamma}. The cases $i = 0$ and $i = 1$ follow from \eqref{thm:equational-characterisation:eqn:x1-to-gamma} and \eqref{thm:equational-characterisation:eqn:nu-cut}, respectively. All remaining cases $i \geq 2$ are proven by \eqref{thm:equational-characterisation:eqn:gamma-cut} and linearity.

It still remains to point out how to cause a contradiction: Recall that both $\gamma[\vec S]$ and $\nu[\vec S]$ support only expressions in which $X$ occurs. Consequently, $(\nu + \frac{1 - |\nu|}{|\gamma|} \cdot \gamma)[\vec S]$ fulfils this property, too. Additionally, we know that $F \not\unguarded \dirac X$, thus, \eqref{thm:equational-characterisation:eqn:x1-to-nu-gamma} contradicts \ref{thm:equational-characterisation:itm:prop}, which is guaranteed to hold for $\mathcal T$ by induction.

We continue by proving that \ref{thm:equational-characterisation:itm:prop} holds for $\mathcal S$, so let $E \not\unguarded V$. Again, we prove the statement by contradiction: Suppose there is some transition
\begin{equation}
	\dirac{X_1} \wtrans{}_{\mathcal S} \mu, \label{thm:equational-characterisation:eqn:x1-to-nu}
\end{equation}
such that each $Y \in V$ is contained in an expression supported by $\mu[\vec S]$. Observe that $X \not\in V$, as $X$ does not occur in $\mathcal S$, anymore. Besides, note that each $Y \in V$ also occurs in some expression supported by $\mu[\vec T]$, since every variable other than $X$ occurs in $S_j$ if and only if it occurs in $T_j$. We know there exists some cut
\begin{equation}
	\dirac{X_1} \wtrans{}_{\mathcal T} \pi + \gamma \label{thm:equational-characterisation:eqn:gamma-cut:2}
\end{equation}
through \eqref{thm:equational-characterisation:eqn:x1-to-nu}, where $\pi \leq \mu$ and $\gamma \synid \sum_{i \in \Nat} \gamma_i$. Analogously to before, we find that each expression supported by $\gamma[\vec T]$ contains $X$. In addition to that, we may assume w.l.o.g.\ that each $Y \in V$ occurs in some expression supported by $\pi[\vec T]$. Suppose there was some $Y$ contradicting this assumption. Then the branch reaching $Y$ in the original transition \eqref{thm:equational-characterisation:eqn:gamma-cut:2} has been cut off, so we continue cutting the respective continuation until some expression containing $Y$ is reached.

Now, distinguish two cases: First, let $|\gamma| = 1$. Then $\dirac{X_1} \wtrans{}_{\mathcal T} \gamma$, where all expressions in the support of $\gamma[\vec T]$ contain $X$. Thus, \ref{thm:equational-characterisation:itm:prop} entails $F \unguarded X$, which contradicts $E$ being guarded.

So let $|\gamma| < 1$. Then all variables $Y \in V \union \{X\}$ occur in some expression supported by $(\pi + \gamma)[\vec T]$. Again, \ref{thm:equational-characterisation:itm:prop} implies $F \unguarded V \union \{X\}$, so by definition:
\[
	\AxiomC{$F \unguarded V \union \{X\}$}
	\RightLabel{$V \union \{X\} \neq \{X\},$}
	\UnaryInfC{$\rec X F \unguarded V$}
	\DisplayProof
\]
i.e.\ $E \unguarded V$, contradicting our initial assumption. This finishes the proof.
\end{proof}

\section{Unique solution of equations} \label{apx:unique-solution}

\noindent
Up to this point, we exclusively considered standard ESs. This section's theorem destructures a given ES, such that we need to settle with a weaker property: An ES $\mathcal S: \vec X = \vec S$ is called \emph{semi-standard} if, within every subexpression of an expression $S_j$, a formal variable $X_i$ occurs only within subexpressions of the form $\alpha\prefix (\psingle{X_i} \pchoice p P)$. Thus, we still prohibit jumping between variables without taking transitions in between, however, we now permit nested prefixing and recursions in the defining expressions $\vec S$.

For some expression $E$, we define $\Reach{E}$ as the set of all expressions reachable from $E$ with positive probability. Let $\mathcal S: \vec X = \vec S$ be a semi-standard ES. We define $\Reach{\mathcal S}$ to collect all expressions reachable from some $S_j$\footnote{Strictly speaking, we only include expressions reachable from $S_j$, if $X_j$ is reachable from $X_1$. So hereinafter, we implicitly assume that every formal variable is reachable from $X_1$, otherwise simply drop the respective defining equation.}. Moreover, we define a transition relation ${\trans{}_{\mathcal S}} \subseteq \Reach{\mathcal S} \times \Act \times \Distr{\Reach{\mathcal S}}$ as follows:
\[
	E \trans\alpha_{\mathcal S} \mu \quad\text{iff}\quad
	\begin{cases}
		S_i \trans\alpha \mu, &\text{if $E \synid X_i$,}\\
		E \trans\alpha \mu, &\text{otherwise}.
	\end{cases}
\]
Note that this transition relation coincides with the earlier characterisation (\autoref{def:es-standardness}) if $\mathcal S$ is standard: Then $\Reach{\mathcal S} \subseteq \tilde X$, so the second case never occurs.

After defining $\wtrans\alpha_{\mathcal S}$ as usual, we finally generalise our understanding of guardedness: A semi-standard ES $\mathcal S$ is called (probabilistically) guarded if there is no $i$, such that $\dirac{X_i} \wtrans\tau_{\mathcal S} \dirac{X_i}$.

\uniquesolutionequ*
\begin{proof}
The theorem is proven by showing a stronger claim: Contrary to what is required above, we let $\mathcal S: \vec X = \vec S$ be not necessarily standard, but semi-standard. We wish to find expressions $\vec E = (E_1, \ldots, E_n)$, such that $\vec E = \vec S \subst{\vec E}{\vec X}$. However, we do not require $E :\synid E_1$ to be a perfect solution of $\mathcal S$ (after all, we have not introduced an adapted definition), but rather show the following property:
\begin{enumerate}[label=($*$)]
\item there is a bijection $b: \Reach{\mathcal S} \to \Reach E$, with $b(S_1) = E$,
such that for all $G \in \Reach{\mathcal S}$, $b(G) \trans\alpha \mu$ if and only if $G \trans\alpha_{\mathcal S} \mu \comp b$. \label{thm:unique-solution-equ:itm:bijection}
\end{enumerate}
Here, $\comp$ denotes functional composition. Intuitively, \ref{thm:unique-solution-equ:itm:bijection} states that the PA induced by $E$ and $\mathcal S$, respectively, are isomorphic. In the case that $\mathcal S$ is standard, \ref{thm:unique-solution-equ:itm:bijection} implies that $E$ is a perfect solution of $\mathcal S$.

We perform an induction on the number of equations $n$.

If $n = 1$, then $\mathcal S$ contains the single equation $X_1 = S_1$. We choose $E \synid E_1 \synid \rec{X_1} S_1$. Then $E$ is a valid solution of $\mathcal S$, since
\[
	E \synid \rec{X_1} S_1 = S_1 \subst{\rec{X_1} S_1}{X_1} \synid S_1 \subst E {X_1},
\]
due to \axiom{R1}. Obviously, $\Reach E = \{ G \subst{\rec{X_1} S_1}{X_1} \mid G \in \Reach{\mathcal S}\}$, so we assign $b(G) \synid G \subst{\rec{X_1} S_1}{X_1}$. It is easy to check that \ref{thm:unique-solution-equ:itm:bijection} holds.

Now let $F$ satisfy $\mathcal S$, too, i.e.\ $F = S_1 \subst F {X_1}$. We know that $\mathcal S$ is guarded, so $X_1$ is guarded in $S_1$. Hence, \axiom{R2} yields $F = \rec{X_1} S_1 \synid E$.

So let $n > 1$ and let $\mathcal S$ contain the equations $\vec X = \vec S$ and $X_n = S_n$. We are now constructing expressions $E_i$, such that:
\begin{align*}
	\vec E &= \vec S \substtwo{\vec E}{\vec X}{E_n}{X_n}, &
	E_n &= S_n \substtwo{\vec E}{\vec X}{E_n}{X_n}.
\end{align*}
To obtain these, we define $\vec T \synid \vec S \subst{\rec{X_n} S_n}{X_n}$ at first. Consider the ES $\mathcal T: \vec X = \vec T$ containing $n - 1$ equations. One can easily check that $\mathcal T$ is semi-standard and for now, assume that $\mathcal T$ is guarded; we will show this fact later. Then, by induction, we obtain $n - 1$ expressions $\vec E$, such that $\vec E = \vec T \subst{\vec E}{\vec X}$. We continue by assigning $E_n \synid \rec{X_n} S_n \subst{\vec E}{\vec X}$. By using \axiom{R1} and some straight-forward identities involving substitution, we conclude that
\begin{align*}
	\vec E &= \vec T \subst{\vec E}{\vec X} & E_n &\synid \rec{X_n} S_n \subst{\vec E}{\vec X} \\
	&\synid \vec S \subst{\rec{X_n} S_n}{X_n} \subst{\vec E}{\vec X} & &= S_n \subst{\rec{X_n} S_n}{X_n} \subst{\vec E}{\vec X} \\
	&\synid \vec S \subst{\rec{X_n} S_n \subst{\vec E}{\vec X}}{X_n} \subst{\vec E}{\vec X} & &\synid S_n \subst{\rec{X_n} S_n \subst{\vec E}{\vec X}}{X_n} \subst{\vec E}{\vec X} \\
	&\synid \vec S \subst{E_n}{X_n} \subst{\vec E}{\vec X} & &\synid S_n \subst{E_n}{X_n} \subst{\vec E}{\vec X} \\
	&\synid \vec S \substtwo{\vec E}{\vec X}{E_n}{X_n}, & &\synid S_n \substtwo{\vec E}{\vec X}{E_n}{X_n}.
\end{align*}
It remains to assert \ref{thm:unique-solution-equ:itm:bijection}. Note that $\Reach{\mathcal T} = \{ G \subst{\rec{X_n} S_n}{X_n} \mid G \in \Reach{\mathcal S}\}$. So we define a bijection $b': \Reach{\mathcal S} \to \Reach{\mathcal T}$ with $b'(G) \synid G \subst{\rec{X_n} S_n}{X_n}$ and assert that $b'(G) \trans\alpha \mu$ iff $G \trans\alpha \mu \comp b'$: Distinguish the following cases for $G \in \Reach{\mathcal S}$.

If $G \synid X_n$, then $b'(X_n) \synid \rec{X_n} S_n$. Obviously, $\rec{X_n} S_n \trans\alpha_{\mathcal T} \mu$ iff $S_n \trans\alpha \mu \comp b'$ iff $X_n \trans\alpha_{\mathcal S} \mu \comp b'$.

If $G \synid X_i \not\synid X_n$, then $b'(X_i) \synid X_i$. Hence, $X_i \trans\alpha_{\mathcal T} \mu$ iff $T_i \trans\alpha \mu$ iff $S_i \subst{\rec{X_n} S_n}{X_n} \trans\alpha \mu$ iff $X_i \trans\alpha_{\mathcal S} \mu \comp b'$.

If $G \not\synid X_i$ for all $i$, then $b'(G) \synid G \subst{\rec{X_n}{S_n}}{X_n}$. Since $\mathcal T$ is semi-standard, we conclude that the substitution does not introduce fresh transitions. Thus, $G \subst{\rec{X_n}{S_n}}{X_n} \trans\alpha_{\mathcal T} \mu$ iff $G \trans\alpha_{\mathcal S} \mu \comp b'$.

By induction, we obtain some $b'': \Reach{\mathcal T} \to \Reach E$ satisfying \ref{thm:unique-solution-equ:itm:bijection}. So we choose $b$ as the composition $b'' \comp b'$, which clearly meets the condition \ref{thm:unique-solution-equ:itm:bijection}. Moreover, the bijection $b'$ entails that $\mathcal T$ is guarded: Suppose there was some transition $\dirac{X_i} \wtrans\tau_{\mathcal T} \dirac{X_i}$, then we could use $b'$ to translate each participating strong transition into a strong transition in $\mathcal S$. So we would find a transition $\dirac{X_i} \wtrans\tau_{\mathcal S} \dirac{X_i}$, which contradicts $\mathcal S$ being guarded.

Next, assume there are expressions $\vec F$ and $F_n$ satisfying $\mathcal S$, i.e.
\begin{align*}
	\vec F &= \vec S \substtwo{\vec F}{\vec X}{F_n}{X_n}, &
	F_n &= S_n \substtwo{\vec F}{\vec X}{F_n}{X_n}.
\end{align*}
Since $S_n \subst{\vec F}{\vec X}$ guards $X_n$ and $F_n = S_n \subst{\vec F}{\vec X} \subst{F_n}{X_n}$, \axiom{R2} implies $F_n = \rec {X_n} S_n \subst{\vec F}{\vec X}$. Moreover, we refine the $n - 1$ equations:
\[
	\vec F = \vec S \subst{F_n}{X_n} \subst{\vec F}{\vec X} = \vec S \subst{\rec{X_n} S_n}{X_n} \subst{\vec F}{\vec X},
\]
such that $\vec F = \vec T \subst{\vec F}{\vec X}$, i.e.\ $\vec F$ satisfies $\mathcal T$. It follows that $\vec E = \vec F$, by induction. Lastly, we obtain the remaining equality:
\\\mbox{}\hfill
$	F_n = \rec{X_n} S_n \subst{\vec F}{\vec X} = \rec{X_n} S_n \subst{\vec E}{\vec X} = E_n.$
\end{proof}

\bibliographystyle{eptcsini}
\bibliography{references}

\end{document}